\documentclass[12pt,reqno]{amsart}
\usepackage[letterpaper,margin=1.25in]{geometry}                
\usepackage{multirow}
\usepackage{setspace}
\usepackage{graphicx}
\usepackage{amsmath,amssymb,amsthm}
\usepackage{epstopdf}
\usepackage[authoryear]{natbib}
\usepackage{color}
\usepackage{comment}
\allowdisplaybreaks

\newtheorem{theorem}{Theorem}[section]	
	
\newtheorem{prop}{Proposition}[section]	
\newtheorem{assumption}{Assumption}[section]
\newtheorem{lem}{Lemma}[section]
\theoremstyle{definition}
\newtheorem{algo}{Algorithm}[section]

\newtheorem{remark}{Remark}[section]

\DeclareMathOperator{\plim}{plim}

\newcommand{\hF}{\widehat{ F}}
\newcommand{\QTE}{\mathsf{QE}}
\newcommand{\CI}{\mathsf{CI}}

\newcommand{\E}{\mathsf{E} }

\newcommand{\tr}{\mathsf{tr}}
\newcommand{\Var}{\mathsf{Var}}
\newcommand{\Cov}{\mathsf{Cov}}
\newcommand{\rank}{\mathsf{rank}}

 \newcommand{\vecc}{\mathsf{vec}}
 \newcommand{\vbeta}{\boldsymbol{\beta}}
 
 \newcommand{\vtheta}{\boldsymbol{\theta}}
 \newcommand{\veta}{\boldsymbol{\eta}}
\newcommand{\vgamma}{\boldsymbol{\gamma}}
 
 \newcommand{\vx}{\boldsymbol{x}}
 \newcommand{\vz}{\boldsymbol{z}}
 \newcommand{\vw}{\boldsymbol{w}}
 \newcommand{\vb}{\boldsymbol{b}}
 \newcommand{\vv}{\boldsymbol{v}}

\newcommand{\vZ}{\boldsymbol{Z}}

\numberwithin{figure}{section}
 \numberwithin{equation}{section}
 \numberwithin{table}{section}

\usepackage{xr-hyper}
\makeatletter
\newcommand*{\addFileDependency}[1]{
  \typeout{(#1)}
  \@addtofilelist{#1}
  \IfFileExists{#1}{}{\typeout{No file #1.}}
}
\makeatother

\newcommand*{\myexternaldocument}[1]{
    \externaldocument{#1}
    \addFileDependency{#1.tex}
    \addFileDependency{#1.aux}
}

\myexternaldocument{supp_revise}

\begin{document}

\global\long\def\a{\alpha}%

\global\long\def\b{\beta}%

\global\long\def\g{\gamma}%

\global\long\def\d{\delta}%

\global\long\def\e{\epsilon}%

\global\long\def\l{\lambda}%

\global\long\def\t{\theta}%

\global\long\def\o{\omega}%

\global\long\def\s{\sigma}%

\global\long\def\G{\Gamma}%

\global\long\def\D{\Delta}%

\global\long\def\L{\Lambda}%

\global\long\def\T{\Theta}%

\global\long\def\O{\Omega}%

\global\long\def\R{\mathbb{R}}%

\global\long\def\N{\mathbb{N}}%

\global\long\def\P{\mathbb{P}}%
\global\long\def\Q{\mathbb{Q}}%

\global\long\def\X{\mathbb{X}}%

\global\long\def\U{\mathbb{U}}%

\global\long\def\E{\mathbb{E}}%
\global\long\def\N{\mathbb{N}}%

\global\long\def\Q{\mathbb{Q}}%

\global\long\def\cX{\mathscr{X}}%

\global\long\def\cY{\mathscr{Y}}%

\global\long\def\cA{{\cal A}}%

\global\long\def\cB{\mathscr{B}}%

\global\long\def\cM{\mathscr{M}}%
\global\long\def\cG{\mathscr{G}}%
\global\long\def\cF{\mathscr{F}}%

\global\long\def\es{\emptyset}%

\global\long\def\mc#1{\mathscr{#1}}%

\global\long\def\ind{\mathbf{1}}%

\global\long\def\any{\forall}%

\global\long\def\ex{\exists}%

\global\long\def\p{\partial}%

\global\long\def\cd{\cdot}%

\global\long\def\Dif{\nabla}%

\global\long\def\imp{\Rightarrow}%

\global\long\def\iff{\Leftrightarrow}%

\global\long\def\up{\uparrow}%

\global\long\def\down{\downarrow}%

\global\long\def\arrow{\rightarrow}%

\global\long\def\rlarrow{\leftrightarrow}%

\global\long\def\lrarrow{\leftrightarrow}%

\global\long\def\abs#1{\left|#1\right|}%

\global\long\def\norm#1{\left\Vert #1\right\Vert }%

\global\long\def\rest#1{\left.#1\right|}%

\global\long\def\bracket#1#2{\left\langle #1\middle\vert#2\right\rangle }%

\global\long\def\sandvich#1#2#3{\left\langle #1\middle\vert#2\middle\vert#3\right\rangle }%

\global\long\def\turd#1{\frac{#1}{3}}%

\global\long\def\ellipsis{\textellipsis}%

\global\long\def\sand#1{\left\lceil #1\right\vert }%

\global\long\def\wich#1{\left\vert #1\right\rfloor }%

\global\long\def\sandwich#1#2#3{\left\lceil #1\middle\vert#2\middle\vert#3\right\rfloor }%

\global\long\def\abs#1{\left|#1\right|}%

\global\long\def\norm#1{\left\Vert #1\right\Vert }%

\global\long\def\rest#1{\left.#1\right|}%

\global\long\def\inprod#1{\left\langle #1\right\rangle }%

\global\long\def\ol#1{\overline{#1}}%

\global\long\def\ul#1{\underline{#1}}%

\global\long\def\td#1{ \widetilde{#1}}%

\global\long\def\upto{\nearrow}%

\global\long\def\downto{\searrow}%

\global\long\def\pto{\overset{p}{\longrightarrow}}%

\global\long\def\dto{\overset{d}{\longrightarrow}}%

\global\long\def\asto{\overset{a.s.}{\longrightarrow}}%

\title[Dynamic Heterogeneous Distribution Regression]{Dynamic Heterogeneous Distribution Regression Panel Models, with an Application to Labor Income Processes$^*$}
\author[Fern\'andez-Val, Gao, Liao, and Vella]{Iv\'an Fern\'andez-Val, Wayne Yuan Gao, Yuan Liao, and Francis Vella$^\dag$}





\date{This draft: \today}                                           

\thanks{$^*$  We thank the editor Stephane Bonhomme, three anonymous referees, Manuel Arellano, Dmitry Arkhangelsky,  Kirill Evdokimov, Matt Hong, Koen Jochmans,  Hiro Kaido, Roger Koenker, Dennis Kristensen, Robert Moffitt,  Pierre Perron, Zhongjun Qu, Enrique Sentana, Youngki Shin, Allan Timmermann, Xi Wang, Chaowen Zheng, and seminar participants at 6th IAER Econometrics Workshop, 2021 IESR Microeconometrics Workshop, 2022 Econometric Society North America Winter Meeting, BU, Cemfi,  Erasmus, Glasgow, Oxford, Tsinghua, UPF and York for comments.}

\thanks{$^\dag$ Fern\'andez-Val: Department of Economics, Boston University; Gao: Department of Economics, University of Pennsylvania; Liao: Department of Economics, Rutgers University; Vella: Department of Economics, Georgetown University}

\begin{abstract}
We introduce a dynamic distribution regression panel data model with heterogeneous coefficients across units. The objects of primary interest are functionals of these coefficients, including predicted one-step-ahead and stationary cross-sectional distributions of the outcome variable. Coefficients and their functionals are estimated via fixed effect methods. We investigate how these functionals vary in response to counterfactual changes in initial conditions or covariate values. We also identify a uniformity problem related to the robustness of inference to the unknown degree of coefficient heterogeneity, and propose a cross-sectional bootstrap method for uniformly valid inference on function-valued objects. 
We showcase the utility of our approach through an empirical application to individual income dynamics. 
Employing the annual Panel Study of Income Dynamics data, we establish the presence of substantial coefficient heterogeneity. We
then highlight some important empirical questions that our methodology can address. First, we quantify the impact of a negative labor income shock on the distribution of future labor income.
Second, we demonstrate the existence of heterogeneity in income mobility, and its implications for an individuals' incidence to be trapped in poverty. Simulation evidence confirms that our procedures work well in small samples.

\end{abstract}

\maketitle

 
 
   \onehalfspacing

 \section{Introduction}
The panel data literature typically features a somewhat limited treatment of parameter heterogeneity \citep[c.f.,][]{browning2010heterogeneity}. Although 
 random coefficient panel models allow heterogeneous coefficients between units, and some recent developments incorporate heterogeneous coefficients within units, relatively few studies incorporate heterogeneous coefficients both between units (individual heterogeneity) and within units (nonlinearity).\footnote{Exceptions include \cite{chetverikov2016iv}, \cite{okui2019panel}, \cite{zhang2019quantile} and \cite{chen2021quantile}.} 
This paper employs fixed effects distribution regression (DR) to estimate a  
dynamic panel model with coefficient heterogeneity between and within units. The model captures within-unit heterogeneous relationships between outcome and covariates through function-valued coefficients, and between-unit heterogeneity via coefficients which can vary across units in an unrestricted fashion. 
The model facilitates the analysis of various new functionals of the coefficients, including linear projections on unit covariates and predicted distributions.
We can also consider interesting counterfactual scenarios by manipulating the values of the initial conditions of the outcome variable, or the covariates, to examine their impact on these functionals. We can also consider both one-period-ahead and stationary counterfactual distributions to measure the short and long term effects of these changes.


Our flexible treatment of heterogeneity is partially motivated by 
the absence of consensus regarding the degree of heterogeneity required to 
model labor income dynamic processes.  
 For example,  \cite{abowd1989covariance} and \cite{macurdy1982use} considered models with a limited allowance  for heterogeneity between units, whereas \cite{browning2007heterogeneity} and \cite{browning2010modelling}  found some features of income processes, such as variances of the shocks, differ considerably between units. Moreover, \cite{browning2007heterogeneity} and \cite{browning2010modelling}  noted that allowing for between unit heterogeneity has drastic implications for specific empirical questions. Others, including \cite{arellano2017earnings},  allow for flexible heterogeneity within units to capture nonlinear persistence, but restrict the heterogeneity between units.
 
 Modeling the correct degree of heterogeneity is also important for inference as 
 some inference procedures are 
 only valid under specific circumstances. 
We provide a procedure that is 
uniformly  valid over the degree of unknown heterogeneity of the coefficients. 
This covers ``homogeneous", ``partially heterogeneous" (i.e. heterogeneity concentrated on subpopulations or parts of the distribution), and ``completely heterogeneous" models  as  special cases.   The 
theoretical challenge is that the rate of convergence of the estimators depends on the potentially unknown degree of heterogeneity. 
We establish that 
standard analytical plug-in methods are not valid for inference uniformly with respect to the degree of  heterogeneity.  We address 
this via a cross-sectional bootstrap scheme that resamples from the empirical distribution of the estimated coefficients. We show this bootstrap is valid uniformly over various degrees of heterogeneity. Note that a similar uniformity problem arises 
for average partial effects in nonlinear panel data models. \cite{fernandez2016individual}, for example,  bypassed this problem by assuming strong heterogeneity on the partial effects.


We also establish a relationship between dynamic DR models with discrete outcomes and finite-state Markov chains.\footnote{\cite{browning2007heterogeneity,browning2010heterogeneity,browning2014dynamic} previously established a related connection between dynamic binary response models with random coefficients and two-state Markov chains.} One can then 
express objects such as  stationary distributions, mobility probabilities and recurrence times  as functionals of the model coefficients. 

\subsection{The empirical study on labor income dynamics}

Our methodology is applicable to many empirical settings and we employ it here to examine labor income dynamics.  This is  
a large literature, starting with \cite{champernowne53}, \cite{hart1976}, \cite{shorrocks1976} and \cite{lillard1978dynamic}, including many papers featuring econometric innovations.   We 
employ data from the Panel Study of Income Dynamics (PSID)
to perform economically interesting experiments.

First, we  consider how a \textit{ceteris paribus} reduction in annual labor income in a given year, implemented via a negative shock, affects future annual labor income. We find that the predicted effect on the cross-sectional distribution of labor income after one period varies substantially depending on whether 
we account for heterogeneity in the level and persistence of income. Our model predicts substantially smaller effects than existing autoregressive models that restrict between and/or within heterogeneity by imposing several forms of homogeneous coefficients. 


Second, we address the existence of poverty traps. We 
model the 
conditional probabilities of individuals being in poverty in a specific year given they were in poverty in the previous year. We establish that the substantial cross-sectional heterogeneity in the level and persistence of annual labor income has important implications for an individual's tendency to remain in a certain location of the income distribution. 


\subsection{Relationship with existing literature}
From a theoretical perspective, our paper is related to \cite{chernozhukov2013inference} (CFM) and \cite{chernozhukov2018network} (CFW). The former studies DR for cross-sectional data and the latter for panel data with fixed effects. Both flexibly model and estimate counterfactual distributions. We introduce two substantial and important departures from this earlier work. First, whereas all coefficients in CFM and CFW except for the intercept are fixed, we treat all coefficients as random. This
facilitates the analysis of many economically interesting functionals
which cannot be analyzed in the CFM and CFW frameworks. Moreover, our evidence below indicates this coefficient heterogeneity is empirically important to study labor income dynamics. It also introduces  the theoretical challenge of how to perform inference that remains uniformly valid with respect to the degree of coefficient heterogeneity. These issues were not considered in CFM and CFW.  Second, our model is dynamic, whereas those in CFM and CFW are static. This allows us to estimate economically interesting objects 
 related to persistence.

Our model differs from the traditional random coefficients models of \cite{swamy1970}, \cite{hsiao2008random}, \cite{arellano2012identifying}, \cite{fernandez2013panel} and \cite{su2016}, among others, as we allow for heterogeneous coefficients both between and within units. Moreover,  existing distribution and quantile regression models with fixed effects  often allow the intercepts to vary across units but restrict the slopes to be homogeneous; e.g., \cite{Koenker2004}, \cite{Galvao2011}, \cite{GalvaoKato2016smooth}, \cite{KatoGalvaoMontes-Rojas2012}, \cite{ArellanoWeidner2017},   and \cite{chernozhukov2018network}.       \cite{chetverikov2016iv} and \cite{chen2021quantile} develop models similar to ours, but  focus on  projections of coefficients as the objects of interest in static quantile regression models.  Other related recent works are \cite{okui2019panel} and \cite{zhang2019quantile} noting that their models and objects of interest differ from ours.




Bias correction methods based on large-$T$ asymptotic approximations for fixed effects estimators of dynamic and nonlinear panel models have been previously studied in \cite{Nickell81}, \cite{Phillips:1999p733}, \cite{Hahn:2004p882}, \cite{fernandez2009fixed}, \cite{HahnKuersteiner2011}, \cite{dhaene2015split}, and \cite{fernandez2016individual}, among others (see \cite{ArellanoHahn2007} and \cite{fernandez2018fixed} for reviews). We extend these debiasing methods to new functionals of the coefficients.

Inference 
robust to unknown heterogeneity 
is studied 
by \cite{liao2018unifddorm} and \cite{lu2022uniform} 
for linear random coefficient panel models estimated by least squares. Two main differences arise in our approach. First, our DR model is nonlinear and the coefficients are estimated by conditional maximum likelihood methods. This requires treatment of the resulting
incidental parameter problem. Second, 
the model coefficients are infinite-dimensional, whereas those in \cite{liao2018unifddorm} and \cite{lu2022uniform} are finite dimensional. Inference is more challenging as the coefficient functions in our model might exhibit different degrees of heterogeneity at different points of their domain. 
Similar to  \cite{liao2018unifddorm}, we propose a cross-sectional panel bootstrap to make inference that is robust to the degree of coefficient heterogeneity. This method was previously used for panel data as a resampling scheme that preserves the dependence in the time series dimension, e.g.,  \cite{kapetanios2008bootstrap}, \cite{kaffo2014bootstrap}, and \cite{gonccalves2015bootstrap}. We demonstrate that it also has robustness properties in models with heterogeneous coefficients.



Although our empirical work is related to the literature on labor income and earnings processes, many aspects of our results are novel. 
This literature has typically focused on allocating the total error variances into transitory and permanent components. A summary is provided in \cite{moffitt2018income} and three important recent innovations are \cite{arellano2017earnings,arellano2018nonlinear} and \cite{hu2019semiparametric}. The first two examined nonlinear persistence in the permanent component and how it varies over the earnings distribution. The third allowed for a flexible representation of the distributions of both components. Our approach is not intended to supersede these methodologies. Rather, we 
illustrate how our approach can complement the existing literature.  The approach most similar to ours is  \cite{arellano2017earnings,arellano2018nonlinear}, which provided evidence of  nonlinearity in income dynamics.  While  they considered nonlinear persistence that can vary by location in the 
distribution, they do not allow heterogeneity  between units. 
We incorporate income persistence  that can vary  by location in the earnings distribution and also across units. Moreover, we allow persistence to be a function of both observed and unobserved individual characteristics and target different objects including counterfactual distributions and mobility probabilities.  We acknowledge that in many settings it is important to distinguish between permanent and transitory income. Perhaps the leading example is where transitory income reflects measurement error. However, in certain instances one may be primarily concerned with observed income. Particularly when this is the economic object of relevance to the individual or the policy maker. This includes situations in which the transitory component captures an income or macro shock.   

The representation of the income mobility  model as a finite-state Markov chain is motivated by \cite{champernowne53} and \cite{shorrocks1976}, which previously used homogeneous Markov chains.
We 
estimate a separate Markov chain for each unit to allow for unrestricted unit heterogeneity. We can recover the associated transition probabilities and apply standard tools for Markov chains to study stationary distributions and recurrence times. Other related work on income dynamics includes \cite{hirano2002} and \cite{gu2017}, which estimated autoregressive labor income processes using flexible semiparametric Bayesian methods, and  \cite{hoffmann2019hip}, which studied the robustness of model parameters across specifications of the earnings dynamics. Finally, \cite{chamberlain2022feedback} and \cite{lee2025identificationestimationdynamicrandom} study identification and estimation of dynamic random coefficient models in short panels. In that setting, coefficients and their functionals are only partially identified when the time dimension is held fixed.

\subsection{Notation} Let $\mathcal{F}_{it}$ be the filtration defined in Section \ref{subsec:model}. We make use of several expectations. For a sequence of random variables $\{y_{it}: 1 \leq i \leq N, 1 \leq t \leq T\}$, where $i$ indexes cross-sectional units and $t$ time periods, we denote the expectation with respect to the distribution of $y_{it}$ conditional on $\mathcal{F}_{it}$ by $\mathbb E_{it} y_{it}  := \mathbb E( y_{it} \mid \mathcal{F}_{it})$, the cross-sectional expectation at $t$ as  $\mathbb E_{t} y_{it}  := \text{plim}_{N \to \infty} N^{-1} \sum_{i=1}^N y_{it}$, and the cross-sectional and temporal expectation as $\mathbb E y_{it} := \text{plim}_{N,T \to \infty} (NT)^{-1} \sum_{i=1}^N \sum_{t=1}^T y_{it}$, provided that they exist. In what follows we shall assume without qualification that an expectation exists whenever it is used. For two deterministic sequences $a_T, b_T$,  we use the notation $a_T\ll b_T$ and $b_T\gg a_T$ if $a_T=o(b_T)$.

\subsection{Outline} Section \ref{sec:model} presents the model and objects of interest. Section \ref{sec:est} discusses estimation and inference.
We present the empirical application 
in Section \ref{sec:emp} and Section \ref{sec:theory} establishes the associated asymptotic theory.
Section
 \ref{sec:mc} reports simulation evidence. 
Proofs and additional results are gathered in the Appendix. 

\section{The model and objects of interest}\label{sec:model}

 \subsection{The model}\label{subsec:model}
We observe panel data $\{(y_{it},\vx_{it}) : 1 \leq i \leq N, 1 \leq t \leq T\}$, where $i$ indexes observational units and $t$ indexes time periods. The scalar $y_{it}$ represents the outcome; and $\vx_{it}$ is a $d_x$-vector of covariates, which includes a constant, lagged outcome values, and other predetermined covariates denoted by $\boldsymbol{v}_{it}$. That is, 
$$
\vx_{it} =(1, y_{i(t-1)},...,y_{i(t-L)}, \vv_{it}')'.
$$

Let $\mathcal{F}_{it}$ be a filtration  to which   $\vx_{it}$ and any time invariant variable for unit $i$ are adapted. We model the distribution of $y_{it}$ conditional on $\mathcal{F}_{it}$ as, for any $y \in \mathbb{R}$,
\begin{equation}\label{eq:hdr}
\Pr(y_{it} \leq y \mid \mathcal{F}_{it}) = F_{y_{it}}(y \mid \vx_{it}) = \Lambda(- \vx_{it}'\vbeta_i(y)), \quad 1 \leq t \leq T, \quad 1 \leq i \leq N,
\end{equation}
where $\Lambda: \mathbb R\mapsto[0,1]$ is a known, strictly increasing, and four times  continuously  differentiable  link function (e.g., the standard normal or logistic CDF), and $y \mapsto - \vx_{it}'\vbeta_i(y)$ is increasing almost surely (a.s).\footnote{We can replace $\vx_{it}$ by $P(\vx_{it})$ where $P$ is a vector of transformations with good approximation properties such as polynomials, splines or interactions. Our theory cover cases where the dimension of $P$ is fixed with the sample size. One could allow the dimension of $P$ to grow with the sample size and $\Lambda$ to be unknown using semiparametric methods, but we do not pursue those extensions here.}  This is a DR model for panel data with heterogeneous coefficients, which we call a heterogeneous DR model (HDR).


By iterating expectations, the cross-sectional distribution of the observed outcome at time $t$ can be written in terms of the model coefficients as
  \begin{equation}\label{eq:acdf}
   F_{t}(y) := \E_t 1\{y_{it} \leq y  \} = \E_t \E_{it} 1\{y_{it} \leq y  \}  = \E_t \Lambda(- \vx_{it}'\vbeta_i(y)).
  \end{equation}
   This representation serves several purposes. First, as the basis for a specification test of the model where an estimator of $F_{t}(y)$ based on the right hand side of \eqref{eq:acdf} is compared with the cross-sectional empirical distribution of $y_{it}$. Second, when $\vx_{it}$ only includes lagged values of $y_{it}$, we can construct one-period-ahead predicted distributions by setting $t=T+1$. These distributions are useful for forecasting. Third, we can analyze dynamics of the distribution of $y_{it}$ over time. 
   Fourth, we can consider the impact of interventions by comparing the counterfactual distribution after changing $\vx_{it}$ or $\vbeta_i(y)$ with the actual distribution. 

\subsection{Heterogeneous coefficient functions}
  

The main innovation of \eqref{eq:hdr} is that all model coefficients are random functions,
$$
y \mapsto \vbeta_i(y),
$$
where the variation across $i$ and $y$ captures between-unit heterogeneity and within-unit heterogeneity (nonlinearity), respectively. We assume $\vbeta_i(y)$ does not vary over time. 
However, we explore if the heterogeneity is associated with observed unit characteristics using linear projections. Let $\vw_{i}, \vz_{i} \in \mathcal{F}_{i1}$ denote time invariant covariates such that $\dim(\vw_{i})\geq \dim(\vz_{i})$ and $\E(\vw_{i} \vz_{i}')$ has full column rank.  Consider  a linear regression
  \begin{equation}\label{eq:projection}
  \vbeta_i(y) =\vtheta(y) \vz_{i}  +\vgamma_i(y),\quad  \E(  \vgamma_i(y) \mid \vw_{i} )=0,
  \end{equation}
   which covers the standard linear projection when $\vw_{i} = \vz_{i}$, and instrumental variables when $\vw_i$ is the instrument.  The coefficient $\vtheta(y)$ informs which covariates are associated with the heterogeneity in $\vbeta_i(y)$ across $i$, where we allow these relationships to vary within the distribution as indexed by $y$.  

\subsection{Counterfactual distributions}
  
  We can construct counterfactual distributions resulting from changing the values of the covariates and coefficients
   \begin{equation} \label{eq:cdI}
      G_t(y) = \E_t \Lambda(- h_{it}(\vx_{it})'\vbeta^g_i(y)),
  \end{equation}
  where  $h_{it}$ is a possibly data dependent transformation, and $ \vbeta_i^g(y)$ is a transformation of the random coefficients. 
  Specifically, we consider 
  $$
  \vbeta_i^g(y) = \vtheta(y)g(\vz_{i}) + \vgamma_i(y) = \vbeta_i(y) + \vtheta(y)[g(\vz_{i})-\vz_{i}],
  $$
  for a known transformation $g$ of the time invariant covariates $\vz_i$. This transformation allows us to study the effect of  changing  the values of the covariates on the cross-sectional distribution through their impact on the random coefficients, provided that $\vgamma_i(y)$ is policy invariant.  
  
  For instance, consider a hypothetical scenario where  at time $t$ we increase the number of years of schooling to 12 for any worker who has less. If $\vz_{i} = (z_{1i},\vz_{-1,i}')'$, where $z_{1i}$ is the observed years of schooling of worker $i$ and $\vz_{-1,i}$ includes the remaining components of $\vz_i$, this counterfactual scenario is implemented via the transformation
\begin{equation}\label{eq:hs_counter}
g(\vz_i) = (\max(z_{1i},12),\vz_{-1,i}).
\end{equation}
$G_t(y)$ would then represent the counterfactual distribution of labor income at $t$ after the change.  Another example is 
\begin{equation*}\label{eq:one_counter}
g(\vz_{i}) = (z_{1i}+1,\vz_{-1,i}),
\end{equation*}
which corresponds to giving an additional year of schooling to all workers.  

We can also study the impact of shocks in  dynamic models. For example, suppose at time $t-1$ a shock reduces income by $100*\kappa \%$ for individuals with income higher than certain (known) threshold $\tau_0$. This corresponds to the transformation
    \begin{equation}\label{eq:prop_tax}
  h_{it}(\vx_{it}) = (1,  y_{i(t-1)}+1\{y_{i(t-1)}> \log \tau_0\}\log(1-\kappa))',
  \end{equation}
  where $y_{i(t-1)}$ is measured in logarithmic scale. $G_t(y)$ now represents the counterfactual income distribution resulting from this income shock at time $t$.
We consider both  continuous and discrete $y_{i,t}$, enabling an investigation of quantile effects and stationary distributions respectively. 

  \subsection{Quantile effects}

  When the actual and counterfactual cross-sectional distributions of $y_{it}$ are continuous, 
we assume that both distributions are continuously differentiable. It allows us to  consider quantiles of  both  distributions, and define  quantile effects as their difference. 
Given a univariate distribution $F$, the quantile (left-inverse) operator is 
$$
\phi(F,\tau) := \inf\{y \in \mathbb{R}:  F(y)\geq\tau\}, \quad \tau \in [0,1],
$$
with the convenction $\inf \{ \emptyset \} = + \infty$. 
We apply this operator to the cross-sectional distributions defined above to obtain the quantile  effects 
as
$$
\QTE_{t}(\tau) :=  \phi(G_{t}, \tau) -\phi(F_t, \tau),   \quad \tau \in [0,1].
$$
The quantile effect measures the  contemporaneous impact of the hypothetical policies at different parts of the outcome distribution, and  is  based on comparisons between counterfactual and actual marginal distributions.

\subsection{Stationary distributions } \label{subsec:stationary}
When   $y_{it}$  is discrete with finite support, one can model the distribution of $y_{i,t}$ using a discrete Markov chain with a unique stationary distribution.\footnote{We focus on discrete outcomes when analyzing long run effects for theoretical reasons, as addressing the discretization error that arises in constructing the Markov chains becomes substantially more involved when the outcome variable is continuous.} Specifically,  suppose  the process $\{y_{i1}, \ldots, y_{iT}\}$ is ergodic for each $i$.  The  distribution of $y_{it}$ conditional on $\vx_{it}$  can   be represented by a time-homogeneous finite-state Markov chain for each unit. The cross-sectional stationary distribution can be characterized by aggregating the transition matrices of all units.

Let the discrete support of $y_{it}$  be $\mathcal{Y}_i = \{y_i^1 < \cdots < y_i^K\}$, noting it may differ for each unit, and $\vx_{it}$ include only the first lag of $y_{it}$, i.e. $\vx_{it} =(1, y_{i(t-1)},\vv_{it}')'$.   For each $i$, let $\boldsymbol{P}_i$ be the $K \times K$  transition matrix with $\vv_{it}$ fixed at a value $\vv_i$, which might be different for each unit. The typical element of this matrix can be expressed as
\begin{multline}\label{eq:tmat}
P_{i,jk} = \Pr(y_{it} = y_i^j \mid y_{i(t-1)} = y_i^k, \vv_{it} = \vv_i) \\ 
= \Lambda\left(- \vx_{i}^{k'}\vbeta_i(y_i^j)\right) - 1(j > 1)\Lambda\left(- \vx_i^{k'}\vbeta_i(y_i^{j-1})\right),    
\end{multline}
where $\vx_i^k = (1,y_i^k,\vv_i)'$. By standard theory for Markov Chains, see, e.g., \cite[p. 684]{hamilton2020time}, the ergodic probabilities $\boldsymbol{\pi}_i = (\pi_{i1}, \ldots, \pi_{iK})$ are
$$
\boldsymbol{\pi}_i = (\boldsymbol{A}_i'\boldsymbol{A}_i)^{-1} \boldsymbol{A}_i'\boldsymbol{e}_{K+1}, \quad \boldsymbol{A}_i = \left(\begin{array}{c}
     \boldsymbol{I}_K - \boldsymbol{P}_i \\
     \boldsymbol{1}_K' 
\end{array}\right),
$$
where $\boldsymbol{I}_K$ is the identity matrix of size $K$, $\boldsymbol{1}_K$ is a $K$-vector of ones, and $\boldsymbol{e}_{K+1}$ is the $(K+1)$th column of $\boldsymbol{I}_{K+1}$. The cross-sectional stationary  distribution is 
$$
F_{\infty}(y) = \plim_{N\to \infty}  \frac{1}{N} \sum_{i=1}^N F_{i,\infty}(y), \quad F_{i,\infty}(y) = \sum_{k: y_i^k \leq y} \pi_{ik},
$$
where $F_{i,\infty}$ is a step function with steps at the elements of $\mathcal{Y}_i$.\footnote{$F_{i,\infty}$ is measurable because we assume that the cross-sectional population indexed by $i$ is countable.  }


Stationary counterfactual distributions can be formed by replacing $\vbeta_i(y_i^j)$ by $\vbeta_i^g(y_i^j)$ in \eqref{eq:tmat}. That is
\begin{equation*}
P^g_{i,jk} = \Lambda\left(- \vx_{i}^{k'}\vbeta^g_i(y_i^j)\right) - 1(j > 1)\Lambda\left(- \vx_i^{k'}\vbeta^g_i(y_i^{j-1})\right).
\end{equation*}
We denote the resulting cross-sectional stationary distribution as $G_{\infty}$.  Note that changes in $y_{i(t-1)}$ do not affect the stationary distribution by the ergodicity assumption. The stationary distribution is useful for analyzing dynamics of the distribution of $y_{it}$ in the long 
run.
In practical applications where the actual distribution of $y_{it}$ may be continuous, we discretize the support $\mathcal Y_i$ separately for each unit. 

\subsection{Comparison with Heterogeneous Quantile Regression}\label{seccomdrqr}


    When the distribution of $y_{it}$ is continuous, the model parameters can be related to derivatives of conditional quantiles. Let $Q_{y_{it}}(u \mid \vx_{it})$ denote the $u$-th conditional quantile of $y_{it}$ given $\vx_{it}$, defined as the left-inverse of the conditional distribution function $y \mapsto F_{y_{it}}(y \mid \vx_{it})$:
\[
Q_{y_{it}}(u \mid \vx_{it})
:=
\inf \{ y \in \mathbb{R} : F_{y_{it}}(y \mid \vx_{it}) \ge u \}.
\]
If $y \mapsto F_{y_{it}}(y \mid \vx_{it})$ is strictly increasing over the support of $y_{it}$, and if $y \mapsto \vbeta_i(y)$ is differentiable with derivative $\dot{\vbeta}_i(y) := \mathrm{d}\vbeta_i(y)/\mathrm{d}y$, then the derivative of the conditional quantile with respect to $\vx_{it}$ under HDR satisfies
\begin{equation}\label{eq:qder}
\frac{\partial Q_{y_{it}}(u \mid \vx_{it})}{\partial \vx_{it}}
=
\left.
\frac{\vbeta_i(y)}{-\, \vx_{it}' \dot{\vbeta}_i(y)}
\right|_{y = Q_{y_{it}}(u \mid \vx_{it})}.
\end{equation}
This shows that the HDR coefficient $\vbeta_i(y)$ is proportional to the quantile derivative. However, we do not recommend using the representation \eqref{eq:qder} to estimate quantile derivatives. Estimating $\dot{\vbeta}_i(y)$ and $Q_{y_{it}}(u \mid \vx_{it})$ is challenging  because it requires numerical approximation, which introduces an additional source of bias. 
In contrast, quantile derivatives are in principle straightforward to estimate using heterogeneous quantile regression (HQR), which assumes that the slope parameter $\vbeta_i(u)$ satisfies
\[
P\big(y_{it} \le \vx_{it}'\vbeta_i(u) \mid \vx_{it}\big) = u,
\qquad u \in (0,1).
\]
Under this specification,
\[
\frac{\partial Q_{y_{it}}(u \mid \vx_{it})}{\partial \vx_{it}}
=
\vbeta_i(u),
\]
which can be directly computed from the estimated HQR coefficients \citep[e.g.,][]{galvao2020unbiased}. Therefore, when the primary objective is to estimate heterogeneous quantile derivatives, HQR is a more convenient approach than HDR.

However, HDR often leads to simple estimators of  distribution-related quantities. For example, consider the derivative of the conditional distribution function. Under HDR, this derivative admits a simple closed form:
\[
\frac{\partial F_{y_{it}}(y \mid \vx_{it})}{\partial \vx_{it}}
=
-\dot{\Lambda}\big(-\vx_{it}'\vbeta_i(y)\big)\vbeta_i(y),
\]
where $\dot{\Lambda}(t) = \mathrm{d}\Lambda(t)/\mathrm{d}t$, so it is straightforward to construct a plug-in estimator. In contrast, under HQR and assuming that $u \mapsto \vbeta_i(u)$ is differentiable with derivative $\dot{\vbeta}_i(u) := \mathrm{d}\vbeta_i(u)/\mathrm{d}u$, 
\[
\frac{\partial F_{y_{it}}(y \mid \vx_{it})}{\partial \vx_{it}}
=
\left.
\frac{\vbeta_i(u)}{-\, \vx_{it}' \dot{\vbeta}_i(u)}
\right|_{u = F_{y_{it}}(y \mid \vx_{it})},
\]
which requires numerical approximation to estimate  $ \dot{\vbeta}_i(u)$ and $F_{y_{it}}(y \mid \vx_{it})$.

HDR and HQR are powerful tools for analyzing quantile and distribution related objects. HQR is particularly convenient for estimating functions of the conditional quantile function. HDR often provides simpler expressions for functionals of the conditional distribution function. The choice between the two methods should be guided by the specific objective of the empirical application.

\section{Estimation and Inference}\label{sec:est}

  \subsection{Estimation}
  Our estimation procedure can be conducted in two stages. 
In stage 1, we estimate the coefficients by DR applied separately to the time series dimension of each unit 
and debias the resulting estimates.
\footnote{Unit-by-unit estimation of random coefficient models without bias correction has been previously considered in the literature; see, e.g., \citet[Chap.~6]{hsiao2012} and \citet[Chap.~28]{pesaran2015time}.}  In stage 2, we estimate functionals via the plug-in method and further debias if the functionals are nonlinear.

  \subsubsection{First stage: Model coefficients} 
We start with the HDR estimator of  $\vbeta_i(y)$. That is
  $$
  \widetilde\vbeta_i(y)=\arg\max_{\beta \in \mathbb{R}^{d_x}}Q_{y,i}(\beta),\quad y \in \mathcal{Y}_i, \quad i=1,...,N, 
  $$
  where 
  $$
  Q_{y,i}(\beta)=\sum_{t=1}^T1\{y_{it}\leq y\}\log \Lambda(-\vx_{it}'\beta)+\sum_{t=1}^T1\{y_{it} > y\} \log [1-\Lambda(-\vx_{it}'\beta)],
  $$
and $\mathcal{Y}_i$ is the set of observed values of the outcome for unit $i$, i.e. $\mathcal{Y}_i := \{y_{i1}, \ldots, y_{iT}\}$. If $\Lambda$ is the standard normal or logistic link, these are standard logit or probit estimators.
We then obtain $\widetilde\vbeta_i(y)$ for other values of $y$ noting that $y \mapsto \widetilde\vbeta_i(y)$ is a vector of step functions with steps at the elements of $\mathcal{Y}_i$.

Two complications arise in this first stage. First, $\widetilde\vbeta_i(y)$ is well-defined only if $ y \in [\underline{y}_i, \overline{y}_i)$, where $\underline{y}_i = \min_{1 \leq t \leq T} y_{it}$ and $\overline{y}_i = \max_{1 \leq t \leq T} y_{it}$.
Let $N_0(y)$ be the number of indexes $i$ for which $y \geq \overline{y}_i$, $N_1(y)$ be the number of indexes $i$ for which $y < \underline{y}_i$, and $N_{01}(y) = N - N_0(y) - N_1(y)$ denote
the number of indexes $i$ for which  $\widetilde\vbeta_i(y)$ exists. Without loss of generality we rearrange the index $i$ such that $\widetilde\vbeta_i(y)$ exists for all $i = 1,\ldots,N_{01}(y)$. 

 Second, the first stage estimator $\widetilde\vbeta_i(y)$  is   $\sqrt{T}$-consistent  but possesses a nonlinear bias of order $T^{-1}$. It is necessary to remove the bias from the coefficients so that
it does not affect  the subsequent functionals that employ them.  We debias  $\widetilde\vbeta_i(y)$ using analytical methods. That is 
 \begin{equation} \label{eq:bcdr}
  \widehat\vbeta_i(y)= \widetilde\vbeta_i(y) - \frac{\widehat B_{i,T}(y)}{T}, \quad i = 1,\ldots, N_{01}(y),
  \end{equation}
  where $\widehat B_{i,T}(y)$ is a consistent estimator of the bias of $\widetilde\vbeta_i(y)$.  The specific expressions of the bias and its estimator are presented in the Appendix, 
where we also consider alternative debiasing methods based on the Jackknife \citep{dhaene2015split,jochmans2024inference}.  While our theory applies to both analytical and Jackknife methods, we focus on analytical methods as they have less demanding data requirements and perform better in our numerical simulations. By removing the $T^{-1}$-bias we can deal with the setting in which $T/N\to\rho\in[0,\infty]$. More precisely, we consider asymptotic sequences where $N=o(T^2)$.

\subsubsection{Second stage: Functionals}\label{sec:step2}
We provide estimators for all the functionals of interest. 



\paragraph{\textbf{Projections of coefficients}} A plug-in estimator of $\vtheta(y)$ corresponds to applying two-stage least squares to \eqref{eq:projection} replacing  $\vbeta_i(y)$ by $\widehat\vbeta_i(y)$. This yields,
\begin{equation}\label{eq:lp}
\widehat\vtheta(y)= \sum_{i=1}^{N_{01}(y)} \widehat\vbeta_i(y) \widehat{\vz}_i(y)' \left(\sum_{i=1}^{N_{01}(y)} \widehat{\vz}_i(y) \widehat{\vz}_i(y)' \right)^{-1}, 
\end{equation}
where
$$
\widehat{\vz}_i(y) := \sum_{j=1}^{N_{01}(y)} z_j \vw_{j}' \left(\sum_{j=1}^{N_{01}(y)} \vw_{j} \vw_{j}' \right)^{-1} \vw_{i}.
$$
When $\vw_{i} = \vz_{i}$, the estimator simplifies to the OLS estimator with $\widehat{\vz}_i(y) = \vz_{i}$.

\medskip

\paragraph{\textbf{Actual and counterfactual distributions}}
The plug-in estimators of the actual and counterfactual distributions are
\begin{eqnarray}\label{eq:ecd}
\widehat F_{t}(y)&=& \frac{1}{N}\sum_{i=1}^{N_{01}(y)} \Lambda(-\vx_{it}'\widehat\vbeta_i(y)) + \frac{N_1(y)}{N} -
\frac{\widehat B(y)}{T},\cr
\widehat G_{t}(y)&=
&\frac{1}{N}\sum_{i=1}^{N_{01}(y)} \Lambda(-h(\vx_{it})'\widehat\vbeta^g_i(y)) + \frac{N_1(y)}{N} -
\frac{\widehat B_{G}(y)}{T},
\end{eqnarray}
where 
\begin{eqnarray*}
\widehat\vbeta_i^g(y)&=& \widehat\vbeta_i(y) + \widehat\vtheta(y)[g(\vz_{i})-\vz_{i}],
\cr
\widehat B(y)&=&\frac{1}{2}\frac{1}{N}\sum_{i=1}^{N_{01}(y)}\tr\left(\ddot\Lambda(-\vx_{it}'\widehat\vbeta_i(y)) \vx_{it} \vx_{it}'\widehat \Sigma_i(y)^{-1} \right) \cr
\widehat B_{G}(y)&=&\frac{1}{2}\frac{1}{N}\sum_{i=1}^{N_{01}(y)}\tr\left(\ddot\Lambda(-h(\vx_{it})'\widehat\vbeta^g_i(y))h(\vx_{it})h(\vx_{it})'\widehat \Sigma_i(y)^{-1} \right).
\end{eqnarray*}
Here $\widehat B(y)$ and $\widehat B_G(y)$ are  estimators of the first-order bias coming from the nonlinearity of $F_{t}$ and $G_t$ as a functional of $\vbeta(y)$, 
$\widehat \Sigma_i(y)^{-1}$ is an estimator of the asymptotic variance matrix of $\sqrt{T}(\widetilde\vbeta_i(y) - \vbeta_i(y))$, $\tr$ is the trace operator, and $\ddot\Lambda$ is the second derivative of $\Lambda$.
For units for which $\widehat\vbeta_i(y)$ is not well-defined we set $\Lambda(-\vx_{it}'\vbeta_i(y)) = \Lambda(-h(\vx_{it})'\vbeta^g_i(y)) = 1$ if $y < \underline{y}_i$ and $\Lambda(-\vx_{it}'\vbeta_i(y)) = \Lambda(-h(\vx_{it})'\vbeta^g_i(y)) = 0$ if $y \geq \overline{y}_i$. 

\medskip

\paragraph{\textbf{Quantile effects}} 

   The estimators of the  quantile effects are:
\begin{eqnarray}\label{eq:qe}
\widehat\QTE_t(\tau)= \widetilde\phi(\widehat{ G}_{t},\tau) -\widetilde\phi(\widehat{ F}_t,\tau), 
\end{eqnarray}
where $\widetilde\phi$ is the generalized inverse or rearrangement operator
$$
\widetilde\phi(F,\tau) = \int_0^{\infty} 1\{F(y) \leq  \tau\} \mathrm{d} y - \int_{-\infty}^0 1\{F(y) \geq  \tau\} \mathrm{d} y
$$
which monotonizes $y \mapsto F(y)$ before applying the inverse operator.

\medskip

\paragraph{\textbf{Stationary distributions}} 

We start with the empirical transition matrix as a preliminary plug-in estimator of  $\boldsymbol{P}_i$,  which we modify  to enforce that all entries are non-negative and the rows add to one. More precisely, we define the $K \times K$ matrix $\widehat{\boldsymbol{Q}}_i$ with typical element
\begin{equation}\label{eq:tmathat}
\widehat Q_{i,jk} = 1(j=K) + 1(j<K)\Lambda\left(- \vx_i^{k'}\widehat \beta_i(y_i^j)\right).
\end{equation}
For each column of $\widehat{\boldsymbol{Q}}_i$, we sort (rearrange) the elements in increasing order 
to form the matrix $\check{\boldsymbol{Q}}_i$ with typical element $\check Q_{i,jk}$. We then construct the empirical transition matrix $\widehat{\boldsymbol{P}}_i$  with typical element
$$
\widehat P_{i,jk} = \check Q_{i,jk} - 1(j>1) \check Q_{i,(j-1)k}.
$$
The empirical ergodic probabilities $\widehat{\boldsymbol{\pi}}_i = (\widehat \pi_{i1}, \ldots, \widehat \pi_{iK})$ are now
$$
\widehat{\boldsymbol{\pi}}_i = (\widehat{\boldsymbol{A}}_i'\widehat{\boldsymbol{A}}_i)^{-1} \widehat{\boldsymbol{A}}_i'\boldsymbol{e}_{K+1}- \frac{1}{T} \widehat B_{\boldsymbol{\pi}_i}, \quad \widehat{\boldsymbol{A}}_i = \left(\begin{array}{c}
     \boldsymbol{I}_K - \widehat{\boldsymbol{P}}_i \\
     \boldsymbol{1}_K' 
\end{array}\right),
$$
where $\widehat B_{\boldsymbol{\pi}_i}$ is an estimator of the bias due to the nonlinearity of $\boldsymbol{\pi}_i$ as a functional of $(\vbeta_i(y_i^1), \ldots, \vbeta_i(y_i^K))$. The expression of $\widehat B_{\boldsymbol{\pi}_i}$ is provided in the Appendix.  

The estimator of the  stationary distribution  is
$$
\widehat F_{\infty}(y) = \frac{1}{N} \sum_{i=1}^{N} \widehat F_{i,\infty}(y), \quad \widehat F_{i,\infty}(y) = \sum_{k: y_i^k \leq y} \widehat \pi_{ik}.
$$
Estimators of stationary counterfactual distributions can be formed by replacing  $\widehat \vbeta_i(y_i^j)$ by  $\widehat \vbeta_i^g(y_i^j)$ and modifying the bias estimator, $\widehat B_{\boldsymbol{\pi}_i}$, in \eqref{eq:ecd}. The modified expression of the bias estimator is  given in the Appendix. The resulting estimator of $G_{\infty}$ is denoted by $\widehat G_{\infty}$.

  \subsection{Inference}
  


Inference under flexible heterogeneity
is accompanied by a  novel uniformity challenge associated with the incidental parameter problem. 
 The rate of convergence becomes unknown and variable, affecting the asymptotic distribution. 
 We illustrate this through a simple linear model with  a scalar coefficient.


\subsubsection{Inference problem}

 Consider the model
 $$
  y_{it}  = \beta_i+ e_{it}, \quad \mathbb E(e_{it}\mid \beta_i)=0, \quad \mathbb E(\beta_{i}) = \theta,
 $$
  where $\Var(\beta_i)\in[0,C]$, $C>0$, with zero as an admissible value. This class of data generating processes captures different degrees of heterogeneity.
  For simplicity, we assume $e_{it}$ and $\beta_i$ are both i.i.d. sequences in both $i$ and $t$ and mutually independent and $\Var(e_{it})>0$. The estimator of $\theta$ is 
  $$
  \widehat\theta= \frac{1}{N}\sum_{i=1}^N\widehat\beta_i,\quad \widehat\beta_i=\frac{1}{T}\sum_{t=1}^Ty_{it} = \beta_i + \frac{1}{T}\sum_{t=1}^T e_{it}.
  $$
   The goal is inference for $\theta$ based on $\widehat\theta$ that remains uniformly valid over $\Var(\beta_i)\in[0,C]$.
   
   Let $\overline{\beta} = \sum_{i=1}^N \beta_i/N$.  The asymptotic distribution of $\widehat\theta$ is determined by two components:
   $$
   \widehat\theta-\theta= (\widehat \theta - \overline{\beta}) + (\overline{\beta} - \theta),
   $$
   where
   \begin{eqnarray*}
   \widehat \theta - \overline{\beta} &=&\frac{1}{NT}\sum_{i=1}^N\sum_{t=1}^T e_{it},\quad 
   \overline{\beta} - \theta =\frac{1}{N}\sum_{i=1}^N(\beta_i - \mathbb E(\beta_i)).
   \end{eqnarray*}
 While both terms admit central limit theorems, they may have different rates of convergence.     The rate of convergence of $\overline{\beta} - \theta$ depends on the degree of heterogeneity, $\Var(\beta_i)$, which is unknown and this gives rise to a new issue  related  to estimating the incidental parameters 
$\beta_1,...,\beta_N$.     To illustrate this, consider three special cases:
\begin{enumerate}
    \item \textit{Strong heterogeneity}: If $T\Var(\beta_i)\to\infty$, then the term $\overline{\beta} - \theta $ converges slower than $(NT)^{-1/2}$, and  dominates the expansion, yielding
$$
\frac{ \sqrt{N}(\widehat\theta-\theta)}{\sqrt{\Var(\beta_i)}}\to^d \mathcal N(0, 1).
$$
The rate of convergence of $\widehat\theta-\theta$ is also  slower than  $(NT)^{-1/2}$.
 This case covers $\Var(\beta_i) = c > 0$, which is often assumed in practice \citep[e.g.,][]{fernandez2016individual}. 
\item \textit{Moderate heterogeneity}: If  $T \Var(\beta_i) \to c >0$, then the two terms of the expansion have the same order, yielding  
 $$
 \sqrt{NT}(\widehat\theta- \theta)\to^d \mathcal N(0,\Var(e_{it})+c).
 $$
 This knife edge case is useful for theoretical purposes as the asymptotic variance contains elements from the two terms of the expansion.
 
 \item \textit{Local or no heterogeneity}: If $T \Var(\beta_i) \to 0$, $\widehat \theta - \overline{\beta}$ becomes the dominating term, yielding 
 $$
 \sqrt{NT}(\widehat\theta- \theta)\to^d \mathcal N(0,\Var(e_{it})).
 $$
 We refer to this 
 as ``local or no heterogeneity" as it corresponds to the case when the degree of heterogeneity is small relative to the sample size as formalized by $\Var(\beta_i) = o(T^{-1})$. This may arise in empirical applications where the degree of heterogeneity is unknown and the time dimension is only moderately large. 
\end{enumerate}  

Any degree of heterogeneity between the above three  cases would lead to an unknown rate of convergence $\widehat\theta-\theta=O_P(\xi_{NT})$ where $\xi_{NT}\in[(NT)^{-1/2}, N^{-1/2}]$. Moreover, this unknown rate of convergence has consequences for the properties of standard inferential methods.  Note that
 \begin{equation} 
 \Var(\widehat\theta)=\frac{1}{NT} \Var(e_{it}) + \frac{1}{N}\Var(\beta_i).
 \end{equation}
A common method to estimate this variance  is to plug in sample analogs of  $\Var(e_{it})$ and  $\Var(\beta_{i})$,
$$
\widehat{\Var}(\widehat\theta)=\frac{1}{TN} \widehat{\Var}(e_{it}) + \frac{1}{N}\widehat{\Var}(\beta_i),
$$
where
$$
\widehat{\Var}(e_{it})  = \frac{1}{TN} \sum_{i=1}^N \sum_{t=1}^T (y_{it} - \widehat\beta_i)^2 = \Var(e_{it}) + o_P(1),
$$
and
$$
\widehat{\Var}(\beta_i) = \frac{1}{N} \sum_{i=1}^N (\widehat\beta_i - \widehat \theta)^2 = \Var(\beta_i) + O_P\left( \frac{1}{T}  \vee \sqrt{\frac{\Var(\beta_i)}{T}}\right).
$$

In the moderate and local heterogeneity cases
$$
\Var(\widehat \theta) = O\left( \frac{1}{NT}\right), \quad \widehat\Var(\widehat \theta)-\Var(\widehat \theta) =  O_P\left( \frac{1}{NT}\right).
$$
This leads to  incorrect inference of the standard confidence intervals
$$
\CI_{1-p}(\theta) = \widehat \theta \pm \Phi^{-1}(1-p/2) \sqrt{\widehat\Var(\widehat \theta)} =  \widehat \theta \pm \Phi^{-1}(1-p/2) \sqrt{\Var(\widehat \theta) + O_P((NT)^{-1})},
$$
because $\CI_{1-p}(\theta)$ is scaled by a quantity of the same order as the length of the interval leading to asymptotic distortion
$$
\Pr(\theta \in \CI_{1-p}(\theta)) = 1 - p + O(1).
$$
The source of the problem is that the estimation error of $\widehat{\Var}(\beta_i)$ does not adapt to degree of heterogeneity because $\widehat{\Var}(\beta_i) - \Var(\beta_i) = O_P(T^{-1})$ when $\Var(\beta_i) = o(T^{-1})$. Note that the solution of setting $\widehat \Var(\beta_i) = 0$  leads to asymptotic under-coverage in the strong heterogeneity case. 
In the next section, we propose a bootstrap method that is 
robust to the degree of heterogeneity and is convenient for simultaneous inference on function-valued parameters.

 \subsubsection{The cross-sectional bootstrap}
 
We now develop a simple cross-sectional bootstrap scheme that is uniformly valid over a large class of data generating processes that include 
both local and strong heterogeneity. We introduce the method in the context of the example from the previous section and provide implementation algorithms for the functionals of interest in Appendix \ref{sec:bootstr}. The formal theoretical results on the validity of cross-sectional bootstrap are given in Theorem \ref{th4.2}. 
  
The cross-sectional bootstrap is based on resampling with replacement of the estimated coefficients $\widehat \beta_i$ instead of the observations $y_{it}$. We call this a cross-sectional bootstrap because it is equivalent to resampling the entire time series $\{y_{i1}, \ldots, y_{iT}\}$ of each cross-sectional unit.  Let $\{\widehat\beta_i^*: i=1,..., N\}$ be random sample with replacement from  $\{\widehat\beta_i: i=1,..., N\}$. The bootstrap draw of $\widehat \theta$ is
$$
\widehat\theta^* =\frac{1}{N}\sum_{i=1}^N\widehat\beta_i^*.
$$
We approximate the asymptotic distribution of $\widehat \theta - \theta$ by the bootstrap distribution of $\widehat\theta^* - \widehat\theta$. 



\begin{figure}[htbp!]
	\begin{center}
     \includegraphics[width=8cm]{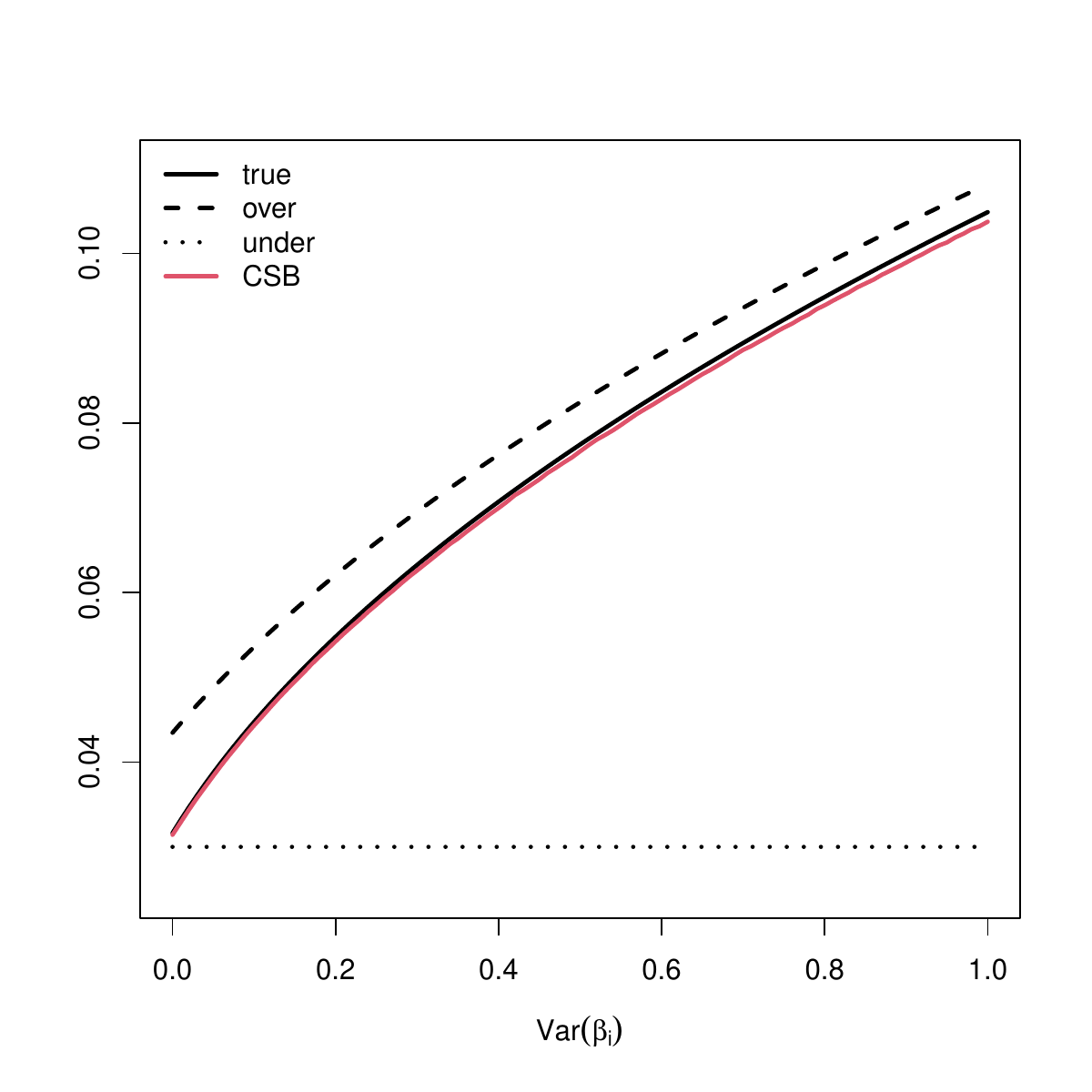}		
		\caption{\small Comparison of analytical and cross-sectional bootstrap estimators of standard deviation of $\widehat \theta$ in this example.}
		\label{fig:toy_example} 
	\end{center}
\end{figure}

Figure~\ref{fig:toy_example} provides a numerical comparison of analytical and cross-sectional bootstrap estimators of the standard deviation of $\widehat \theta$ using a design where $e_{it} \sim \mathcal{N}(0,1)$, $\beta_i \sim \mathcal{N}(\theta, \Var(\beta_i))$, $\Var(\beta_i) \in \{0, 0.1, \ldots, 1\}$, $\theta=1$, $N=100$, and $T=10$.  It reports the (true) standard deviation of $\widehat \theta$, based on  $\Var(\widehat \theta) = \Var(e_{it})/(NT) + \Var(\beta_i)/N$, as a function of $\Var(\beta_i)$; together with averages over $5,000$ simulations of the following estimators:
(1) Standard plug-in: based on
    $$
    \widehat{\Var}(\widehat \theta) = \frac{1}{N^2T^2} \sum_{i=1}^N \sum_{t=1}^T (y_{it} - \widehat\beta_i)^2 + \frac{1}{N^2} \sum_{i=1}^N (\widehat\beta_i - \widehat \theta)^2.
    $$
    This estimator is labeled as ``over". (2)
    Plug-in that omits the heterogeneity in $\beta_i$, based on the first term of the previous expression. This estimator is labeled as ``under". (3)
  Cross-sectional bootstrap interquartile range rescaled by the interquartile range of the standard normal based on $1,000$ draws.

We find that the standard analytical plug-in estimator overestimates the standard error for any degree of heterogeneity, whereas the analytical plug-in estimator that omits the heterogeneity in $\beta_i$ underestimates the standard error in the presence of any heterogeneity.  As predicted by the asymptotic theory, the mean of cross-sectional bootstrap estimator is very close to the standard error uniformly for all the degrees of heterogeneity considered.

\subsubsection{Simultaneous inference}  
The   bootstrap algorithms for the model functionals presented in Appendix \ref{sec:bootstr} are designed to construct confidence bands that cover the functionals simultaneously over the region of points of interest. For example, if we are interested in the scalar function $y \mapsto \xi(y)$ over $y \in \mathcal{Y}$, the asymptotic $p$-confidence band $\CI_p(\xi(y)) := [\widehat \xi_l(y), \widehat \xi_u(y)]$ is defined by the data dependent end-point functions $y \mapsto \widehat \xi_l(y)$ and $y \mapsto \widehat \xi_u(y)$ that satisfy
$$
\Pr\left(\widehat \xi_l(y) \leq \xi(y) \leq \widehat \xi_u(y), y \in \mathcal{Y}\right) \to p \text{ as } N,T \to \infty.
$$
We illustrate in Section \ref{sec:emp} how these confidence bands can be used to test multiple hypotheses about the sign and shape of the functionals. Pointwise confidence intervals are special cases obtained by setting the region $\mathcal{Y}$ to include only one point.
  

  


 \section{The Dynamics of Labor Income}   \label{sec:emp}
 
       We employ the HDR model to study income dynamics using the Panel Study of Income Dynamics  (PSID) data. A vehicle for studying  income dynamics is the permanent-transitory model, i.e., \cite{abowd1989covariance,lochner2014understanding, arellano2017earnings}. It 
  decomposes individual income into a permanent component reflecting long-run earning capacity and a transitory component capturing short-run  shocks,  linking income dynamics to underlying skill shocks and heterogeneity. Unlike this model, we do not specifically model the permanent component, but  assume that income is first-order Markov conditioning on the heterogeneity in the HDR framework. 
  In addition,  the  permanent-transitory model captures  classical  measurement error, as it   focuses on the  latent components rather than observed income. If one is interested in modeling permanent income and the transitory component accounts for measurement error, working with raw labor income might render our estimates inconsistent as our model and estimators are nonlinear. A possible solution to this problem is to separate the permanent component from the transitory component using deconvolution methods and working with the permanent component. 
\cite{lee2025identificationestimationdynamicrandom}  adopts this approach and  finds  similar results for persistence with raw labor income and the extracted permanent component using a dynamic random coefficient model. However, the exercise performed  relies on  assumptions  such as the distribution of transitory shocks being Gaussian and known. 

The HDR  and permanent-transitory models are not nested.  The permanent-transitory model might predict income better over long periods, while a   Markovian model with individual heterogeneity  might fit short periods  better. 
Extending our approach to account for   heterogeneity within a nonlinear permanent-transitory model    is
an important extension which we defer to future work.

\subsection{Data}       
 We employ  PSID data for the years 1967 to 1996 \citep{survey2020panel}. The sample selection follows
 \cite{hu2019semiparametric} which restricts the sample to male heads of household working a minimum of 40 weeks.\footnote{This sample is commonly employed in this literature as it
represents full time full year workers.} We drop the worker-year observations where labor income is above the 99th sample percentile or below the 1st sample percentile, and keep workers observed for a minimum of 15 years. This selection results in an unbalanced panel with 1,629 workers and 33,338 worker-year observations.

The variables used in the analysis include measures of labor income, years of schooling, number of children, marital status, year of birth, survey year and an indicator denoting the individual is white. The years of schooling variable is constructed from the categorical variable highest grade completed with the following equivalence: 0-5 grades =  5 years, 6-8 grades = 7 years, 9-11 grades = 10 years, 12 grades = 12 years, some college = 14 years, and college degree = 16 years. Following the literature on labor income processes, we construct the  outcome, $y_{it}$, as the residuals of the pooled regression of the logarithm of annual real labor income in 1996 US dollars, deflated
by the CPI-U-RS price deflator, on indicators for marital status, number of children, year of birth and survey year. We refer to these residuals as labor income.\footnote{We   acknowledge that using residuals as the outcome variable  follows the approach in  \cite{hu2019semiparametric} and \cite{arellano2017earnings} to facilitate comparability, although it is not ideal. Our method can be extended to work with the original income data including additional covariates and time effects in the distribution regressions of the first stage. This extension would require dealing with the incidental parameter bias introduced by the estimation of the time effects and a longer panel than  we have in the application. We leave this extension to future research.}

    \subsection{Projections of coefficients} We estimate the HDR model \eqref{eq:hdr} with $\vx_{it} = (1, y_{i,t-1},\vv_{it})'$, where $\vv_{it}$ is  the age of the worker $i$ at time $t$. Denote the model coefficients by $\vbeta_i(y) = (\alpha_i(y), \rho_i(y), \pi_i(y))'$,  where we refer to $y \mapsto \alpha_i(y)$ as the intercept or level function and $y \mapsto \rho_i(y)$ as the slope or  persistence function.  We explore if specific worker characteristics are associated with the heterogeneity in the level and persistence coefficients using projections. 
Specifically, we apply \eqref{eq:lp} with $\vz_i$ including a constant, the initial labor income,  years of schooling, and a white indicator and $\vw_i=\vz_i$. 

    \begin{figure}[htbp!]
	\begin{center}
    \includegraphics[width=2.6in]{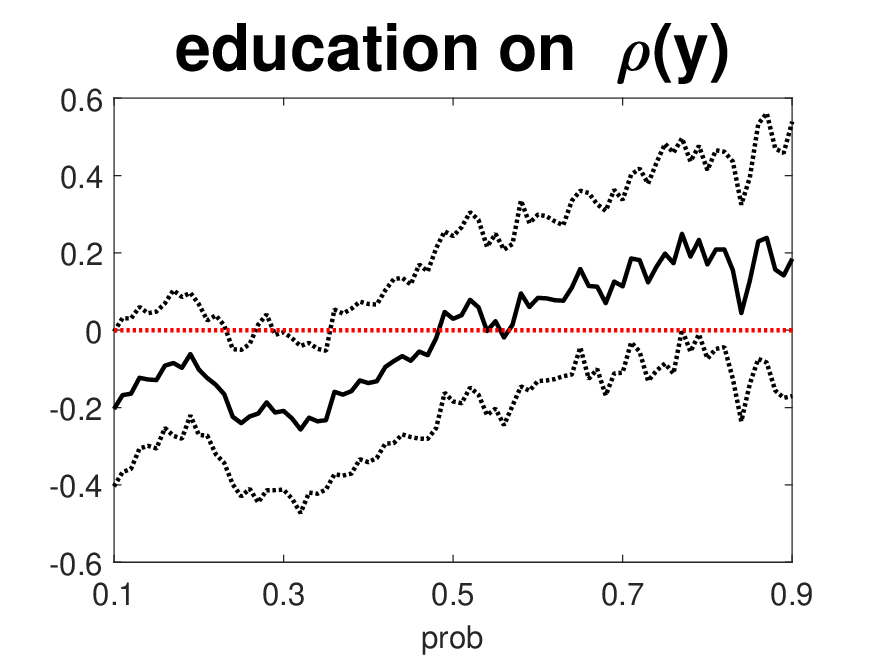}
     \includegraphics[width=2.6in]{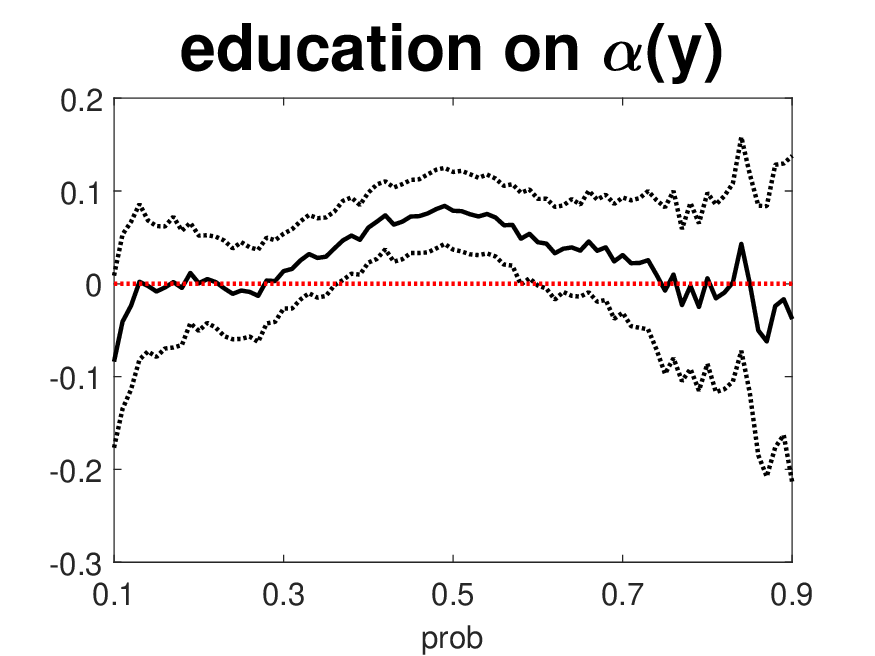}
		\caption{\small Projection coefficients of   $\vbeta_i(y)=(\rho_i(y), \alpha_i(y))$ on worker education levels. The confidence bands are obtained by cross-sectional bootstrap using Algorithm \ref{alg5.1} with $B=500$.}
		\label{fig:lp_slopes} 
	\end{center}
	    
\end{figure}

Figure  \ref{fig:lp_slopes} reports the estimates and 90\% confidence bands of the projection coefficient function $y \mapsto \vtheta(y)$ for education over a region $\mathcal{Y}$ that includes all the sample percentiles of the pooled sample of $y_{it}$ with probability levels $\{0.10, 0.11, \ldots, 0.90\}$, plotted with respect to these probability levels.   We find the education level is associated with coefficient heterogeneity at some locations of the distribution. For example, the persistence parameter  $\rho_i(y)$ is negatively associated with education at the bottom of the distribution, whereas the level parameter $\alpha_i(y)$ is positively associated with education  in the middle of the distribution. The effect of education on $\rho_i(y)$ is increasing with $y$, although this pattern should be interpreted carefully as the function is not very precisely estimated, as reflected by the width of the confidence band.

  \subsection{Impact of an income shock}\label{sec:ctp}

    An implication of the HDR representation of labor income is that an individual's location in the income distribution in a specific time period partially depends on his location in previous periods. Moreover, the dependence level varies by worker. This indicates that a shock to current labor income will determine the path of future income.

    To further illustrate the presence and heterogeneity of this dependence we examine the impact on future income resulting from a negative shock to initial income.
        We implement the shock by reducing labor income in 1985 by 25 percent simultaneously for all individuals.\footnote{We choose 1985 as the base year as it has the largest number of observations in the dataset.}  We interpret this as an unanticipated shock in that we change the level of initial income but keep all other aspects of the model constant. Specifically,  we 
    estimate the counterfactual distribution \eqref{eq:cdI} for the transformation   $$h_{it}(\vx_{it}) = (1,  y_{i(t-1)}+\log(1-\kappa), \vv_{it})'$$  with $\kappa = 0.25$. 
 This transformation 
 yields a counterfactual distribution of labor income in $t=1986$. We also estimate the actual distribution  and the corresponding  quantile effects.  We compare our estimates with those from the following alternative models: 

(a) Het.DR: the proposed model: $
  \Pr(y_{it}\leq y \mid \mathcal{F}_{it}) = \Lambda (-\vx_{it}'\vbeta_i(y)  )$.
 
  (b) Hom.DR: the homogeneous DR: $
  \Pr(y_{it}\leq y \mid \mathcal{F}_{it}) = \Lambda (-\vx_{it}'\vbeta(y)  )$.

  (c) Het.AR: the heterogeneous AR model:
  $
  \Pr(y_{it}\leq y \mid \mathcal{F}_{it}) = \Lambda ((y-\vbeta_i'\vx_{it})/\sigma  ).$
  
  (d) Hom.AR:   the homogeneous AR model:
  $
  \Pr(y_{it}\leq y \mid \mathcal{F}_{it}) = \Lambda ((y-\vbeta'\vx_{it})/\sigma  ).$
 
(e) AR-fixed effect:  the AR model with fixed effects: $
  \Pr(y_{it}\leq y \mid \mathcal{F}_{it}) = \Lambda ((y-\vbeta'\vx_{it} - \alpha_i)/\sigma  ).$


 The parameters of the AR models are estimated by least squares, the parameters of the  DR models are estimated  with $\Lambda$ equal to the standard logistic distribution. \footnote{We do not implement heterogeneous quantile regressions, because it would require numerical integrations to compute conditional CDF, whose debias is also an open theoretical question.} 

    
        \begin{figure}[htbp!]
	\begin{center}
     \includegraphics[width=5in]{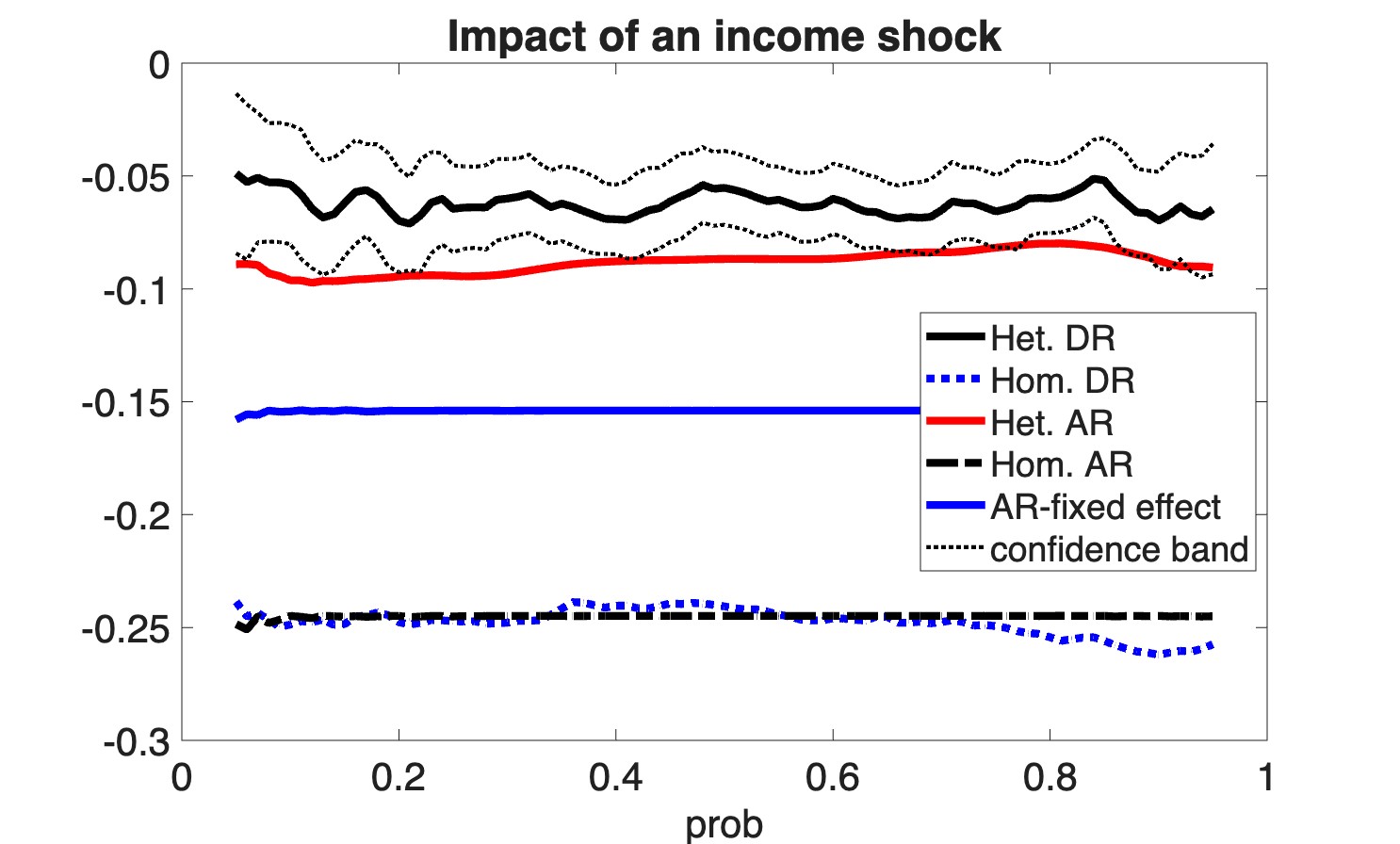}		
		\caption{\small Quantile effects of counterfactual distribution of unanticipated income shock.  The 90\% confidence band for the estimation is based on the  heterogeneous DR model.}
		\label{fig:tax} 
	\end{center}
	
\end{figure}


Figure \ref{fig:tax} reports estimates and 90\% confidence bands of the quantile effects from Hetero.DR, together with the estimates obtained from the alternative models. The confidence bands are computed using Algorithm \ref{alg5.4} with $p=.90$, $B=500$ and $\mathcal{T} = \{.05, .06, \ldots, .95\}$. The estimates show that the fully homogeneous location-shift and DR models predict that the income shock reduces next period income almost on a one-for-one basis throughout the distribution. The linear AR models with fixed effects lower the effect to about 15\% and 10\%, whereas the HDR model further reduces  it to about  5\%. Quantile regression (QR)  provides estimates similar to the fully homogeneous models.
The confidence bands   of the HDR model indicate there is no evidence of heterogeneous effects across 
the distribution. Moreover, they do not fully cover the estimates of the other three models. In results not reported, we find that joint confidence bands from these models do not fully overlap with the confidence bands of the HDR model.\footnote{The confidence level of the joint bands is corrected  by the union bound to $97.5\%$, in order to preserve the joint coverage to at least 90\%.} We can formally reject the homogeneity restrictions imposed by the alternative models. From the comparison of the difference estimators, we conclude that while between-unit heterogeneity matters the most, nonlinearity also plays a significant role in assessing the impact of the negative shock. 


\subsection{Dynamic aspects of relative poverty} 


We now analyze  labor income mobility and the existence of ``relative poverty" traps. We evaluate the probability of remaining in lower locations of the residual distribution noting that we refer to this as relative poverty as we acknowledge that the total income level may not be below the poverty line. We do so via the model from Section \ref{subsec:stationary}, where the conditional distribution is represented by a  Markov chain. We treat income as discrete and set the states for each worker as the observed values of $y_{it}$, that is  $\mathcal{Y}_i:=\{y_{it}: t=1,...,T_i\}$, where $T_i$ is the number of years available for worker $i$, and set $\vv_{it}$ to the median value of age in the sample ($\vv_i = 36$ for all $i$).  

Following \cite{hu2019semiparametric},  consider the following probabilities to describe mobility
	 $$
	 P_i(p,q,h):= \Pr(y_{i(t+h)}< y_p \mid y_{it}< y_q, \mathcal{F}_{it}  ),\quad i=1,...,N,
	 $$
	 where $y_p$ and $y_q$ are the $p$-quantile and $q$-quantile of the distribution of labor income. These probabilities correspond to the following experiment: If we exogenously set labor income below $y_q$ at time $t$, then $P_i(p,q,h)$ is the probability labor income is below $y_p$ after $h$ years.\footnote{The probability $P_i(p,q,h)$ is identified if $y_{it}$ is observed below $y_p$ for some $t$. We restrict the sample to workers that satisfy this condition in the sample period to estimate these probabilities.} For example, if we define the poverty line as the $10$-percentile, then $P_i(0.1,0.1,5)$ is the probability that worker $i$ would remain in poverty after 5 years if he falls below the poverty line due to, for example, a negative income shock.
	 
	 Our model allows the probabilities $P_i(p,q,h)$ to be heterogeneous across workers. To summarize this heterogeneity, we can examine the average probability 
	 \begin{eqnarray*}
	 \bar P(p,q,h) &=&\frac{1}{N}\sum_{i=1}^N P_i(p,q,h).
	 \end{eqnarray*}
  For instance, $\bar P(0.3, 0.1, 1)$ is the probability that a randomly chosen worker is below the 30-percentile if in the previous year he was below the 10-percentile. We also examine quantiles of the  probabilities such as 
$$ Q_{\tau}(p, q, h) $$
which denotes the $\tau$-quantile of  $\{P_i(p, q,h):, i=1,...,N\}$ for fixed $(p, q, h)$. For example, $Q_{0.25}(0.3, 0.1, 1)$ is the first quartile of the probability that a worker is below the 30-percentile if in the previous year he was below the 10-percentile.

The  upper panel of Figure \ref{fig:poverty} plots $p \mapsto \bar P(p,q,h)$ for $p \in [0,0.5]$, $q \in \{0.1,0.25,0.5\}$ and $h \in \{1,2,5\}$. We find heterogeneity with respect to the initial condition that vanishes with time due to the ergodicity of the process. The probability that a randomly selected worker remains below the 10-percentile after one year is more than 50\%, whereas this probability decreases by about half if the worker was initially below the median. This difference in probabilities reduces after two years and almost vanishes after five years. The lower panel of Figure \ref{fig:poverty} plots  $p \mapsto  Q_{\tau}(p,q,h)$ for $p \in [0,0.5]$, $q =0.1$, $h \in \{1,2,5\}$ and $\tau \in \{0.1,0.5,0.9\}$. We uncover significant heterogeneity across workers that is hidden in the 
analysis of the mean worker. Even after 5 periods the deciles of the probability of remaining below the 10-percentile range from $0$ to over $0.9$. This illustrates the importance of accounting for heterogeneity in understanding the probability of escaping poverty.

 \begin{figure}[htbp!]
	\begin{center}
    \includegraphics[width=6in]{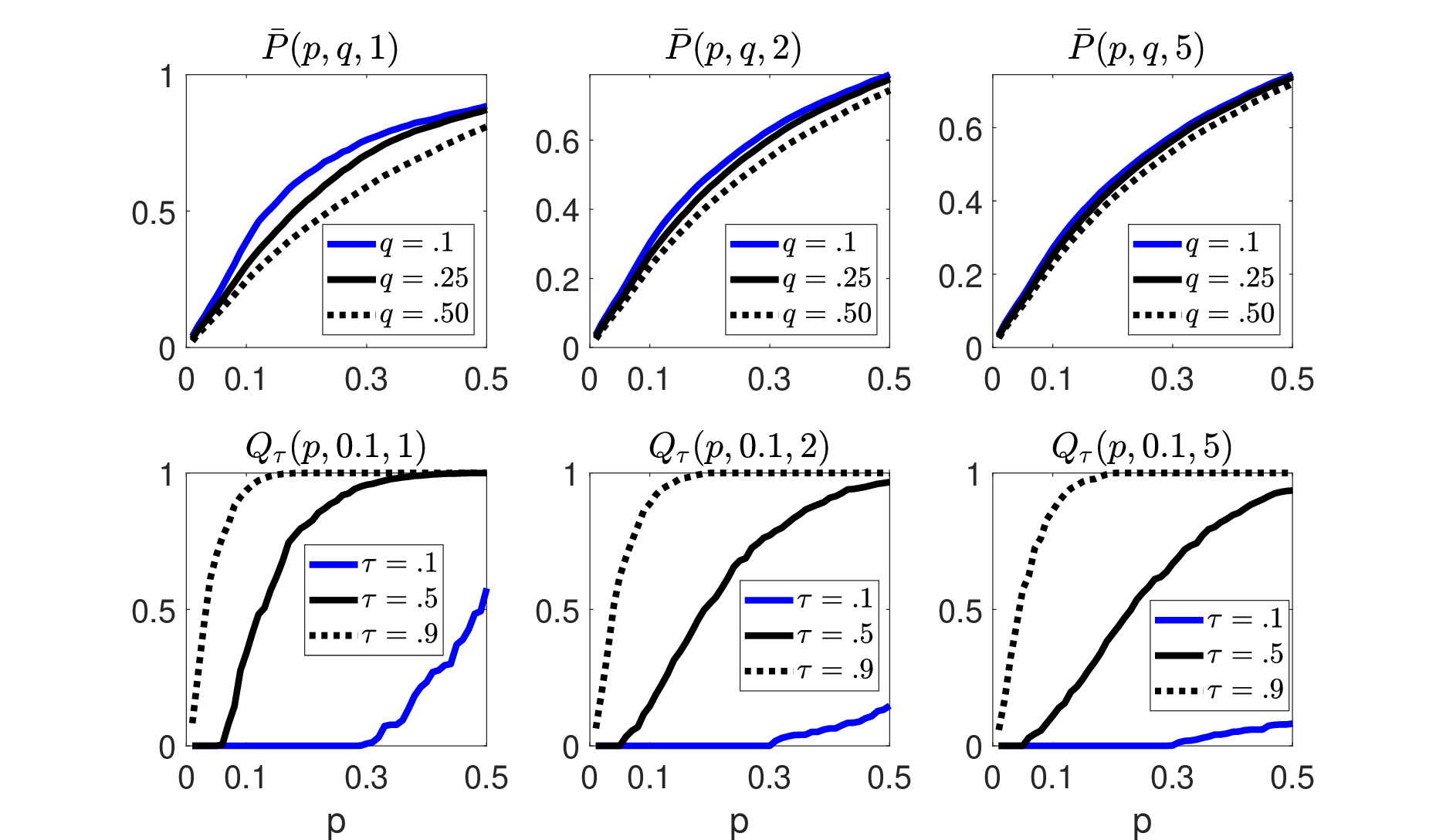}	
		\caption{\small Means and quantiles of probabilities of income mobility. The upper panels report $\bar P(p,q,h)$ and the lower panels report $Q_{\tau}(p,q,h).$}
		\label{fig:poverty} 
	\end{center}
	\end{figure}

Let $h_i(p)$ denote the recurrence time of $y_p$. That is, starting from  $\{y_{it} < y_p\}$, the number of years $h$ until the first occurrence of $\{y_{i(t+h)} > y_p\}$. For example, if $y_{0.10}$ is the poverty line, $h_i(0.10)$ is a random variable that measures the number of years that worker $i$ takes to escape from poverty.
Then, 
$$
\Pr(h_i(p) = h) = \Pr(y_{i(t+h)}>y_p, y_{i(t+h-1)}<y_p,...,y_{i(t+1)}<y_p \mid y_{it}<y_p, \mathcal{F}_{it}),
$$
which can be expressed as a functional of the parameters of the HDR model. 
Another interesting quantity is 
$$H_i(p) =\sum_{h} h \Pr(h_i(p) = h),$$
which gives the expected recurrence time for each individual. In the previous example,  $H_i(0.10)$ gives the expected number of years that worker $i$ would take to escape from poverty. Figure \ref{fig:poverty} plots a histogram of the estimated $H_i(0.10)$. More than 60\% of the workers would escape from the poverty in two or less years, but about 10\% of the workers would stay for more than 20 years.   Table \ref{t4.1} reports several quantiles  of the estimated  $H_i(0.1)$ for  groups stratified by education and race. We find substantial heterogeneity between workers associated with education and race. Whereas the deciles of the expected recurrence time range from 1 to 7 years for workers with at least high school, the corresponding value of 176 years indicates there are more than 10\% of workers with less than high school that would never escape poverty. The large value indicates that for these individuals, poverty is an absorbing state. The distribution of the expected recurrence time also differs by race. The upper decile of the expected recurrence time is about 20 years higher for nonwhite than for white workers. This heterogeneity in the persistence of poverty has clear implications for the design of poverty alleviation policies. As they employ a different sample to ours and employ a different definition of ``relative poverty" we do not directly compare these results to \cite{lillard1978dynamic}. However, in addition to confirming the dependence in labor income documented in their study, we illustrate the remarkable difficulty facing some workers in escaping relative poverty.


\begin{figure}[htbp!]
	\begin{center}
    \includegraphics[width=3in]{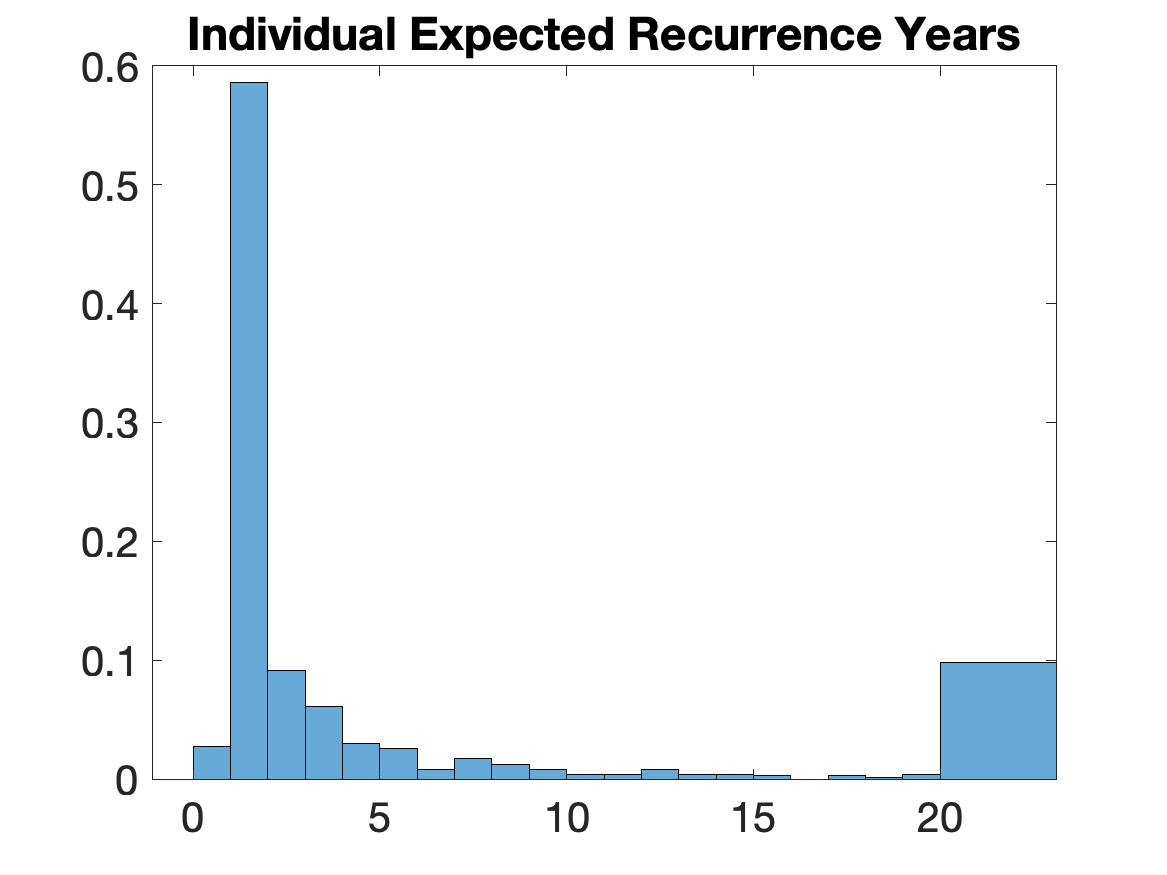}	
		\caption{\small Histogram of expected recurrence time out-of-poverty in years, $H_i(0.10)$}
		\label{fig:poverty} 
	\end{center}
	\end{figure}

\begin{table}[htbp]

	\begin{center}
		\caption{\small  Quantiles of expected recurrence time out-of-poverty in years, $H_i(0.10)$, by education and racial groups}
		\begin{tabular}{c|ccccc}
			\hline
				\hline
 &\multicolumn{5}{c}{Quantiles}    \\
 &   0.10 & 0.25 &  0.50 &  0.75 &  0.90   \\
\hline
All & 1.00 &   1.00 &   1.47&    3.63 &  19.45\\

Edu$<12$ years& 1.00   & 1.35 &   2.92 &   9.75 & 175.8\\
 
Edu$\geq 12$ years   &  1.00 &   1.00   & 1.20   & 2.39  &  7.37\\
    
    White &  1.00   &    1.00   &    1.27   &    3.12   &  13.88\\
    
    non-White &   1.00   &    1.11   &    1.81     & 5.52    &  33.91\\
\hline
 					
		\end{tabular}
			\end{center}
	 
	\label{t4.1}
\end{table}%

\subsection{Goodness of fit} 
\begin{figure}[htbp!]
	\begin{center}
    \includegraphics[width=6in]{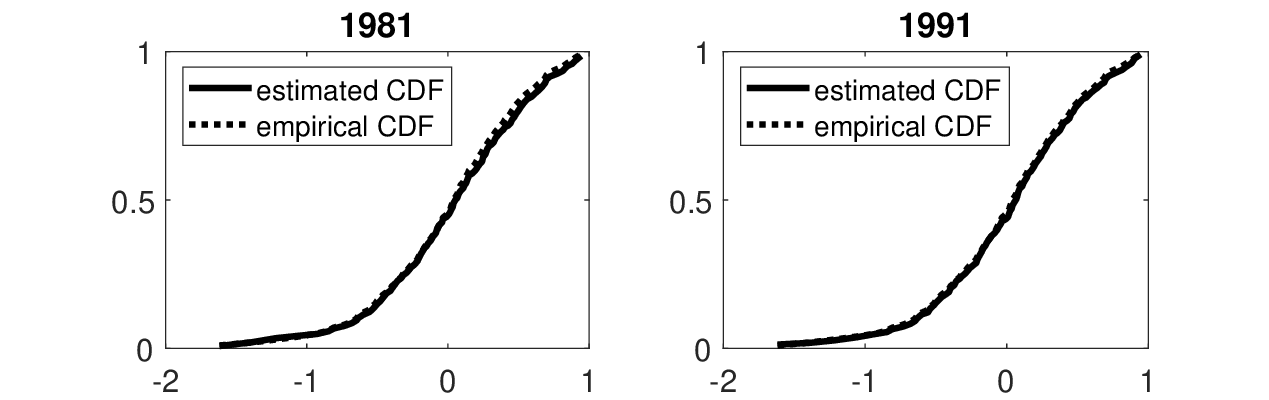}		
		\caption{\small Empirical and predicted actual distributions, $F_t$, $t\in\{1981,  1991\}$.}
		\label{fig:goodnessfit} 
	\end{center}
	\end{figure}

We examine the capacity of the HDR model to fit the PSID data. Figure \ref{fig:goodnessfit}   compares the empirical distributions of $y_{it}$ in 1981 and 1991 with those predicted by the HDR model.   
The model provides a remarkably close fit to the empirical distribution for all the values of $y$, including the tails. In results not reported, the HDR model also provides a good fit of income dynamics by comparing model-based and empirical estimates of the autocorrelation of income.

  \section{Asymptotic Theory}\label{sec:theory}

This section develops asymptotic theory for the estimators of the functionals of interest.   We start by introducing some notation.  Recall that the loss function for the estimation of the coefficients is: $Q_{y,i}(b)=T^{-1}\sum_{t=1}^Tq_{y,it}(b)$, where
$$
  q_{y,it}(b) := 1\{y_{it}\leq y\}\log \Lambda(-\vx_{it}'b)+1\{y_{it} > y\}\log[1-\Lambda(-\vx_{it}'b)].
  $$
  Let
 \begin{eqnarray}\label{eq5.1}
 	\psi_{it}(y)&:=&\nabla q_{y,it}(\vbeta_i(y))\cr
 	 \varpi_{it}^d(y)&:=& \nabla^d q_{y,it}(\vbeta_i(y))-\E\nabla^d q_{y,it}(\vbeta_i(y)),\quad d=1, 2,3. \cr
	 \mathbb A_{1i}(y)&:=& [ \E\nabla^2 q_{y,it}(\vbeta_i(y))]^{-1}, \quad
 		\mathbb A_{2i}(y) :=  \E\nabla^3 q_{y,it}(\vbeta_i(y)),
 \end{eqnarray}
where all terms   are evaluated at the true value of $\vbeta_i(y)$.
For a generic function $q(\vbeta)$ where $\vbeta$ is a $d_{\beta}$-dimensional vector and $d_{\beta}:=\text{dim}(\vbeta)$,
$\nabla q(\vbeta)$ denotes the gradient $d_{\beta}$-dimensional vector 
whose components are the partial derivatives of $\vbeta \mapsto q(\vbeta)$; and $\nabla^2 q(\vbeta)$  is the Hessian matrix. In addition,  $\nabla^3 q(\vbeta)$ is a $d_{\beta}\times d_{\beta}^2$ matrix, defined as $(\nabla B_1(\vbeta),...,\nabla B_{d_{\beta}}(\vbeta))$, where $\nabla B_j(\vbeta)$ is the $d_{\beta}\times d_{\beta}$ Jacobian   of the $j$ th row of $\nabla^2 q(\vbeta)$.

\subsection{Assumptions}\label{sec:assum}  The following assumptions relate to the properties of the sampling process. Recall that $\mathcal F_{i1}\subset...\subset \mathcal F_{iT}$ is the sequence of filtrations over time, and $ \vx_{it}$ is updated to $\mathcal F_{it}$.  The underlying  probability space is equipped with a probability measure $P_T$. Let $\mathcal P$ denote   the collection of all DGPs $P_T$ where our model holds
\begin{eqnarray*}
    \Pr(y_{it} \leq y \mid \mathcal{F}_{it}) &=& \Lambda(- \vx_{it}'\vbeta_i(y)), \cr 
    \vbeta_i(y) &=&\vtheta(y) \vz_{i}  +\vgamma_i(y),\quad  \E(  \vgamma_i(y) \mid \vw_{i} )=0.
\end{eqnarray*}
We assume that the following assumptions hold on $P_T$.
Throughout these assumptions, we  let $c$ and  $C$ be absolute constants, which means they do not  depend on the specific DGP  $P_T\in \mathcal P$.  Under different $P_T$, the degree of heterogeneity will vary and lead to different rates of convergence.   We  aim to establish inference results which are uniformly valid in $\mathcal P$.



We will allow  the distribution of $y_{it}$ to be either continuous or discrete. When it is continuous,  let $\mathcal Y$ be a compact subset of the support of $y_{it}$ on which the density of $y_{it}$ conditional on $\vx_{it}$ is bounded away from zero. When $y_{it}$ is discrete, let $\mathcal Y$ be the set of discrete values of the support.

\begin{assumption}[Cross-section dimension]\label{dgpcs}

(i)  For any $y_1, y_2\in\mathcal Y$ and $i=1,..., N,$

$\E(\vgamma_i(y_1) \mid \vw_{i},\{\psi_{it}(y_2), t\leq T\})=0$ 
and  $\E[\psi_{it}(y_1) \mid \vbeta_i(y_2),  \vx_{it}, \vz_{i}, \vw_{i}, \vgamma_i(y_2)]=0$.

 (ii)     The filtrations $\mathcal F_{iT}$ are independent across $i=1,...,N$.
 
 (iii)  $\{(  Y_{it}, \vx_{it},  \vw_i): t\leq T\}$ are  identically distributed   across $i=1,...,N$.

  \end{assumption}


 \begin{assumption}\label{assemp}


  
 Recall $\varpi^d_{it}(y)$ as defined in (\ref{eq5.1}). 
There are  universal constants $C,c>0$ such that for all $d=1,2,3,$ almost surely, 
\begin{eqnarray*}
  && \max_{i\leq N}\E\left[\sup_{y\in\mathcal Y}\|\frac{1}{\sqrt{T}}\sum_t\varpi^d_{it}(y)\|^{8+c}  \right]<C.
\end{eqnarray*}

 \end{assumption}

 For a given integer $M>0$, let 
  $Y_M=(y_1,...,y_M)'$ be an arbitrary $M$-dimensional vector on $\otimes_{i=1}^M\mathcal Y.$ 
   Let  $S_{wz}:= C_{1}^{-1}C_{2}(C_{2}'C_{1}^{-1}C_{2})^{-1}$ where $C_{1}=\E \vw_{i}\vw_{i}'$, $C_{2}=\E \vw_{i}\vz_{i}'$, and 
     \begin{eqnarray}\label{eq4.1}
 V_{\psi}(y_k, y_l)&:=& \E\left\{   (S_{wz}'\vw_{i}\vw_{i}'S_{wz})
 \otimes \left[ \mathbb A_{1i}(y_k) \E( \frac{1}{T}\sum_{s,t\leq T}     \psi_{it}(y_k)  \psi_{it}(y_l)' |\vw_{i})\mathbb A_{1i}(y_l)  \right]\right\},\cr
  V_{\vgamma}(y_k, y_l)&:=&  \E\left\{ (S_{wz}'\vw_{i}\vw_{i}'S_{wz})
 \otimes \E(\vgamma_i(y_k)\vgamma_i(y_l)'     \mid \vw_{i})\right\}, \cr
\Sigma_{NT}(y_k,y_l)&:=& \frac{1}{NT}V_{\psi}(y_k,y_l)+   \frac{1}{N}V_{\vgamma}(y_k,y_l), \cr
\Sigma_{NT}(y)&:=&\Sigma_{NT}(y,y).
 \end{eqnarray}
Consider a covariance kernel is  given by the limit of the elements of the following $M\times M$ matrix 
$$
H_{\eta,NT}=(H_{\eta, NT}(y_k, y_l))_{M\times M}
$$
which is an $M\times M$ matrix with the $(k,l)$ element as: 
$$
H_{\eta,NT}(y_k, y_l)=\frac{ \eta'  \Sigma_{NT}(y_k, y_l)\eta}{  [ \eta'  \Sigma_{NT}(y_k)\eta]^{1/2}   [ \eta'  \Sigma_{NT}(y_l)\eta]^{1/2}   }
$$
and $\eta\in\mathbb R^{\dim(\vecc\vtheta)}$. We make the following assumption regarding this covariance kernel:
  
\begin{assumption}[Covariance kernel]\label{kernel}
For any  $\eta\in\mathbb R^{\dim(\vecc\vtheta)}$ and $\|\eta\|>c>0$, 
any  integer $M>0$, and any  $M$-dimensional vector
  $Y_M=(y_1,...,y_M)'$    on $\otimes_{i=1}^M\mathcal Y$,  
there is an $M\times M$ matrix $H_\eta$, such that almost surely,  
 \begin{equation}\label{eq4.2}
  \lim_{N,T\to\infty} H_{\eta,NT}=H_\eta.
 \end{equation}
 In addition, there is $c_{Y_M,\eta}>0$ such that 
  \begin{equation}\label{eq4.3}
 \lambda_{\min}(H_\eta)> c_{Y_M,\eta}.
 \end{equation}
 Here $c_{Y_M,\eta}$ may depend on $Y_M, M$ and $\eta.$

\end{assumption}

For a generic estimator $\widehat F(y)$ of $F(y)$, which is either $F_t$ or $G_t$, one can show that it has the following expansion (proved in (\ref{eqe.1})):
$$
\widehat F(y)-F(y) =\frac{1}{N }\sum_{i=1}^N[\frac{1}{\sqrt{T}}d_{\psi,i}(y) + d_{\vgamma,i}(y)]+ o_P(\zeta_{NT}(y))
$$
where $\zeta_{NT}(y)= (NT)^{-1/2}\Var_t(d_{\psi,i}(y))^{1/2}+ N^{-1/2}\Var_t(d_{\vgamma,i})^{1/2}$, and the two leading terms $d_{\psi,i}(y) $ and $d_{\vgamma,i}(y)$ are asymptotically independent, and respectively capture the sampling variation from the first-stage and second-stage. 
The quantile effect has similar expansions 
\begin{eqnarray}\label{eq4.888}
\widehat \QTE_t(\tau)-\QTE_t(\tau) &=&\frac{1}{N }\sum_{i=1}^N[\frac{1}{\sqrt{T}}p_{\psi,i}(\tau) + p_{\vgamma,i}(\tau)]+ o_P(\bar\zeta_{NT}(\tau))
\end{eqnarray}
where $\bar \zeta_{NT}(\tau)= (NT)^{-1/2}\Var_t(p_{\psi,i}(\tau))^{1/2}+ N^{-1/2}\Var_t(p_{\vgamma,i}(\tau))^{1/2}$, and  $p_{\psi,i}(\tau) $ and $p_{\vgamma,i}(\tau)$ are zero-mean uncorrelated terms.  The formal definitions of $(d_{\psi,i}, d_{\vgamma, i},p_{\psi,i}, p_{\vgamma, i})$ depend on the specific $F\in\{F_t, G_{t}\}$, which are given in the Appendix.


\noindent The following condition bounds the moments. For notational simplicity, we write
 $$
   V_{\vgamma}(y ):=  V_{\vgamma}(y, y),\quad    V_{\psi}(y ):=  V_{\psi}(y, y).
 $$
 Recall $\Lambda(s)$ denotes the link function of the distribution regression. 
Let $\dot{\Lambda}(s)= \mathrm{d}\Lambda(s)/\mathrm{d} s$ and  $\ddot{\Lambda}(s)=\mathrm{d}^2\Lambda(s)/\mathrm{d} s^2$.

 \begin{assumption}[Moment bounds]\label{ass4.3}  There are  universal constants $C,c>0$  so that 
 
 (i)  
 \begin{eqnarray*}
\E\left[ \sup_{y\in\mathcal Y}\left(\frac{\|   \vgamma_i(y)\vw_{i}'  \|  }{ \lambda^{1/2} _{\min}(V_{\vgamma}(y))}\right)^{4}   \right]<C. \quad \E_t\left[\sup_{y\in\mathcal Y}\left(\frac{d_{\vgamma, i}(y)}{\Var_t(d_{\vgamma, i}(y))^{1/2}}\right)^{4}\right]<C .
  \end{eqnarray*}
 
 (ii)  Let $\Theta$ be the parameter space for $\{\vbeta_1(y),...,\vbeta_N(y): y\in\mathbb R\}$.  The following moment bounds hold:

  (a)  $\max_{i\leq N}\sup_{y\in\mathcal Y}[\|\mathbb A_{1i}(y)\|+\|
\mathbb A_{2i}(y)\|]<C$
 
 (b)   $\sup_y\sup_{b\in\Theta}\max_{i\leq N}[ \|\nabla ^3Q_{y,i}(b)\|+ \|\nabla ^4Q_{y,i}(b)\|+ 
 \|  ( \nabla^2 Q_{y,i}( b ))^{-1} \|] 
 =O_P(1) $
 
 (c) $\max_{i\leq N}\E \|\vw_{i}\|^{8+c}+\E_t \| h_{it}( \vx_{it})\|^8 + \E \|\vx_{it}\|^8\|g(\vz_{i})-\vz_{i}\|^8<C $. 

 (d) $\sup_s |\dot\Lambda(s)| +\sup_s|\ddot\Lambda(s)|<C  $.

 (iii)  Let   $	S_{\psi,i}(y)=	\Var\left(  \frac{1}{ \sqrt{T}}  \sum_{t=1}^T   \psi_{it}(y)  |\vw_{i}  \right) $. Then 
 $$
 \min_i\inf_{y\in\mathcal Y} \lambda_{\min}(S_{\psi,i}(y))>c, \text{ almost surely.}\quad  \inf_y\lambda_{\min}(\Var_t(d_{\psi,i}(y)))>c. 
 $$
 In addition, 
 all eigenvalues of $C_{1}$ and $C_{2}'C_{2}$ are bounded away from zero and infinity, where
 $C_{1}=\E \vw_{i}\vw_{i}'$ and $C_{2}=\E \vw_{i}\vz_{i}'$, with $\rank(C_2)\geq \dim(\vz_{i})$.

 	(iv) $\frac{1}{T}\sum_t\mathbb E_i \vx_{it}\vx_{it}'$ is of full rank for each $i$, where the expectation $\mathbb E_i$ is taken with respect to the joint density of $(\vx_{i1},...,\vx_{iT})$ conditional on $\mathcal F_{i1}$.

 	 \end{assumption}

\begin{assumption}[Continuity when $\mathcal Y$ is Uncountable]\label{ass4.7} Let  $\widehat F(y)$ be the estimator of $F(y)\in\{F_t(y), G_t(y)\}$.  
Recall that $(d_{\psi,i}, d_{\vgamma, i},p_{\psi,i}, p_{\vgamma, i})$  are the leading terms in the asymptotic expansions, whose   formal definitions are given in (\ref{eqb.2n}) in the Appendix.  The following conditions hold if $y_{it}$ is continuous:  There is a universal constant $C>0$ such that 

(i) for all $y_1,y_2\in\mathcal Y$,   $$
\|V_{\psi}(y_1) -   V_{\psi}(y_2)\|+\max_{i\leq N}\|	\mathbb A_{d,i}(y_1)-	\mathbb A_{d,i}(y_2)\|<C|y_1-y_2| , \quad d=1,2. 
	$$

(ii) for all $\epsilon>0,$ and fixed $\eta\neq 0$, let \begin{eqnarray*}
    z_{i,1}(y)&=&\frac{\eta'\vecc(\vgamma_i(y)\vw_i'S_{wz})}{[\eta'(\frac{1}{T}V_{\psi}(y)+V_{\gamma}(y))\eta]^{1/2}} \cr 
    z_{i,2}(y)&=&\frac{d_{\gamma,i}(y)}{[\frac{1}{T}\Var_t(d_{\psi,i}(y))+\Var_t(d_{\gamma,i}(y))]^{1/2}}\cr 
     z_{i,3}(y)&=&\frac{p_{\gamma,i}(y)}{[\frac{1}{T}\Var_t(p_{\psi,i}(y))+\Var_t(p_{\gamma,i}(y))]^{1/2}}.
\end{eqnarray*} 
Then
\begin{eqnarray*}
    \frac{1}{N}\sum_{i=1}^N \E\sup_{|y_1-y_2|<\epsilon} \left|z_{i,d}(y_1)-z_{i,d}(y_2)\right|^2 <C\epsilon^{1/2},\quad d=1,2,3.
\end{eqnarray*}


(iii)
There is $C>0$ for any $\epsilon>0$,
\begin{eqnarray*}
   &&\max_{i\leq N}\E \left[\sup_{|y_1-y_2|\leq\epsilon}\|\frac{1}{\sqrt{T}}\sum_t\varpi^d_{it}(y_1)-\frac{1}{\sqrt{T}}\sum_t\varpi^d_{it}(y_2)\|^8  \right]<C\epsilon^2, 
\end{eqnarray*}
for $d=1,2,3.$
  
(iv) There are  $C>0$ and  $k\geq 4$, for any $y_1, y_2\in\mathcal Y$, and   $\widetilde \vx_{it}\in\{\vx_{it}, h_{it}(\vx_{it})\}$, 
$\widetilde \vbeta_{i}\in\{\vbeta_{i}^g, \vbeta_{i}\}$, \begin{eqnarray*}
&&\mathbb E_t  | \ddot  \Lambda(-\widetilde \vx_{it}'\widetilde\vbeta_i(y_1))-\ddot  \Lambda(-\widetilde \vx_{it}'\widetilde\vbeta_i(y_2))   |^k \|\widetilde \vx_{it}\|^{2k}
 \leq C|y_1-y_2|^k \cr
 &&\mathbb E_t  | \dot  \Lambda(-\widetilde \vx_{it}'\widetilde \vbeta_i(y_1))-\dot  \Lambda(-\widetilde \vx_{it}'\widetilde \vbeta_i(y_2))   |^k \|\widetilde \vx_{it}\|^{k}
 \leq C|y_1-y_2|^k\cr
 && \E_t  | \dot  \Lambda(- \widetilde \vx_{it}'\vbeta^g_i(y_1))-\dot  \Lambda(- \widetilde \vx_{it}'\vbeta^g_i(y_2))   | \| \widetilde \vx_{it}[g(\vz_{i})-\vz_{i}]'\| 
 \leq C|y_1-y_2|
 \end{eqnarray*}
where $ \vbeta_i^g(y)= \vtheta(y)[g(\vz_{i})-\vz_{i}] +\vbeta_i(y)$.

\end{assumption}

 Assumption \ref{dgpcs} requires that the data are cross-sectionally independent. 
 Assumption \ref{assemp} imposes conditions regarding serial dependence. We  impose two high level conditions regarding the empirical process for weakly dependent data. 
 It requires some primitive conditions, e.g., mixing conditions, so that $\{(Y_{it}, \vx_{it}): t\leq T\}$  is serially weakly dependent.    Additionally, we \textit{do not assume}   stationarity when the analytical debias is used to address the incidental parameter problem.  

Assumption \ref{kernel} is used to establish the finite dimensional distribution (f.i.d.i.) of $  \eta'\vecc(\widehat\vtheta(\cdot)- \vtheta(\cdot)) $, which is required for a given $Y_M, M$ and $\eta$. Therefore, the constant $c_{Y_M,\eta}$ is allowed to depend on these parameters.  To show that Assumption \ref{kernel} is reasonable even though the variance of $\vgamma_i(y)= \vbeta_i(y) - \vtheta(y)\vz_{i}$ may vary across $y$ in the second-stage regression, we consider the following model
\begin{eqnarray}\label{eq4.4}
\vgamma_i(y) &=& \xi_{NT}(y)\bar\vgamma_i(y),\quad \forall y\in\mathcal Y
, \forall i\leq N\cr
 V_{\vgamma}(y_k, y_l)&=& \xi_{NT}(y_k) \xi_{NT}(y_l) V_{\bar \vgamma}(y_k, y_l) ,\quad 
 \inf_y\lambda_{\min} (V_{\bar\vgamma}(y,y)   )>c.
\end{eqnarray}
Here $\xi_{NT}(y)$ is a bounded non-stochastic sequence  that may converge to zero, whose rate depends on $y$; $\bar \vgamma_i(y)$ is a random vector of ``normalized" $\vgamma_i(y)$ , so 
$V_{\bar \vgamma}(y,y)$ can be understood as  a  normalized covariance matrix. 
Hence the strength  of $\vgamma_i(y)$ is determined by the rate of convergence of $\xi_{NT}(y)$.
Given this setting, consider the following special cases:

\begin{description}
\item[Case 1] $\xi_{NT}(y_k)\ll T^{-1/2}$  and $\xi_{NT}(y_l)\gg T^{-1/2}$. Here the  explanatory power of $\vw_{i}$  is strong for $\vbeta_i(y_k)$, but relatively weak  for $\vbeta_i(y_l)$.  Then 
$$\lim_{N,T\to\infty}H_{\eta, NT}(y_k, y_l)=0.
$$
Note that the  case of $\xi_{NT}(y_k)\gg o(T^{-1/2})$  and $\xi_{NT}(y_l)=T^{-1/2}$ is also covered. 
\item[Case 2] Both $\xi_{NT}(y_k), \xi_{NT}(y_l)\gg T^{-1/2}$. Here  the  explanatory power of $\vw_{i}$  is strong for both $\vbeta_i(y_k)$ and $\vbeta_i(y_l)$. Then 
$$
\lim_{N,T\to\infty}H_{\eta, NT}(y_k, y_l)=\lim_{N\to\infty}\frac{ \eta'  V_{\bar \vgamma}(y_k,y_l) \eta}{  [ \eta'  V_{\bar \vgamma}(y_k,y_k) \eta]^{1/2}   [ \eta' V_{\bar \vgamma}(y_l,y_l) \eta]^{1/2}   } ,
$$
where the limit of the right hand side is assumed to exist.
\item[Case 3] Both $\xi_{NT}(y_k), \xi_{NT}(y_l)\ll T^{-1/2}$. Here the  explanatory power of $\vw_{i}$  is relatively weak for both $\vbeta_i(y_k)$ and $\vbeta_i(y_l)$. Then 
$$
\lim_{N,T\to\infty}H_{\eta, NT}(y_k, y_l)=\lim_{N\to\infty}\frac{ \eta'  V_{\psi}(y_k,y_l) \eta}{  [ \eta'  V_{\psi}(y_k,y_k) \eta]^{1/2}   [ \eta' V_{\psi}(y_l,y_l) \eta]^{1/2}   } ,
$$
where the limit of the right hand side is assumed to exist.
\end{description}
Thus each element has a limit given on the right hand side.
With sufficient variation across $y_k$,  the  limit of  the matrix $H_{\eta, NT}$ is non-degenerate and satisfies  (\ref{eq4.3}).

 Assumption \ref{ass4.3} (i)  requires that the fourth moments of $\vgamma_i(y)$ and $d_{\vgamma, i}(y)$ are  bounded by their second moment up to a constant, uniformly in $y$. To see the plausibility of this condition, again consider  model (\ref{eq4.4}).  Then the left hand side of condition (i) becomes 
	$$
 \E\left[ \sup_{y\in\mathcal Y}\left(\frac{\|   \vgamma_i(y)\vw_{i}'  \| ^2 }{ \lambda _{\min}(V_{\vgamma}(y))}\right)^{2}   \right] =
   \frac{ \frac{1}{N}\sum_{i=1}^N \E( \sup_y \|  \bar\vgamma_i(y)\vw_{i}'  \| ^{4}) }{  \inf_{y\in\mathcal Y}\lambda _{\min}^{2}(V_{\bar \vgamma}(y,y))}   ,
   $$ 
	 which is upper bounded by a constant provided $\frac{1}{N}\sum_{i=1}^N\E( \sup_y \|  \bar\vgamma_i(y)\vw_{i}'  \| ^{4})<C.$  
	 	Other conditions of this assumption are standard.  Condition (ii) requires higher moments to be bounded. For instance, we need $\E \|\vw_{i}\|^{8+c}<C$ where $\vw_i$ is a vector of characteristics such as initial labor income, years of schooling and race. These are standardized so it is plausible to assume they have high moments.

	 Conditions (iii) and (iv)    identify the parameters $\vtheta(y)$ and  $\vbeta_i(y)$. To see this, note that the model implies 
	 	$$
	 -\frac{1}{T}\sum_{t=1}^T\mathbb E_i\vx_{it}\Lambda^{-1}\left(	\Pr(y_{it}\leq y|\mathcal F_{it})\right)= \left(\frac{1}{T}\sum_{t=1}^T\mathbb E_i\vx_{it}\vx_{it}'\right)\vbeta_i(y).
	 	$$
	 Inverting $\frac{1}{T}\sum_{t=1}^T\mathbb E_i\vx_{it}\vx_{it}'$ leads to the identification of $\vbeta_i(y)$. In addition, $\rank(C_2)\geq \dim(\vz_{i})$ implies the identification of $\vtheta(y).$

When the support of $y_{it}$ is continuous, Assumption \ref{ass4.7} imposes continuity of  moment, the link and the  $\vbeta_i(y)$ functions. In particular, as the condition is imposed on rescaled functions in Condition (ii), so that it is not affected by the unknown strength of $\vgamma_i(y)$ and $(d_{\vgamma,i}(y),p_{\vgamma,i}(y))$.

\subsection{Theoretical Results}

  In the next theorem,  $L$ denotes the number of lags used for the Newey-West truncation for long-run variance, which is needed for analytical bias corrections. 
\begin{theorem}[Projection Coefficients]\label{th4.1} 
Suppose  $N=o(T^2)$ and $NL^2=o(T^3)$. 
Also, if $\vbeta_i(y)$ is estimated using Jackknife-debias, then we additionally assume Assumption \ref{assa.1}.  Also let 
$$
\Sigma_{NT}(y):= \frac{1}{NT}V_{\psi}(y)+   \frac{1}{N}V_{\vgamma}(y).
$$

(i) If $\mathcal Y$ is continuous, 
  Assumptions \ref{dgpcs}-\ref{ass4.7}  hold.   Then for any $\eta$ such that $\|\eta\|>c>0$,
   $$
 \frac{\eta'\vecc(\widehat\vtheta(\cdot)- \vtheta(\cdot))}{  \left[\eta'\Sigma_{NT}(\cdot)\eta\right]^{1/2} }  \Rightarrow \mathbb G(\cdot)
 $$
  where
 $\mathbb G(\cdot)$ is a centered Gaussian process  with a covariance function $H(y_k,y_l)$ as the $(k,l)$ element of $H_\eta$.

(ii) If $\mathcal Y$ is discrete with finite support, Assumptions  \ref{dgpcs}-\ref{ass4.3} hold. Then for any $y\in\mathcal Y$, 
$$
\Sigma_{NT}(y)^{-1/2}\vecc(\widehat\vtheta(y)- \vtheta(y))\to^d\mathcal N(0,I).
$$
 
\end{theorem}

\begin{remark} While we  assume $N=o(T^2)$, it is possible to allow   larger $N$   by analytically removing  biases of higher orders. 
\end{remark}

 \begin{remark}
Both $V_{\psi}(y)$ and $V_{\vgamma}(y)$ contribute to the asymptotic variance but, as we discussed, the order of the latter is unknown and can vary including 
$V_{\vgamma}(y)=0$ as a special case.  Suppose dim$(\vtheta(y))=1$ for ease of discussion.      If $V_{\vgamma}(y)\ll T^{-1}$, then $\Sigma_{NT}(y)=\frac{1}{NT}V_{\psi}(y)  (1+o(1))$, and only $V_{\psi}(y)$   contributes to the asymptotic variance.  If $V_{\vgamma}(y)\gg T^{-1}$, then $\Sigma_{NT}(y)=\frac{1}{N}V_{\vgamma}(y)  (1+o(1))$, and  only $V_{\vgamma}(y)$   contributes to the asymptotic variance.  If $V_{\vgamma}(y)\asymp T^{-1}$, then $\Sigma_{NT}(y)=\frac{1}{NT}V_{\psi}(y)  +\frac{1}{N}V_{\vgamma}(y)$, both components contribute to the asymptotic variance in the same order. 

\end{remark}




 





 The theorems below  additionally require  Assumption   \ref{ase.48}, which are based on some additional notation for the stationary distribution.  They are presented in the appendix.
 
 Theorem \ref{th4.20} refers to the estimated distributions. It allows the support of $y_{i,t}$ to be either continuous or discrete with finitely-many states such that the stationary distribution can be modeled using   Markov chains with finite states. For the processes $F_n(\cdot)$ and $\mathbb G(\cdot)$, we define $F_n(\cdot)\Rightarrow \mathbb G(\cdot)$ in $\ell^{\infty}(\mathcal Y)$ as the weak convergence  in the set of bounded functions on $\mathcal Y$.

 \begin{theorem}[Predicted Distributions]\label{th4.20} Suppose Assumption \ref{ase.48} hold. 
Let 
$$
v_{NT}^2(y):= \frac{1}{NT}\Var_t(d_{\psi,i}(y))+   \frac{1}{N}\Var_t(d_{\vgamma,i}(y))
$$
where  the  formal definitions of $(d_{\psi,i}, d_{\vgamma, i})$  are  given in (\ref{eqb.2n}) in the Appendix.

(i) If $\mathcal Y$ is continuous, suppose the assumptions of Theorem \ref{th4.1} (i). Let $F\in\{F_t, G_{t}\}$ and $\widehat F\in\{\widehat  F_t, \widehat G_{t} \}$. 
 Then  
$$ 
\frac{   \widehat F(\cdot)-F(\cdot)}{ \upsilon_{NT}(\cdot)   }\Rightarrow \mathbb G(\cdot) \text{ in } \ell^{\infty}(\mathcal Y),$$
  where $\mathbb G(\cdot)$ is a zero-mean Gaussian process  with covariance kernel function 
   $$
\lim_{N,T \to \infty}
 \frac{ v^{2}_{NT}(y_k, y_l)}{v_{NT}(y_k)v_{NT}(y_l)    } ,\quad
v^{2}_{NT}(y_k, y_l):= \frac{1}{NT}\E_td_{\psi,i}(y_k)d_{\psi,i}(y_l)+\frac{1}{N}\E_td_{\vgamma,i}(y_k)d_{\vgamma,i}(y_l),
  $$
 assuming that the limit exists for each pair $(y_k, y_l)$.

(ii) If $\mathcal Y$ is discrete with finite support, suppose assumptions of Theorem \ref{th4.1} (ii) hold. Let 
$F\in\{F_t, G_{t}, F_{\infty}, G_{\infty}\}$ and $\widehat F\in\{\widehat  F_t, \widehat G_{t},  \widehat F_{\infty}, \widehat G_{\infty}\}$.    Then for each $y\in\mathcal Y$,
$$ 
\frac{   \widehat F(y)-F(y)}{ \upsilon_{NT}(y)   }\to^d\mathcal N(0,1).$$

 \end{theorem}

 Theorem \ref{th4.30} refers to the estimated quantile effect  when  we  further require  the conditional distribution of $y_{it}$ given $\vx_{it}$ is continuously differentiable. We do not consider the quantile effect of the stationary distribution. Let $\mathcal T\subset  (0,1)$ be a set of quantile indices, such that $\{\phi(F_t,\tau): \tau\in\mathcal T\}\subset \mathcal Y$.

 \begin{theorem}[Quantile Effects]\label{th4.30} 
Suppose Assumption \ref{ase.48} and  assumptions of Theorem \ref{th4.20} (i) hold. Assume also
for $ F\in \{ F_{t},   G_{t}\}$, $F$ is continuously differentiable, whose density (denoted by $\dot{ F}$) satisfies $
 	\inf_\tau\inf_{|y-\phi( F,\tau)|<C} \dot{  F}(y)>c
 	$  for some  $C, c>0$.   Assumption \ref{ass4.7} also holds.  Then, 
$$ 
\frac{   \widehat \QTE_t(\cdot)-\QTE_t(\cdot)}{ J_{NT}(\cdot)   }\Rightarrow \mathbb G_{\QTE}(\cdot) \text{ in } \ell^{\infty}(\mathcal T),
$$
  where $J_{NT}^2(\tau):=J_{NT}^2(\tau,\tau)$, with
 $$
 J^{2}_{NT}(\tau_k, \tau_l):= \frac{1}{NT}\E_tp_{\psi,i}(\tau_k)p_{\psi,i}(\tau_l)+\frac{1}{N}\E_tp_{\vgamma,i}(\tau_k)p_{\vgamma,i}(\tau_l),
 $$
 and $\mathbb G_{\QTE}(\cdot)$ is a zero-mean Gaussian process  with covariance kernel function 
   $$
\lim_{N,T \to \infty}
 \frac{ J^{2}_{NT}(\tau_k, \tau_l)}{J_{NT}(\tau_k)J_{NT}(\tau_l)    } ,
  $$
 assuming that  the limit exists for each pair $(\tau_k, \tau_l)$.

 \end{theorem}

  Theorem \ref{th4.2} shows the  uniform validity of cross-sectional bootstrap   over a large class of data generating processes with varying degrees of coefficient heterogeneity.

\begin{theorem}[Bootstrap Inference]\label{th4.2} 
Suppose  the assumptions of Theorem \ref{th4.1} hold for all probability sequences  $\{P_T: T\geq 1\}\subset\mathcal P$, where the universal constants do not depend on the specific choice of $P_T$. 
Then uniformly for all $\{P_T: T\geq 1\}\subset\mathcal P$,

(i) For the significance level $0 < a < 1$, 
 $$
  P_T\left(\eta'\vecc(\vtheta(y))\in \CI_{a}(y), \forall y\in\mathcal Y\right)\to 1-a,
           $$
            where $\CI_a(y)
  =\{m:  |\eta'\widehat \vtheta(y)-m|\leq q_as^*(y)\},
  $ and $q_a$ and $s^*$ are defined corresponding to $(\vtheta,\widehat \vtheta)$ using the cross-sectional bootstrap 
 Algorithm \ref{alg5.1} in Appendix \ref{sec:bootstr}.

  (ii)  For $(F, \widehat F)\in\{(F_{t}, \widehat F_{t}),(G_{t}, \widehat G_{t}),(F_{\infty}, \widehat F_{\infty}), (G_{\infty}, \widehat G_{\infty}) \}$
  $$
  P_T\left(F(y)\in \CI_{a}(y), \forall y \in\mathcal Y\right)\to 1-a.
  $$
  where $\CI_a(y)
  =\{m:  |\widehat F(y)-m|\leq q_as^*(y)\},
  $ and $q_a$ and $s^*$ are defined corresponding to the specific $(F,\widehat F)$ using the cross-sectional bootstrap 
Algorithm \ref{alg5.1} in Appendix \ref{sec:bootstr}.
  
  (iii)
  $$
  P_T\left(\QTE_t(\tau)\in \CI_{a}(\tau), \forall \tau\in\mathcal T\right)\to 1-a.
  $$
  where $\CI_a(\tau)
  =\{m:  |\widehat \QTE_t(\tau)-m|\leq q_as^*(\tau)\},
  $ and $q_a$ and $s^*$ are defined  using the cross-sectional bootstrap 
Algorithm \ref{alg5.1}  in Appendix \ref{sec:bootstr}.
\end{theorem}

\vspace{-.1cm}


\vspace{-.3cm}

\section{Simulation Evidence}\label{sec:mc}
    
    We now provide some simulation evidence documenting the finite-sample performance of our method.   The online appendix includes additional simulation results.

\subsection{Dynamic DR model}  \label{SimDR_uncalib}
  
Consider the dynamic DR model: 
  \begin{eqnarray*}
  \Pr(y_{it}\leq y \mid \mathcal{F}_{it})&=&\Phi(y_{i(t-1)}\beta_i(y)),\cr
  \beta_i(y)&=&\theta(y)w_{i}+\theta(y)\bar\gamma_i,\quad \E(\bar\gamma_i \mid w_{i})=0.
  \end{eqnarray*}
with  $$\theta(y)= 3 \text{ sgn}(y-2)(y-2)^2, \  y\in \mathcal Y.$$
 We set  $\mathcal Y= \{1.7, 1.8,..., 2.3\}$, where the two endpoints of $\mathcal Y$ are chosen to avoid the estimation of extreme quantiles. The marginal probabilities $\Pr(y_{it}<1.7)$ and $\Pr(y_{it}>2.3)$ are both approximately 0.1.  We generate the simulated data by independently drawing
  $(e_{it}, w_{i}, \bar\gamma_i)$   from:
$$
e_{it}\sim\mathcal N(0,1),\quad   w_{i} \sim \text{Uniform}(1.5,2.5), \quad \bar\gamma_i\sim \text{Uniform}(-0.5,0.5).$$
Finally, $y_{it}$ is initialized by $ y_{i0}\sim  \text{Uniform}(0.52,1.52)$, and iteratively generated via
$$
y_{it}=\theta^{-1}\left( \frac{e_{it}}{y_{i(t-1)}(w_{i}+\bar\gamma_i)}\right).
$$
 The parameters of this DGP are chosen so that   $y_{i(t-1)}(w_{i}+\bar\gamma_i)>0$ for all $t$ with high probability. Therefore, $  \Pr(y_{it}\leq y \mid \mathcal{F}_{it})=\Phi(y_{i(t-1)}\beta_i(y))$  is satisfied.

\begin{figure}[htbp!]
	\begin{center}
	
		\vspace{1em}
\includegraphics[width=11cm]{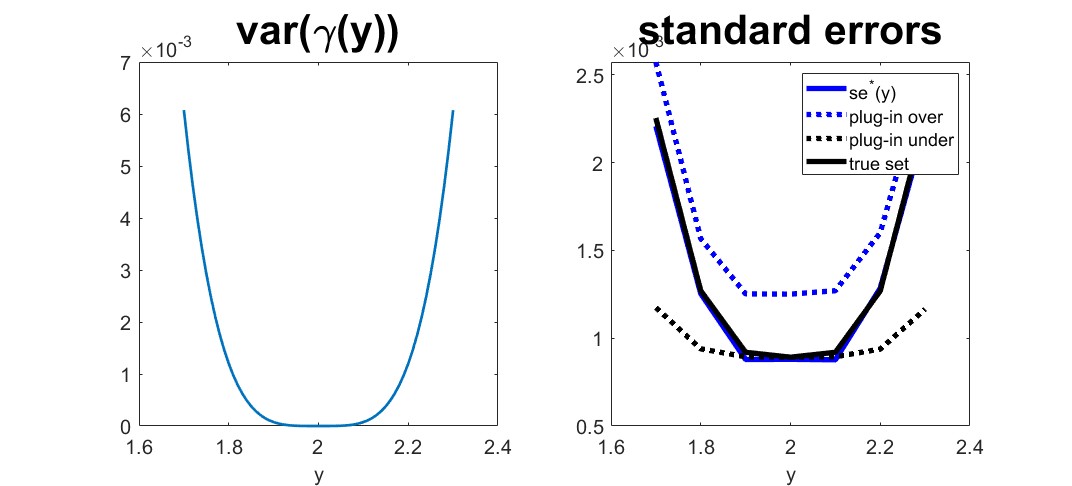}	
		\label{f2dfa} 
	\end{center}
 	\caption{\small Left: $\Var(\gamma_i(y))$. Right: estimated and true standard errors of $\widehat\theta(y) $ for $y\in M$ in the dynamic DR model.  
 	In the right panel, we plot four ``standard errors" for $y\in \mathcal{Y}$ under $N=T=300$. The   true standard error  is calculated as the standard deviation of   $\widehat\theta(y)$ from 1,000 simulations, while the other three are calculated using  a fixed simulation of data. Here $se^*(y)$ is the bootstrap IQR.}
\end{figure}

The object of interest is $\theta(y)$. Figure \ref{f2dfa} plots the variance of $\gamma_i(y) = \theta(y)\bar\gamma_i$, the noise level of $\beta_i(y)$, across $y\in\mathcal Y$. By construction, $\Var(\gamma_i(y))$ degenerates at $y = 2$, and increases as $y$ deviates from 2, which affects the rate of convergence for estimating $\theta(y).$ The right panel
plots the true standard error of the estimator $\widehat\theta(y)$, along with three estimators: the proposed bootstrap interquartile range (IQR) $se^*(y)$ 
 defined as     $ se^*(y) = (q^*_{.75}(y)-q^*_{.25}(y))/(z_{.75}-z_{.25})$, where $q_p^*$ is the bootstrap $p$-quantile of $\theta^*_b(y)-\theta(y)$ and $z_p$ is the  $p$-quantile of the standard normal.   The IQR is a consistent estimator for the asymptotic standard deviation, which is often used to replace the bootstrap variance, as it is challenging to show the latter is consistent in most cases. 
 
 The other two estimators, ``Plug-in-over" and ``Plug-in-under", are defined below.  The plug-in methods are clearly not robust to changes in $\Var(\gamma_i(y))$ across $y$. 

We examine the  coverage properties of $\theta(y)$ and  compare four inferential methods:

(i) \textit{Proposed}: the proposed uniform inference procedure using the interquartile range described in Remark \ref{rem5.1}. 

(ii) \textit{No-debias}: this method does not debias, while all other steps are the same as the proposed method. 

(iii) \textit{Plug-in-over}: this  method   plugs in  the estimated standard error, it uses the estimated 
$V_\psi(y)$ and $V_\gamma(y)$ by:
  \begin{eqnarray*}
  \widehat V_{\psi}(y)&=& \frac{1}{N}\sum_{i=1}^N   (S_{wz}'\vw_{i}\vw_{i}'S_{wz})
  \otimes \left[   \widehat {\mathbb A}_{y,1i} \Xi(y)\widehat {\mathbb A}_{y,1i}  \right].\cr
  \widehat  V_{\gamma}(y)&=&  \frac{1}{N}\sum_{i=1}^N   (S_{wz}'\vw_{i}\vw_{i}'S_{wz})
  \otimes  (\widehat\gamma_i(y)\widehat\gamma_i(y)'    )  
  \end{eqnarray*}
  where computing the estimators $\widehat {\mathbb A}_{y,1i} $ and $\widehat\gamma_i(y)$  is straightforward.  Meanwhile, we apply the Newey-West type estimator $\Xi(y)$  to estimate  $\E( \frac{1}{T}\sum_{s,t\leq T}     \psi_{is}(y_k)  \psi_{it}(y_l)' \mid W)  $. 

(iv) \textit{Plug-in-under}:  this  method also plugs in  the estimated standard error, but replaces $\widehat\Sigma_{NT}(y)$ of the Plug-in-over method with
          $$
          \widetilde\Sigma_{NT}(y) =\frac{1}{NT}\widehat V_{\psi}(y).
          $$

\medskip
 
Table  \ref{t_dr}  summarizes the   coverage probabilities of  $ 
\{\theta(y): y\in {\mathcal{Y}}\} $ 
out of 1,000 replications.
The results are generally as expected. The no-debias method   performs unsatisfactorily when $T\leq N $  due to the incidental parameter bias issue.  The plugin-over method assumes that there is arbitrary heterogeneity in $\beta_i(y)$, so is quite conservative for $T=100,200$; the plugin-under method is the standard treatment in the varying-coefficient literature, which assumes that the heterogeneity in $\beta_i(y)$ can be fully captured by covariates $\vw_i$.  The confidence band resulting from  $         \widetilde\Sigma_{NT}(y)$   undercovers $\theta(y)$.
The Plugin-over is conservative when  $T=200.$







\begin{table}[htbp]
\label{t_dr}
	\begin{center}
		\caption{\small  Coverage Probabilities of  $ 
\{\theta(y): y\in  \mathcal{Y}\} $}
		\begin{tabular}{cc|cccc}
			\hline
			\hline
		 	 &&  \multicolumn{4}{c}{Methods}    \\
$T$ & $N$    &  Proposed & No-debias   &  Plugin-over  &  Plugin-under   \\
\hline

50 & 300 & 0.950 & 0.569 & 0.871 & 0.354 \\
 & 400 & 0.948 & 0.411 &  0.786 & 0.225 \\
100 & 300 & 0.949 & 0.830 & 0.961 & 0.562 \\
 & 400 & 0.949 & 0.762 & 0.923 & 0.458 \\
 200 & 300 & 0.947 & 0.915 & 0.984 & 0.628 \\
 & 400 & 0.945 & 0.889 & 0.980 & 0.603 \\

\hline
 					
		\end{tabular}
			\end{center}

\end{table}%

\subsection{Calibrated DGP}
 
 We simulate a dynamic DR model from a heterogeneous-coefficient autoregressive model with calibrated parameters using the PSID data.
Specifically, we first estimate the following model
\begin{align}
y_{it} & =-\b_{0i}-\b_{1i}y_{i(t-1)}+\s\e_{it},\label{eq:Calib_Model1}
\end{align}
for each in-sample individual $i$ to calibrate $\b_{0i},\b_{1i}$
and $\s$, which we use to calibrate the second-stage model parameters
$\left(\t_{00},\t_{01},\t_{10},\t_{11}\right)$ and $\s_{0},\s_{1}$
by estimating the regressions
\begin{align}
\b_{0,i} & =\t_{00}+\vw_{1i}'\vtheta_{01}+\sigma_{0}\g_{0,i},\label{eq:Calib_Model2}\\
\b_{1,i} & =\t_{10}+\vw_{1i}'\vtheta_{11}+\s_{1}\g_{1,i}.\nonumber 
\end{align}
where $y_{it}$ is the outcome variables (residual log income), and $\vw_{1i}$ is the vector of individual characteristics consisting of the variables initial labor income, years of schooling, a white indicator and year of birth.

Let $\vx_{it}:=\left(1,y_{i(t-1)}\right)'$, $\vw_{i}:=\left(1,\vw_{1i}'\right)'$ and assume $\epsilon_{it}\sim \mathcal N(0,1)$. We 
then rewrite \eqref{eq:Calib_Model1} and \eqref{eq:Calib_Model2}
as following heterogeneous dynamic DR model
\begin{align*}
\Pr\left(y_{it}\leq y \mid \vx_{it}\right) & =\Phi\left(\widetilde\b_{0,i}\left(y\right)+\widetilde \b_{1,i}\left(y\right)y_{i(t-1)}\right)
\end{align*}
 where
\begin{align}
\widetilde{\b}_{0,i}\left(y\right) & :=\frac{y+\b_{0,i}}{\s}=\widetilde{\vtheta}_{0}\left(y\right)'\vw_{i}+\tilde{\s}_{0}\g_{0,i},\label{eq:Calib_DR}\\
\widetilde{\b}_{1,i}\left(y\right) & :=\frac{\b_{1,i}}{\s}=\widetilde{\vtheta}_{1}\left(y\right)'\vw_{i}+\tilde{\s}_{1}\g_{1,i},\nonumber \\
\widetilde{\vtheta}_{0}\left(y\right) & :=\left(\frac{y+\t_{00}}{\s},\frac{\vtheta_{01}'}{\s}\right)',\nonumber \\
\widetilde{\vtheta}_{1}\left(y\right) & :=\left(\frac{\t_{10}}{\s},\frac{\vtheta_{11}'}{\s}\right)',\nonumber \\
\tilde{\s}_0&:=\sigma_0/\sigma, \quad \tilde{\s}_1:=\sigma_1/\sigma.\nonumber
\end{align}
We then: (1) simulate $ y_{it}$
according to models \eqref{eq:Calib_Model1} and \eqref{eq:Calib_Model2}
based on the calibrated values of $\s$, $\s_{0}$, $\s_{1}$, $\theta_{00}$, $\vtheta_{01}$, $\theta_{10}$ and $\vtheta_{11}$; (2)  calculate the implied distribution regression parameters $\widetilde{\b}_{0,i}\left(y\right)$,
$\widetilde{\b}_{1,i}\left(y\right)$, $\widetilde{\vtheta}_{0}\left(y\right)$
and $\widetilde{\vtheta}_{1}\left(y\right)$ based on \eqref{eq:Calib_DR};
and (3) employ our proposed estimation methods to estimate
the quantile treatment effects of a counterfactual increase of the years of schooling variable by 1 year for every unit in the sample.  The size of the panel $N$ and each individual's length of observations $T_i$, are the same as the PSID data. That is, $N=1629$, and the average $T_i$ is 20.5.

Table \ref{tab:QTE_addT2342} compares the MSEs of our proposed estimators using analytical debiasing with the estimator which does not debias. We find that the analytical bias correction yields reductions in the MSE between $21$ and $45\%$ depending on the quantile index. In results not reported, we find that the Jackknife debiasing does not reduce the MSE of the uncorrected estimator as the split-sample estimator uses only a half of the time periods, whose number of time periods in each half sample is only ten. The Jackknife increases finite sample variance due to the estimation over the smaller panels.


\begin{table}[htp]
\caption{\label{tab:QTE_addT2342}QTE of Increasing $edu$ by 1 year}

\centering{}%
\begin{tabular}{ccccccc}
\hline 
\hline
 &  & \multicolumn{5}{c}{Estimator MSE $\times10^{-2}$}\tabularnewline
Quantiles &  & 15\% & 25\% & 50\% & 75\% & 85\%\tabularnewline
\hline 
No-debias &  & 0.028 & 0.019 & 0.024 & 0.022 & 0.016\tabularnewline
Analytical &  & 0.017 & 0.014 & 0.019 & 0.012 & 0.009\tabularnewline
\hline 
\end{tabular}
\end{table}

\section{Conclusion}
  
  We develop  estimation and inference methods for dynamic  distribution regression panel models that incorporate  heterogeneity both within and between units and are applicable to  a large number of economic settings.  An empirical investigation of labor income processes illustrates some economic insights that our approach can provide.  
  


Our model can be extended in several directions. For instance, one could include time fixed effects and covariates with homogeneous coefficients in the first stage.  This is useful for empirical applications which directly model an outcome variable with trends rather than the residuals. To  reduce the number of estimated parameters,  one could   model the individual coefficients in HDR using factor structures as in  \cite{chernozhukov2018inference}. One could also reduce dimensionality by modeling the between and within heterogeneity though a pseudo-factor structure where the value $y$ plays the role of time. 
Alternatively, one could use the  grouped fixed effects approach of 
\cite{bonhomme2015grouped}.  Finally, while our focus here is a panel comprising repeated time series observations on the same unit, our approach could be applied to a network setting in which there is contemporaneous dependence across units. We leave these extensions to future work.

 \appendix
 
 \section{Implementation Algorithms}\label{sec:bootstr}
 
 In this section we introduce the bootstrap algorithm for    confidence bands.
 
  \begin{algo}[Confidence Band for Projections of Coefficients]\label{alg5.1}  \hfill \break
  	\begin{description}

\vspace{-.3cm}

  	   		\item[Step 0] Pick the confidence level $p$, number of bootstrap repetitions $B$, region $\mathcal{Y}$ and a component of the linear projection. This amounts to selecting a vector $\veta$ such that $\eta'\vecc(\vtheta(y))$ over $y \in \mathcal{Y}$ is the function of interest. 
  		\item[Step 1] For $y \in \mathcal{Y}$, obtain the debiased DR coefficient estimates
  		$$\widehat\vbeta(y):=\{\widehat\vbeta_i(y): i=1,...,N_{01}(y)\}$$
  		using  \eqref{eq:bcdr}, and the estimates of the linear projection, $\widehat\vtheta(y)$, using \eqref{eq:lp}.

  		\item[Step 2]  Let  $\{(\widehat\vbeta_i^*(y), \vw_{i}^*, \vz_{i}^* ): i=1,..., N_{01}(y), y \in \mathcal{Y}\}$  be a random sample with replacement from  $\{(\widehat\vbeta_i(y), \vw_{i}, \vz_{i} ): i=1,..., N_{01}(y),y \in \mathcal{Y}\}$.  For $y \in \mathcal{Y}$, compute 
  		\begin{equation*} 
\widehat\vtheta^*(y)= \sum_{i=1}^{N_{01}(y)} \widehat\vbeta^*_i(y) \widehat{\vz}^*_i(y)' \left(\sum_{i=1}^{N_{01}(y)} \widehat{\vz}^*_i(y) \widehat{\vz}^*_i(y)' \right)^{-1}_,
\end{equation*}

$$\widehat{\vz}^*_i(y) := \sum_{j=1}^{N_{01}(y)} \vz^*_j \vw_{j}^{*'} \left(\sum_{j=1}^{N_{01}(y)} \vw^*_j \vw_{j}^{*'} \right)^{-1} \vw^*_i. $$


  		\item[Step 3] Repeat Step 2 for $B$ times to obtain $\{\widehat\vtheta^*_b(y): y \in \mathcal{Y}\}_{b=1}^B$.
  		
  		\item[Step 4]   Let $q_{p} $ be the booststrap $p$-quantile of 
		$$\left\{\sup_{y\in\mathcal Y}\left|\frac{ \eta'\vecc(\widehat\vtheta^*_b(y)-\widehat\vtheta(y)) }{s^*(y)}\right|\right\}_{b=1}^B 
		$$ 
		where $s^*(y)$  is the  rescaled interquartile  range of $\{\eta'\vecc(\widehat\vtheta^*_b(y)\}_{b=1}^B$. See remark \ref{rem5.1} below.
		
	\item[Step 5]  		Compute the asymptotic $p$-confidence band
  		$$
  		\CI_{p}(\eta'\vecc(\vtheta(y))):=[\eta' \vecc(\widehat\vtheta(y)) -q_{p}s^*(y),  \eta'\vecc(\widehat\vtheta(y)) +q_{p}s^*(y) ].
  		$$
  	\end{description}
  \end{algo}
   
      \begin{remark}[Standard Errors]\label{rem5.1}
  The bootstrap interquartile range rescaled with the standard normal distribution is defined as     $ s^*(y) = (q^*_{.75}(y)-q^*_{.25}(y))/(z_{.75}-z_{.25})$, where $q_p^*$ is the bootstrap $p$-quantile of $\eta'\vecc(\widehat\vtheta^*_b(y)-\widehat\vtheta(y))$ and $z_p$ is the  $p$-quantile of the standard normal. 
     \end{remark}

   For the actual and counterfactual distributions, it is convenient to express the estimator in \eqref{eq:ecd} as
   $$
   \widehat G_{t}(y)=\frac{1}{N}\sum_{i=1}^N\Psi_{i}(y; h(\vx_{it}), \widehat\vbeta^g_i(y))
   $$
with 
   \begin{eqnarray*} 
   \Psi_{i}(y; \vx, \vb)  
  & =& 1\{i\leq N_{01}(y)\}\Lambda(-\vx'\vb) +\frac{N_1(y)}{N}\cr &&- 1\{i\leq N_{01}(y)\}\frac{1}{2T}\tr\left(\ddot\Lambda(-\vx'\vb)\vx \vx'\widehat \Sigma_i(y)^{-1} \right),
   \end{eqnarray*}
   to simplify the notation. 
 
  \begin{algo}[Confidence Band for Actual and Counterfactual Distribution]\label{alg5.2}   \hfill \break
  	\begin{description}

\vspace{-.4cm}

  	   		\item[Step 0] Pick the confidence level $p$, number of bootstrap repetitions $B$, and region $\mathcal{Y}$. 
  	   		
		\item[Step 1] For each $y \in \mathcal{Y}$, obtain the debised estimate  $\widehat G_{t}$ from \eqref{eq:ecd}.

		\item[Step 2]  Let  $\{( \vx_{it}^*,\widehat\vbeta_i^*(y), \vw_{i}^*, \vz_{i}^* ): i=1,..., N_{01}(y)\}$ be  a random sample    with replacement from $\{ (\vx_{it},\widehat\vbeta_i(y), \vw_{i}, \vz_{i}): i=1,..., N_{01}(y)\}$.  Compute
			$$
	\widehat G^*_{t}(y) = \frac{1}{N}\sum_{i=1}^N\Psi_{i}(y;h_{it}(\vx_{it}^*),  \widehat\vbeta_i^{g*}(y)), \quad \widehat\vbeta_i^{g*}(y)=\widehat\vbeta^*_i(y) + \widehat\vtheta^*(y)[g(\vz_{i}^*)-\vz^*_i],
		$$
		where $\widehat\vtheta^*(y)$ is defined as in Step 2 of Algorithm \ref{alg5.1}

		\item[Steps 3-5] The same as Steps 3-5  of Algorithm \ref{alg5.1}, with $(\widehat G^*,\widehat  G)$ in place of $(\eta' \vecc(\widehat\vtheta^*), \eta'\vecc(\widehat\vtheta))$.

		
		
	\end{description}
\end{algo}
  
  	The bootstrap inference for the actual distribution $F_t(y)$ is a special case with $h(\vx_{it})=\vx_{it}$ and $g(\vz_i) = \vz_i$. 
Finally,  the algorithm below computes the confidence band for the quantile effects. 

   \begin{algo}[Confidence Bands for Quantile Effect]\label{alg5.4}   \hfill \break
  	\begin{description}

\vspace{-.3cm}

  	   		\item[Step 0] Pick the confidence level $p$, number of bootstrap repetitions $B$, and region of quantile indexes $\mathcal{T}$. 
  	   		
		\item[Step 1] For any $\tau \in \mathcal{T}$, obtain the estimate $\widehat \QTE_t(\tau)$ using \eqref{eq:qe}.


\item[Step 2] 
Compute the bootstrap draws of $\widehat \QTE_t(\tau)$:

(1) Obtain $\widehat F_t^*$ and $\widehat G_{t}^* $ as in step 2 of Algorithm \ref{alg5.2}. For $\widehat F_t^*$ , set $h(\vx_{it})=\vx_{it}$ and $g(\vz_i) = \vz_i$.

(2) For any $\tau \in \mathcal{T}$, calculate 
$$
\widehat\QTE^*_t(\tau)= \widetilde \phi(\widehat{ G}^*_{t},\tau) - \widetilde \phi(\widehat{ F}^*_t,\tau).
$$

		\item[Steps 3-5]  The same as Steps 3-5 of Algorithm \ref{alg5.1}, with $(\widehat{\QTE}_{t}^*, \widehat{\QTE}_{t})$  in place of $(\eta' \vecc(\widehat\vtheta^*), \eta'\vecc(\widehat\vtheta))$.

		
		
	\end{description}
\end{algo}


  \begin{remark}[Computation]
 The most computationally expensive task is the computation of coefficient estimates, which is conducted only in Step 1 of the algorithms.
 \end{remark}
 
 \begin{remark}[Stationary Distributions and Effects]
 The bootstrap algorithms for stationary distributions and quantile effects are omitted because their steps are similar to the corresponding  steps in Algorithms \ref{alg5.2} and \ref{alg5.4}.
 \end{remark}

  \section{Technical Details}
  
    \subsection{Debiased estimators for $\vbeta_i(y)$}
  
  First, recall that  $N_0(y)$ is  the number of indexes $i$ for which $y < \underline{y}_i$, $N_1(y)$ is the number of indexes $i$ for which $y \geq \overline{y}_i$, and $N_{01}(y) = N - N_0(y) - N_1(y)$ is the number of indexes $i$ for which  $\widetilde\vbeta_i(y)$ exists.  In addition, the imposed assumptions ensure that with probability approaching one, the following event holds: 
  
  For all $ y\in\mathcal Y$, and all $i=1,..., N$, 
  we have 	$
 	\min_{t\leq T} y_{it}<y< \max_{t\leq T} y_{it}
 	$. 
 	
 	Under this event, $N_0(y)= N_1(y)=0$ and $N_{01}(y)=N$ for all $y\in\mathcal Y. $ So throughout the technical proofs, we condition on this event, which would not affect the asymptotic results. 
 
   \subsubsection{ Analytical Debias}

 

The initial estimator 
can be expanded as
 $$
 \widetilde\vbeta_i(y)-\vbeta_i(y) =- \mathbb A_{1i} (y)   \nabla Q_{y,i}(\vbeta_i(y))
  -  \frac{1}{T}B_{i,1T}(y)  - \frac{1}{T}B_{i,2T} (y)
 + R_{i}(y)  
 $$
 where $
\mathbb A_{1i}= [\nabla^2 \E Q_{y,i}( \vbeta_i(y))]^{-1}$, $ \nabla Q_{y,i}(\vbeta_i(y))= \frac{1}{T}\sum_{t=1}^T\psi_{it}(y)$ and $R_i(y)$ is the higher order term.
To describe the first-order biases $B_{i,1T}(y)$ and $B_{i,2T}(y)$,  write
   $A_{1i}= [\nabla^2 Q_{y,i}( \vbeta_i(y))]^{-1}$, $
A_{2i} = \nabla^3Q_{y,i}( \vbeta_i(y)),
$
  and $
\mathbb A_{2i} = \nabla^3\E Q_{y,i}( \vbeta_i(y)).
$ 
Then
$$
\mathbb A_{1i}\sqrt{T}\nabla Q_{y,i}(\vbeta_i(y))=  \frac{1}{\sqrt{T}}\sum_t\mathbb A_{1i} \psi_{it}(y),
\quad  \sqrt{T} [A_{1i}^{-1}-\mathbb A_{1i}^{-1}] =\frac{1}{\sqrt{T}}\sum_t\varpi^2_{it}(y).
$$
 Here $\varpi^2_{it}(y)$ is $\dim(\beta_i)\times \dim(\beta_i)$. Let  $\varpi^2_{it,k}(y)$ be its  $k$ th column and 
$$
V_{i,k}(y):=\Var \left[\frac{1}{\sqrt{T}}\sum_t\ell_{it} \right]=\begin{pmatrix}
M_1^i(y) & M_{2,k}^i(y)'\\
M_{2,k}^i(y) & M_{3,k}^i(y)
\end{pmatrix},\quad \ell_{it}:= \begin{pmatrix}
\mathbb A_{1i} \psi_{it}(y)\\
 \varpi^2_{it,k}(y)
\end{pmatrix}.
 $$
 Then 
 \begin{eqnarray}\label{eqbias}
 B_{i,1T}(y)&=&	 
  \frac{1}{2  } \mathbb A_{1i} \mathbb A_{2i}\E[(\mathbb A_{1i}\sqrt{T} \nabla Q_{i}(\beta_i)
  )\otimes ( \mathbb A_{1i}\sqrt{T}\nabla Q_{i}(\beta_i)]=
\frac{1}{2  } \mathbb A_{1i} \mathbb A_{2i} \vecc(M_1^i(y))\cr
B_{i,2T}(y)&=& -  \mathbb  A_{1i} \E [\sqrt{T}(A_{1i}^{-1}-\mathbb  A_{1i}  ^{-1} )\mathbb  A_{1i}\sqrt{T} \nabla Q_{i}(\beta_i)]= -  \mathbb  A_{1i}  
\begin{pmatrix}
\tr(M_{2,1}(y))\\
\vdots\\
\tr(M_{2,\dim(\beta_i)}(y))
\end{pmatrix}.\cr 
 \end{eqnarray}

 Hence  we can estimate $ B_{i,1T}(y)$ and $ B_{i,2T}(y)$ by replacing $V_{i,k}(y)$ by its estimator $\widehat V_{i,k}(y)$; the latter can be obtained by the Newey-West truncation.
  $$
  \widehat V_{i,k}(y)=  \frac{1}{T}\sum_{t=1}^T\widehat \ell_{it} \widehat \ell_{it} '
 +\frac{1}{T}\sum_{h=1}^{L}\sum_{t>h}[\widehat \ell_{it} \widehat \ell_{i(t-h)} '+\widehat \ell_{i(t-h)} \widehat \ell_{it} ' ]:=\begin{pmatrix}
\widehat M_1^i(y) & \widehat M_{2,k}^i(y)'\\
\widehat M_{2,k}^i(y) &\widehat  M_{3,k}^i(y)
\end{pmatrix}.
  $$
  Let $\widehat B_{i,1T}= \frac{1}{2  } \widehat   {\mathbb A}_{1i}\widehat  {\mathbb A}_{2i} \vecc(\widehat M_1^i(y))$ and $\widehat B_{i,2T}   $ be defined as $B_{i,2T}(y)$ with ${\mathbb A}_{1i}$  and $M_{2,k}^i(y)$ replaced with their estimates:
   \begin{equation}  
  \widehat\vbeta_i(y)= \widetilde\vbeta_i(y) + \frac{\widehat B_{i,1T}(y)}{T}+\frac{\widehat B_{i,2T}(y)}{T}, \quad i = 1,\ldots, N_{01}(y),
  \end{equation}
  We use the notation
  $$
  \widehat B_{i,T}(y)=-\widehat B_{i,1T}(y)-\widehat B_{i,2T}(y)
  $$
  so that we can express explicitly as a debiased estimator: $ \widehat\vbeta_i(y)=\widetilde\vbeta_i(y)-\frac{\widehat B_{i,T}(y)}{T}$.
 
When $\Lambda$ is the logit link, $\psi_{it}= -\vx_{it}[1\{y_{it}\leq y\}-\Lambda(-\vx_{it}'\vbeta_i(y))]$, and  $ \varpi^2_{it}(y)=a- \mathbb E a$ where $a=-\vx_{it}\vx_{it}'\Lambda(-\vx_{it}'\vbeta_i(y))[1-\Lambda(-\vx_{it}'\vbeta_i(y))].$ Also $M_{2,k}^i(y)=0.$

  \subsubsection{Jackknife Debias}
 
 Alternative to the analytical debias, we can also employ the sample-splitting Jackknife debias to remove the higher order bias, which was used for instance, by \cite{dhaene2015split, okui2019panel}.

  Randomly split $\{1,...,T\}=\mathcal I\cup\mathcal I^c$, so that $|\mathcal I|= {T/2}$. Let 
  $
  \widetilde\beta_{i,\mathcal I}(y) 
  $ be the same estimated $\vbeta_i(y)$, but using data only for $t\in\mathcal I$. Similarly, let  $
  \widetilde\beta_{i,\mathcal I^c}(y) 
  $ be the estimated $\vbeta_i(y)$, but using data only for $t\in\mathcal I^c$. Let 
  $$
  \bar\beta_{i}(y) = \frac{1}{2}[\widetilde\beta_{i,\mathcal I}(y) +\widetilde\beta_{i,\mathcal I^c}(y) ].
  $$
Then the Jackknife debiased estimator is defined as: $$
  \widehat\vbeta_i(y)= 2\widetilde\vbeta_i(y)-   \bar\beta_{i}(y) .
  $$

\medskip
  
  \subsection{The counterfactual  stationary distribution}

  \subsubsection{The model}
 
We recall that the stationary distribution is defined as 
$
F_{\infty}(y) = \mathbb{E} [F_{i,\infty}(y)],
$ 
where $F_{i,\infty}(y) = \sum_{k: y_i^k \leq y} \pi_{ik}$; the ergodic probabilities $\boldsymbol{\pi}_i = (\pi_{i1}, \ldots, \pi_{iK})$ are
$$
\boldsymbol{\pi}_i = (\boldsymbol{A}_i'\boldsymbol{A}_i)^{-1} \boldsymbol{A}_i'\boldsymbol{e}_{K+1}, \quad \boldsymbol{A}_i = \left(\begin{array}{c}
     \boldsymbol{I}_K - \boldsymbol{P}_i \\
     \boldsymbol{1}' 
\end{array}\right)
$$ and $\boldsymbol{e}_{K+1}$ is the $(K+1)$th column of $\boldsymbol{I}_{K+1}$.
Also, $\boldsymbol{P}_i$ is a $K \times K$ matrix with element
 $$
P_{i,jk} = \Pr(y_{it} = y_i^j \mid y_{i(t-1)} = y_i^k, \mathcal{F}_{it}) = \Lambda\left(- \vx_{i}^{k'}\beta_i(y_i^j)\right) - 1(j > 1)\Lambda\left(- \vx_i^{k'}\beta_i(y_i^{j-1})\right).
 $$
 Hence we can write
 $$
F_{\infty}(y)=\mathbb E  f_i(\vbeta_i, y )
$$
where $\vbeta_i= \vecc( \beta_i(y_i^1),...,\beta_i(y_i^K))$ and 
$$
 f_i(\vbeta_i, y ) =\sum_{k=1}^K 1\{y_i^k \leq y\} \boldsymbol{e}_k' \boldsymbol{G}_i(\vbeta_i),\quad 
\boldsymbol{G}_i(\vbeta_i) =(\boldsymbol{A}_i'\boldsymbol{A}_i)^{-1} \boldsymbol{A}_i'\boldsymbol{e}_{K+1}.
$$

The counterfactual stationary  distribution is defined as 
  $$
  G_{\infty}(y) = \mathbb E f_i(\vbeta_i, \vtheta_i, y),\quad \vtheta_i=\vecc(\vtheta(y_i^1),...,\vtheta(y_i^K)),
  $$
  where
$$
f_i(\vbeta_i,\vtheta_i, y)= \sum_{k=1}^K 1\{y_i^k \leq y\} \boldsymbol{e}_k'  (\boldsymbol{A}_i(\vbeta_i,\vtheta_i)'\boldsymbol{A}_i(\vbeta_i,\vtheta_i))^{-1} \boldsymbol{A}_i(\vbeta_i,\vtheta_i)'\boldsymbol{e}_{K+1}
$$
and 
$\boldsymbol{A}_i(\vbeta_i,\vtheta_i)$ is  defined as 
  $\boldsymbol{A}_i$ but with $\vbeta_i$ replaced by $$\vbeta_i^g= \vecc(\beta_i(y_i^k)+ \vtheta(y_i^k)(g(z_i)-z_i): k=1,...,K).$$

  \subsubsection{Estimation of stationary distributions}

Under the condition that $\Lambda\left(- \vx_{i}^{k'}\beta_i(y_i^j)\right) =1$ for $j=K$, we have
$$
\widehat F_{\infty}(y) =\frac{1}{N}\sum_{i=1}^N  f_i(\widehat\vbeta_i, y ) - \frac{1}{T}\frac{1}{N}\sum_{i=1}^N\widehat B_{\boldsymbol{\pi}_i}
$$
where $\widehat B_{\boldsymbol{\pi}_i}= \frac{1}{2} \tr \left[\partial^2_\beta  f_i( \widehat \vbeta_{i},y )    \frac{1}{T}\sum_t \widehat\vZ_{it}\widehat \vZ_{it}' \right]$ and
$$
\widehat\vZ_{it}=\vecc(\widehat{\mathbb A}_{1i}(y_i^1) \widehat\psi_{it}(y_i^1),...,\widehat{\mathbb A}_{1i}(y_i^K) \widehat\psi_{it}(y_i^K)).
$$
  
Similarly, we estimate $G_{\infty}$ by the following bias-corrected estimator: 
  $$ \widehat G_{\infty}(y)=\frac{1}{N}\sum_{i} f_i(\widehat\vbeta_i, \widehat\vtheta_i, y)-  \frac{1}{2NT}\sum_i\tr \left[
 \partial^2_\beta f_i(\widehat \vbeta_{i}, \widehat\vtheta_i, y)   \frac{1}{T}\sum_t\widehat \vZ_{it}\widehat \vZ_{it}' \right].$$

 \subsection{Definitions of leading terms in expansions}\label{sec:Define}
  
  We shall show that 
  \begin{eqnarray*}
  \widehat F_t(y)-F_t(y) &=&\frac{1}{N }\sum_{i=1}^N[\frac{1}{\sqrt{T}}d^{0}_{\psi,i}(y) + d^{0}_{\vgamma,i}(y)]+ o_P(\zeta_{NT}(y))\cr
  \widehat G_{t}(y)-G_{t}(y) &=&\frac{1}{N }\sum_{i=1}^N[\frac{1}{\sqrt{T}}d^{II}_{\psi,i}(y) + d^{II}_{\vgamma,i}(y)]+ o_P(\zeta_{NT}(y))\cr
   \widehat F_{\infty}(y)-F_{\infty}(y) &=&\frac{1}{N }\sum_{i=1}^N[\frac{1}{\sqrt{T}}d^{\infty}_{\psi,i}(y) + d^{\infty}_{\vgamma,i}(y)]+ o_P(\zeta_{NT}(y))\cr
  \widehat G_{\infty}(y)-G_{\infty}(y) &=&\frac{1}{N }\sum_{i=1}^N[\frac{1}{\sqrt{T}}d^{\infty, II}_{\psi,i}(y) + d^{\infty, II}_{\vgamma,i}(y)]+ o_P(\zeta_{NT}(y))\cr
\widehat \QTE_{t}(\tau)-\QTE_{t}(\tau) &=&\frac{1}{N }\sum_{i=1}^N[\frac{1}{\sqrt{T}}p_{\psi,i}(\tau) + p_{\vgamma,i}(\tau)]+ o_P(\bar\zeta_{NT}(\tau)).
  \end{eqnarray*}

The involved terms are defined as follows. 
  We introduce some notation.  Let 
  $$
   \vZ_{jt,i}=\vecc(\mathbb A_{1j}(y_i^1)\psi_{jt}(y_i^1),...,\mathbb A_{1j}(y_i^K)\psi_{jt}(y_i^K)).\quad \vZ_{it}:= \vZ_{it,i}.
  $$
   In addition,   $q_{\infty, 0}(\tau)=\phi(  F_{\infty}, \tau)$,   and $q_{\infty, II}(\tau)=\phi(  G_{\infty}, \tau),$ and 
	 $
	 \vgamma_{j,i}=\vecc(\vgamma_{j}(y_i^1),...,\vgamma_j(y_{i}^K)).
	 $
\begin{eqnarray}\label{eqb.2n}
d^0_{\psi,i}(y)&:=&\frac{1}{\sqrt{T}} \sum_{t=1}^T   \dot\Lambda(-\vx_{it}'\vbeta_i(y))\vx_{it}'\mathbb A_{1i}(y) \psi_{it}(y) ,\cr
 d_{\psi,i}^{II}(y)&=& \frac{1}{\sqrt{T}} \sum_{t=1}^T   [   \vw_{i}'   S_{wz} \bar G(y) +\dot\Lambda(-h_{it}(\vx_{it})'\vbeta_i^g(y))\vx_{it}' ]   \mathbb A_{1i}(y)   \psi_{it}(y) \cr
 d^0_{\vgamma,i}(y)&=& \Lambda(-\vx_{it}'\vbeta_i(y))-\E_t\Lambda(-\vx_{it}'\vbeta_i(y))\cr
	  d_{\vgamma,i}^{II}(y)&=&  \vw_{i}'   S_{wz} \bar G(y)\vgamma_i(y) +  \Lambda(-h_{it}(\vx_{it})'\vbeta_i^g(y))  -\E_t  \Lambda(-h_{it}(\vx_{it})'\vbeta_i^g(y)),\cr
	   d^{\infty}_{\psi,i}(y)&:=& -\partial_\beta f                 _i(\vbeta_i, y)'  \frac{1}{\sqrt{T}} \sum_{t=1}^T  \vZ_{it} , \cr
	 d^{\infty}_{\vgamma,i}(y)&:=& f(\vbeta_i,y)-\E  f(\vbeta_i,y)\cr
 d_{\psi, j}^{\infty,II}(y)&:=&-\frac{1}{\sqrt{T}}\sum_{t} \partial_\beta f                 _j(y)' \vZ_{jt}
 +\frac{1}{\sqrt{T}}\sum_{t} H_{jt}(y)\vw_j'S_{wz},\quad H_{jt}(y):=\frac{1}{N}\sum_{i}\partial_{\vtheta} f_i( y)'   \vZ_{jt,i}\cr
 d_{\vgamma, j}^{\infty, II}(y) &:=&\frac{1}{N}\sum_{i}\partial_{\vtheta} f_i(\vbeta_i, \vtheta_i, y)' \vgamma_{j,i}\vw_j'S_{wz} +\frac{1}{N}\sum_{i}  f(\vbeta_i, \vtheta_i, y) -\mathbb E  f(\vbeta_i, \vtheta_i, y) , \cr
 \cr
 p_{\psi,i}(\tau)&=&  \kappa^{II}(\tau)d_{\psi,i}^{II}(\phi(  G_{t}, \tau))+ \kappa^{0}(\tau)d_{\psi,i}^0(\phi(  F_{t}, \tau))\cr
  p_{\vgamma,i}(\tau)& =&    \kappa^{II}(\tau)d_{\vgamma,i}^{II}(\phi(  G_{t}, \tau))+ \kappa^{0}(\tau) d_{\vgamma,i}^0(\phi(  F_{t}, \tau))
\end{eqnarray}
where $\bar G(y)= -\E_t \dot\Lambda(-h_{it}(\vx_{it})'\vbeta_i^g(y))\vecc(\vx_{it}(g(\vz_{i})-\vz_{i})'),$  and 
\begin{eqnarray}\label{eq4.10pda}
\kappa^{II}(\tau)&=&\frac{-1}{\dot G_{t}( \phi(  G_{t}, \tau))},\quad 
\kappa^0(\tau)=\frac{1}{\dot F_{t}(\phi(  F_{t}, \tau))}
\end{eqnarray}

\medskip
 	  
\subsection{Further technical conditions}

We further assume the following:
\begin{assumption}[For Jackknife]\label{assa.1}
 (i) For each $i$, $\{(Y_{it}, \vx_{it}): t=1,...,T\}$ is serially strictly stationary. 
	(ii) Long-run covariance:
		write $$
		\mu_{i,T}(y):= \frac{1}{\sqrt{T}}\sum_{t=1}^T(\psi_{it}(y)', \vecc(\varpi_{it}^2(y))').
		$$
		Then almost surely, 
		$\lim_{T\to\infty} \Cov(    \mu_{i,T}(y))$ exists and 
		$$
		\max_i\sup_y\|  \Cov(    \mu_{i,T}(y))-\lim_{T\to\infty} \Cov(   \mu_{i,T}(y))\|=O(T^{-1/2}).
		$$

  \end{assumption}
  
  For the estimation of $\QTE$, we additionally require the following.

 \begin{assumption}[For $\QTE$ and the stationary distribution]\label{ase.48}

There is $C>0$, so that 
\begin{eqnarray*}
\Var_t(d_{\vgamma,i}^0(q_0(\tau)))+\Var_t(d_{\vgamma,i}^{II}(q_{II}(\tau)))&\leq& C\Var_t(\kappa^0(\tau)d_{\vgamma,i}^0(q_0(\tau))+\kappa^{II}(\tau)d_{\vgamma,i}^{II}(q_{II}(\tau))).
 \end{eqnarray*}
 and $\Var_t(d^{\infty}_{\vgamma,i})+\Var_t(d^{\infty, II}_{\vgamma,i})=O(\Var(p_{\vgamma,i}^{\infty, II}))$. 
 \end{assumption}

  \section{Theory for the debiased estimators $\widehat\beta_i$}
  
Using the true value $\beta_i:=\vbeta_i(y)$ (we drop $y$ for notational simplicity), define
\begin{eqnarray}\label{eqR45}
R_{i,4}&=& \frac{1}{2}  \mathbb A_{1i} \mathbb A_{2i}\E[( \mathbb A_{1i}\nabla Q_{i}(\beta_i)
)\otimes ( \mathbb A_{1i}\nabla Q_{i}(\beta_i)) ]
- \frac{1}{2}  \mathbb A_{1i} \mathbb A_{2i}[( \mathbb A_{1i}\nabla Q_{i}(\beta_i)
)\otimes ( \mathbb A_{1i}\nabla Q_{i}(\beta_i)]\cr
R_{i,5}&=&\mathbb  A_{1i}  [(A_{1i}^{-1}-\mathbb  A_{1i}  ^{-1} )\mathbb  A_{1i} \nabla Q_{i}(\beta_i)
  -\E((A_{1i}^{-1}-\mathbb  A_{1i}  ^{-1} )\mathbb  A_{1i} \nabla Q_{i}(\beta_i))]\cr
\end{eqnarray}

 Standard   first-order   Taylor expansion gives
    \begin{equation}\label{eq3.1}
    \widetilde\beta_i-\beta_i= - \nabla^2 Q_{i}( \beta_i)^{-1} \nabla Q_{i}(\beta_i)
    +\Delta_i
    \end{equation}
    where for some $\vbeta_i^*$ between $\widetilde\beta_i$ and $\beta_i$,
    $$ \Delta_{i}=-\nabla^2 Q_{i}( \beta_i)^{-1} [\nabla^2 Q_{i}( \vbeta_i^*)-\nabla^2 Q_{i}( \beta_i)](\widetilde\beta_i-\beta_i).
    $$
Let   $A_{1i}= [\nabla^2 Q_{i}( \beta_i)]^{-1}$, $
A_{2i} = \nabla^3Q_{i}( \beta_i),
$
$
\mathbb A_{1i}= [\nabla^2 \E Q_{i}( \beta_i)]^{-1}$, $
\mathbb A_{2i} = \nabla^3\E Q_{i}( \beta_i).
$ 
 \subsection{Asymptotic expansion for $\widehat\beta_i$}
 Recall the   jackknife debiased estimator $$\widehat\beta_i:= \widetilde\beta_i- (\bar\beta_i-\widetilde\beta_i )$$
 and the analytical debiased estimator is given by
  $$
  \widehat\beta_i= \widetilde\beta_i+\frac{1}{T} [\widehat B_{i,1T}+\widehat B_{i,2T}].
  $$

    \begin{lem}[Jackknife debias]\label{la.1jackdebias} Additionally assume Assumption \ref{assa.1}.  
  	Let $R_{i,d,\mathcal I}$ be similarly defined using data in $\mathcal I$, and $\bar R_{i,d}=\frac{1}{2}[R_{i,d,\mathcal I}+R_{i,d,\mathcal I^c}]$.
  	Then the jackknife estimator satisfies: for some $R_{i,9}$,  (we drop $y$ for notational simplicity)
  	$$
  	\widehat\beta_i-\beta_i=- \mathbb A_{1i}  \frac{1}{T}\sum_t\psi_{it}(y)
  	+R_{i,9}+  2R_{i,4} +2R_{i,5}
  	-\bar R_{i,4}-\bar R_{i,5}
  	$$
  	where $\sup_y\frac{1}{N}\sum_i\|R_{i9}\|^2=O_P(T^{-3})$ and  $\frac{1}{T}\sum_t\psi_{it}(y)=\nabla Q_i(\beta_i)$.
  \end{lem}

  \begin{proof}

By Lemma \ref{la.1},  	for $\sup_y\frac{1}{N}\sum_i\|\Delta_i\|^2=O_P(T^{-3})$,
\begin{eqnarray*}
\widetilde\beta_i-\beta_i&=&  - \mathbb A_{1i} \nabla Q_{i}(\beta_i)
-  \frac{1}{T}B_{i,1T}  - \frac{1}{T}B_{i,2T} 
+ R_{i,4}+R_{i,5} +\Delta_i\cr
&=& - \mathbb A_{1i} \nabla Q_{i}(\beta_i)
-  \frac{1}{T}B_{i}  
+ R_{i,4}+R_{i,5} +\Delta_i+ R_{i,7}
\end{eqnarray*}
 where $R_{i,4}, R_{i,5}$ are defined in (\ref{eqR45}), and 
\begin{eqnarray}\label{eqr7}
B_{i }&=& \lim_{T\to\infty}B_{i,1T} + \lim_{T\to\infty}B_{i,2T}\cr
R_{i,7}&=&   \frac{1}{T} ( \lim_{T\to\infty}B_{i,1T} + \lim_{T\to\infty}B_{i,2T}-B_{i,1T}-B_{i,2T} ).
\end{eqnarray}
Note that the existence of $ \lim_{T\to\infty}B_{i,1T} + \lim_{T\to\infty}B_{i,2T}$ follows from 
 Assumption \ref{assemp} because $B_{i,1T}+B_{i,2T} $
is a function of $ \Cov(    \mu_{i,T}(y)|\vw_{i})$ and $ \mathbb  A_{1i}$;  $ \mathbb  A_{1i}$ does not depend on $T$ due to the serial stationarity.  We introduce $B_{i } = \lim_{T\to\infty}B_{i,1T} + \lim_{T\to\infty}B_{i,2T}$ in the above expansion so that the   higher-order bias $ -  \frac{1}{T}B_{i }$ becomes independent of $T$; in contrast $B_{i,1T} +  B_{i,2T}$ may depend on $T$ due to the weak serial dependence. The fact that $B_i$ is independent of $T$ is required  to apply the jackknife debias device, as we  show below. By Assumption \ref{assemp}
\begin{eqnarray*}
  \frac{1}{N}\sum_i\|R_{i,7}\|^2
 &\leq& O(T^{-3}).
\end{eqnarray*}

 Similar expansion holds for $\widetilde\beta_{i,\mathcal I}$ and $\widetilde\beta_{i,\mathcal I^c} $, whose sample size is $T/2$. For instance, 
 $$
 \widetilde\beta_{i,\mathcal I}-\beta_i=- \mathbb A_{1i} \nabla Q_{i,\mathcal I}(\beta_i)
  -  \frac{1}{T/2}B_{i }
 + R_{i,4\mathcal I}+R_{i,5\mathcal I} +\Delta_{i\mathcal I}+ R_{i,7\mathcal I}.
 $$
Let $\bar\Delta_i=\frac{1}{2}[ \Delta_{i,\mathcal I}+ \Delta_{i,\mathcal I^c}].$  Therefore, with
  $\bar\beta_i=  \frac{1}{2}[\widetilde\beta_{i,\mathcal I}  +\widetilde\beta_{i,\mathcal I^c}  ]$:  
 \begin{eqnarray*}
\bar\beta_i-\beta_i&=&-\mathbb A_{1i}   \frac{1}{2}[\nabla Q_{i,\mathcal I}(\beta_i)+ \nabla Q_{i,\mathcal I^c}(\beta_i)]  -  \frac{2}{T} B_i 
 +  \bar R_{i,4}+\bar R_{i,5} +\bar R_{i,7} +\bar\Delta_i \cr
 &=&-\mathbb A_{1i}  \nabla Q_{i}(\beta_i) -  \frac{2}{T}  B_i 
  +  \bar R_{i,4}+\bar R_{i,5} +\bar R_{i,7} +\bar\Delta_i+    R_{i,8}
\end{eqnarray*}
where we note that the definition of $B_i$ does not depend on the split sample, and 
\begin{eqnarray*}
\bar R_{i,7}&=& \frac{1}{2}[R_{i7,\mathcal I}+R_{i7,\mathcal I^c}]\Rightarrow  \frac{1}{N}\sum_i\|\bar R_{i,7}\|^2= O_P(T^{-3}), \quad (\text{Assumption \ref{assemp}})\cr
 \frac{1}{N}\sum_i\|\bar \Delta_{i}\|^2&=& O_P(T^{-3}) \cr
 R_{i,8}&=&- 1\{T\text{ is odd}\}\frac{\mathbb A_{1i}}{T-1}[ \nabla Q_{i}(\beta_i) - \nabla Q_{i,\mathcal I}(\beta_i) ],\quad \text{ if } |\mathcal I| =(T+1)/2 \text{when $T$ is odd}.
\end{eqnarray*}
Then uniformly in $y$, 
$$
\frac{1}{N}\sum_i\| R_{i,8}\|^2\leq O_P(\frac{1}{T^2})  \frac{1}{N}\sum_i[\| \nabla Q_{i}(\beta_i) \|^2+\| \nabla Q_{i,\mathcal I}(\beta_i)\|^2  ] = O_P(T^{-3}).
$$

Hence $$
\bar\beta_i-\widetilde\beta_i = -\frac{1}{T} {B_i} 
+  \bar \Delta_{i}+ \bar R_{i,4} + \bar R_{i,5} +\bar R_{i,7}+    R_{i,8}
-   ( \Delta_i+R_{i,4} +R_{i,5}+R_{i,7} ).
$$
So  the jackknife debiased estimator  $\widehat\beta_i:= \widetilde\beta_i- (\bar\beta_i-\widetilde\beta_i )$ admits:
$$
\widehat\beta_i-\beta_i=- \mathbb A_{1i} \nabla Q_{i}(\beta_i) 
 +R_{i,9}+  2R_{i,4} +2R_{i,5}
  -\bar R_{i,4}-\bar R_{i,5}
$$
where $R_{i,9}=2\Delta_i -\bar \Delta_i-    R_{i,8}  +2\bar R_{i,7}-\bar R_{i,7}$ and $ \frac{1}{N}\sum_i\| R_{i,9}\|^2= O_P(T^{-3}).$
  \end{proof}

The following lemma characterizes the analytical debias, without assuming time series stationarity. 
\begin{lem}[Analytical debias]\label{la.2anadebias}  
	Use the true value $\beta_i:=\vbeta_i(y)$ (we drop $y$ for notational simplicity).
 The analytical-debiased estimator satisfies: for some $R_{i,9}$, 
	$$
	\widehat\beta_i-\beta_i=- \mathbb A_{1i}  \frac{1}{T}\sum_t\psi_{it}(y)
	+  R_{i,4} +R_{i,5}  +\widetilde\Delta_i
	$$
	where $\sup_y\frac{1}{N}\sum_i\| \widetilde\Delta_i\|^2=O_P(L^2T^{-3})$.
\end{lem}

\begin{proof}
It follows from Lemma \ref{la.1debu} and  Lemma \ref{la.1},
\begin{eqnarray*}
\widetilde\beta_i-\beta_i&=&  -  \frac{1}{T}B_{i,1T}(y)  - \frac{1}{T}B_{i,2T} (y) - \mathbb A_{1i}  \frac{1}{T}\sum_t\psi_{it}(y)
+  R_{i,4} +R_{i,5} +\Delta_i\cr
&=&  -  \frac{1}{T}\widehat B_{i,1T}(y)  - \frac{1}{T}\widehat B_{i,2T} (y) - \mathbb A_{1i}  \frac{1}{T}\sum_t\psi_{it}(y)
+  R_{i,4} +R_{i,5}\cr
&& +\underbrace{\Delta_i- (  \frac{1}{T} \widehat B_{i,1T}(	y)-  \frac{1}{T}B_{i,1T}(y))- (  \frac{1}{T}\widehat B_{i,2T}(y)-\frac{1}{T}B_{i,2T}(y))}_{\widetilde\Delta_i},
\end{eqnarray*}
where $\sup_y\frac{1}{N}\sum_i\|\widetilde\Delta_i\|^2=O_P(L^2/T^3)$.
\end{proof}

Note that Lemma \ref{la.1} below does \text{not } assume the serial stationarity. 
  \begin{lem}[Undebiased estimator]\label{la.1}
Then   for some $\Delta_i$, 
$$
\widetilde\beta_i-\beta_i=  -  \frac{1}{T}B_{i,1T}(y)  - \frac{1}{T}B_{i,2T} (y) - \mathbb A_{1i}  \frac{1}{T}\sum_t\psi_{it}(y)
+  R_{i,4} +R_{i,5} +\Delta_i
$$
where $\sup_y\frac{1}{N}\sum_i\|\Delta_i\|^2=O_P(T^{-3})$ and  $\frac{1}{T}\sum_t\psi_{it}(y)=\nabla Q_i(\beta_i)$.
  \end{lem}
  
  \begin{proof}
  	
  	For notational simplicity,  we drop $y$.  The notation for 
  	higher order matrix derivatives  associated with Taylor expansions is as defined in \cite{rilstone1996second}. For a  real-valued  function $Q(\beta)$, let $\nabla^3 Q(\beta)$ be a $\dim(\beta)\times \dim(\beta)^2$ matrix, whose $j$ th row is given by $[\vecc\nabla^2 (\partial_j Q(\beta)) ]'$. For instance, when $\beta=(x,y)'$, then the first row of $\nabla^3 Q(x, y)$ is given by 
  	$$
  	[\partial_x^2 g,\partial _{xy} g, \partial _{yx} g, \partial_y^2 g],\quad g=\partial_x Q(x,y).
  	$$

  	With this notation, the third-order Taylor expansion leads to 
  	$$
  	\widetilde\beta_i-\beta_i= - \nabla^2 Q_{i}( \beta_i)^{-1} \nabla Q_{i}(\beta_i)
  	- \frac{1}{2}\nabla^2 Q_{i}( \beta_i)^{-1} \nabla^3Q_{i}( \beta_i)[(\widetilde\beta_i-\beta_i)\otimes (\widetilde\beta_i-\beta_i)]
  	+R_{i,1}
  	$$ where $\otimes$ denotes Kronecker product and 
  	$$
  	R_{i,1} = -\frac{1}{6}\nabla^2 Q_{i}( \beta_i)^{-1} \nabla^4Q_{i}( \vbeta_i^*)[(\widetilde\beta_i-\beta_i)\otimes (\widetilde\beta_i-\beta_i)\otimes (\widetilde\beta_i-\beta_i)].
  	$$

  	Substituting from  (\ref{eq3.1}),
  	\begin{eqnarray}\label{eqb.2fadfa}
  		\widetilde\beta_i-\beta_i &=& - A_{1i} \nabla Q_{i}(\beta_i)
  		- \frac{1}{2}A_{1i}A_{2i}[(- A_{1i}\nabla Q_{i}(\beta_i)
  		+\Delta_i
  		)\otimes (- A_{1i}\nabla Q_{i}(\beta_i)
  		+\Delta_i)]
  		+ R_{i,1}\cr
  		&=&- A_{1i} \nabla Q_{i}(\beta_i)
  		- \frac{1}{2}A_{1i}A_{2i}[(A_{1i}\nabla Q_{i}(\beta_i)
  		)\otimes (A_{1i}\nabla Q_{i}(\beta_i)]
  		+R_{i,1}+  R_{i,2} \cr
  		&=&- A_{1i} \nabla Q_{i}(\beta_i)
  		- \frac{1}{2}  \mathbb A_{1i} \mathbb A_{2i}[( \mathbb A_{1i}\nabla Q_{i}(\beta_i)
  		)\otimes ( \mathbb A_{1i}\nabla Q_{i}(\beta_i)]
  		+R_{i,1}+  R_{i,2}+  R_{i,3}\cr
  		&=&- \mathbb A_{1i} \nabla Q_{i}(\beta_i)
  		- \frac{1}{T}B_{i,1T}
  		+\sum_{d=1}^4R_{i,d}  + [\mathbb  A_{1i} - A_{1i}] \nabla Q_{i}(\beta_i)\cr
  		&=&  - \mathbb A_{1i} \nabla Q_{i}(\beta_i)
  		-  \frac{1}{T}B_{i,1T}
  		+\sum_{d=1}^6R_{i,d}   - \frac{1}{T}B_{i,2T} 
  	\end{eqnarray}
  where $R_{i,4}, R_{i,5}$ are defined in (\ref{eqR45}), and 
  	\begin{eqnarray*}
  		R_{i,2}&=&
  		\frac{1}{2}A_{1i}A_{2i}\left\{[(- A_{1i}\nabla Q_{i}(\beta_i)
  		+\Delta_i
  		)\otimes (- A_{1i}\nabla Q_{i}(\beta_i)
  		+\Delta_i)] 
  		- [(A_{1i}\nabla Q_{i}(\beta_i)
  		)\otimes (A_{1i}\nabla Q_{i}(\beta_i)]\right\}\cr
  		R_{i,3}&=&
  		\frac{1}{2}  \mathbb A_{1i} \mathbb A_{2i}[( \mathbb A_{1i}\nabla Q_{i}(\beta_i)
  		)\otimes ( \mathbb A_{1i}\nabla Q_{i}(\beta_i)]
  		-  \frac{1}{2}  A_{1i}   A_{2i}[(  A_{1i}\nabla Q_{i}(\beta_i)
  		)\otimes ( A_{1i}\nabla Q_{i}(\beta_i)].\cr
  		R_{i,6}&=&  \mathbb  A_{1i} [A_{1i}^{-1}-\mathbb  A_{1i}  ^{-1} ](A_{1i}- \mathbb  A_{1i}  ) \nabla Q_{i}(\beta_i).
  	\end{eqnarray*}

By Cauchy-Shwartz and Holder's inequalities,  and Lemma \ref{la.1ini}, Assumption \ref{ass4.3},
  	\begin{eqnarray*}
  		\frac{1}{N}\sum_i\|R_{i,1}\|^2  &\leq&  O_P(1) ({ \frac{1}{N}\sum_{i=1}^N\| \widetilde\beta_i-\beta_i\|^8} )^{3/4}=O_P(T^{-3}).\quad (\text{by Holder  $p=4/3, q=4$}) \cr
  		\frac{1}{N}\sum_i\|R_{i,2}\|^2
  		&\leq& O_P(1)\sqrt{\frac{1}{N}\sum_{i=1}^N\| \nabla Q_{y,i}(\vbeta_i(y))\|^4\frac{1}{N}\sum_{i=1}^N\|  \Delta_i\|^4} +O_P(1) \frac{1}{N}\sum_{i=1}^N\|  \Delta_i\|^4  =O_P(T^{-3}). \cr
  		\frac{1}{N}\sum_i\|R_{i,3}\|^2
  		&\leq& O_P(1)\sqrt{\frac{1}{N}\sum_i \|A_{1i}-\mathbb A_{1i} \|^4+\frac{1}{N}\sum_i \|A_{2i}-\mathbb A_{2i} \|^4}  \sqrt{\frac{1}{N}\sum_i \|  \nabla Q_{y,i}(\vbeta_i(y))\|^8 }\cr
  		&=&O_P(T^{-3}).   \cr
  		\frac{1}{N}\sum_i\|R_{i,6}\|^2
  		&\leq& O_P(1)\sqrt{\frac{1}{N}\sum_i \|A_{1i}-\mathbb A_{1i} \|^4}  \sqrt{\frac{1}{N}\sum_i \|A_{1i}^{-1}-\mathbb A_{1i} ^{-1}\|^2} \sqrt{\frac{1}{N}\sum_i \|  \nabla Q_{y,i}(\vbeta_i(y))\|^4 }\cr
  		&=&O_P(T^{-3}) . 
  	\end{eqnarray*} 
  	Hence for  $\Delta_i:= R_{i,1}+R_{i,2}+R_{i,3}+R_{i,6}  $, we have 
  	$$
  		\widetilde\beta_i-\beta_i=  - \mathbb A_{1i} \nabla Q_{i}(\beta_i)
  		-  \frac{1}{T}B_{i,1T}  - \frac{1}{T}B_{i,2T} 
  		+ R_{i,4}+R_{i,5} +\Delta_i
  	$$
  	and $\sup_y\frac{1}{N}\sum_i\|\Delta_i\|^2=O_P(T^{-3})$.
  	 
  \end{proof}

\medskip

\subsection{Technical lemmas}

Lemmas in this subsection do \text{not } assume the serial stationarity. 
\begin{lem}\label{la.1ini} Uniformly in $y\in\mathcal Y$, 

(i)
 $\frac{1}{N}\sum_{i=1}^N\|\widetilde\beta_i-\beta_i\|^8=O_P(T^{-4})$.

 (ii) $   \frac{1}{N}\sum_{i=1}^N\|  \Delta_i\|^4=O_P(T^{-4}).$
 
 
  (iii) $\frac{1}{N}\sum_i \|A_{1i}-\mathbb A_{1i} \|^4=O_P(T^{-2})$ and  $\frac{1}{N}\sum_i \|A_{2i}-\mathbb A_{2i} \|^4=O_P(T^{-2})$.
\end{lem}

\begin{proof}

For 	notational simplicity,  we drop $y$ in these quantities. We have
 $$
\widetilde\beta_i-\beta_i= - \nabla^2 Q_{i}( b_i)^{-1} \nabla Q_{i}(\beta_i)
 $$
  where $b_i$ is between $\widetilde\beta_i$ and $\beta_i$.
Hence 
\begin{eqnarray*}
	&&\sup_y   \frac{1}{N}\sum_{i=1}^N\|\widetilde\beta_i-\beta_i\|^8
	\leq  O_P(1)\sup_y\frac{1}{N}\sum_{i=1}^N\|   \nabla Q_{i}(\beta_i)\|^8 \cr
	&\leq& O_P(T^{-4}) \max_i\E \sup_y\|   \frac{1}{\sqrt{T}}\sum_t \psi_{it}(y)\|^8 =O_P(T^{-4})
	\end{eqnarray*}
where the first inequality is from: 
$
\sup_y\sup_b\|   \nabla^2 Q_{i}( b )^{-1} \|=O_P(1)
$  (Assumption \ref{ass4.3}).

(ii) Since $\nabla^2 Q_{i}( \beta)$ is differentiable with a uniformly bounded gradient, 
\begin{eqnarray*}
&& \frac{1}{N}\sum_{i=1}^N\|  \Delta_i\|^4\leq \frac{C}{N}\sum_{i=1}^N \|\widetilde\beta_i-\beta_i\|^8=O_P(T^{-4}).
\end{eqnarray*}

(iii) Since $\sup_y\max_i\|A_{1i}\|<C$ almost surely and $\sup_y\max_i\|\mathbb A_{1i}\|<C$, 
\begin{eqnarray*}
\frac{1}{N}\sum_i \|A_{1i}-\mathbb A_{1i} \|^4&\leq& O_P(\frac{1}{T^2})\max_i \E\sup_y\| \frac{1}{\sqrt{T}}\sum_t  \varpi_{it}^2(y)\|^4= O_P(\frac{1}{T^2}).\cr
\frac{1}{N}\sum_i \|A_{2i}-\mathbb A_{2i} \|^4&\leq& O_P(\frac{1}{T^2})\max_i \E\sup_y\| \frac{1}{\sqrt{T}}\sum_t  \varpi_{it}^3(y)\|^4= O_P(\frac{1}{T^2}).
\end{eqnarray*}

\end{proof}

\begin{lem}\label{la.1debu} Suppose $V_{i,k}(y)$ is independent of $W$. In addition, suppose there is $a_{y,it}^d$ so that for $d=1,2,$, $\sup_y\frac{1}{NT}\sum_{it}\|a_{y,it}^d\|^4=O_P(1)$ and  for all $b_1, b_2$, 
	$$
	\|\nabla^d q_{y,it}(b_1)-\nabla^d q_{y,it}(b_2)\|\leq \|a_{y,it}^d\| \|b_1-b_2\|.
	$$
	Also suppose as $N,T,L\to\infty$,  
	$$
\mathbb E	\sup_y  \left\|   \frac{1}{T}\sum_{t=1}^T \ell_{it}  \ell_{it} '
	+\frac{1}{T}\sum_{h=1}^{L} \sum_{t>h}[ \ell_{it}  \ell_{i(t-h)} '+ \ell_{i(t-h)}  \ell_{it} ' ]- \Var(\frac{1}{\sqrt{T}}\sum_t\ell_t) \right\|^2=O_P(T^{-1}).
	$$
Then uniformly in $y\in\mathcal Y$, 
	
	(i)
	$\frac{1}{N}\sum_i\|\widehat B_{i,1T}-B_{i,1T}\|^2=O_P(T^{-1})$.

	(ii)  	$\frac{1}{N}\sum_i\|\widehat B_{i,2T}-B_{i,2T}\|^2=O_P(L^2/T)$.
\end{lem}

\begin{proof}
	(i) By Assumption \ref{ass4.3}, $\|\mathbb A_{1i}\|, \|\mathbb A_{2i}\|$ and $\widehat{\mathbb A}_{2i}\|$ are all bounded uniformly in $i$ and $y$. Then by Lemma \ref{la.1ini} $	\frac{1}{N}\sum_i\|\widehat B_{i,1T}-B_{i,1T}\|^2
	\leq a_1+a_2+a_3$ where
	\begin{eqnarray*}
a_1&=& \frac{1}{N}\sum_i\| \widehat   {\mathbb A}_{1i}- \mathbb A_{1i}\|^2\|\widehat  {\mathbb A}_{2i} \vecc(\widehat M_1^i(y))\|^2\leq   \sqrt{\frac{C}{N}\sum_i\| \widehat   {\mathbb A}_{1i}- \mathbb A_{1i}\|^4 }\sqrt{\frac{C}{N}\sum_i\| \widehat M_1^i(y)\|^4}
\cr
&\leq& O_P(T^{-1}) \cr
a_2&=& \frac{1}{N}\sum_i\|   \mathbb A_{1i}( \widehat  {\mathbb A}_{2i}- {\mathbb A}_{2i}) \vecc(\widehat M_1^i(y))\|^2\leq   \sqrt{\frac{C}{N}\sum_i\| \widehat   {\mathbb A}_{2i}- \mathbb A_{2i}\|^4 }\sqrt{\frac{C}{N}\sum_i\| \widehat M_1^i(y)\|^4}
\cr
&\leq& O_P(T^{-1}) \cr
a_3&=& \frac{1}{N}\sum_i\|   \mathbb A_{1i} {\mathbb A}_{2i}\vecc(\widehat M_1^i(y)- M_1^i(y))\|^2
\leq \frac{C}{N}\sum_i\|\widehat \Sigma_i(y)- \Var(\frac{1}{\sqrt{T}}\sum_t\mathbb A_{1i} \psi_{it}(y))\|^2
	\end{eqnarray*}
 and  $\widehat \Sigma_i(y)=  \frac{1}{T}\sum_t \widehat{\mathbb A}_{1i}\widehat\psi_{it}(y)\widehat\psi_{it}(y)' \widehat{\mathbb A}_{1i}.$  Also, $\widehat M_1^i(y)$ is defined as:
  $$
  \widehat V_{i,k}(y)=  \frac{1}{T}\sum_{t=1}^T\widehat \ell_{it} \widehat \ell_{it} '
 +\frac{1}{T}\sum_{h=1}^{L}\sum_{t>h}[\widehat \ell_{it} \widehat \ell_{i(t-h)} '+\widehat \ell_{i(t-h)} \widehat \ell_{it} ' ]:=\begin{pmatrix}
\widehat M_1^i(y) & \widehat M_{2,k}^i(y)'\\
\widehat M_{2,k}^i(y) &\widehat  M_{3,k}^i(y)
\end{pmatrix}.
 $$

 Note that 
	\begin{eqnarray*}
&& \frac{C}{N}\sum_i\|\widehat \Sigma_i(y)- \Var(\frac{1}{\sqrt{T}}\sum_t\mathbb A_{1i} \psi_{it}(y)|W)\|^2\leq O_P(T^{-1}) + \sqrt{\frac{C}{NT}\sum_{it} \|\widehat\psi_{it}(y)-\psi_{it}(y)\|^4}\cr
&\leq& O_P(T^{-1}) + \sqrt{\frac{C}{NT}\sum_{it} \| a^1_{1,it}\|^4\|\widetilde\vbeta_i(y)-\vbeta_i(y)\|^4}\cr
&\leq& O_P(T^{-1}) + (\frac{C}{NT}\sum_{it} \| a^1_{1,it}\|^8)^{1/4} (\frac{1}{N}\sum_i\|\widetilde\vbeta_i(y)-\vbeta_i(y)\|^8)^{1/4}=O_P(T^{-1}) .
	\end{eqnarray*}

(ii)  $	\frac{1}{N}\sum_i\|\widehat B_{2,1T}-B_{2,1T}\|^2
\leq b_1+b_2$ where
\begin{eqnarray*}
	b_1&=& \frac{1}{N}\sum_i\| \widehat   {\mathbb A}_{1i}- \mathbb A_{1i}\|^2\sum_{k=1}^{\dim(\beta_i)}\| \widehat  M_{2,k}^i(y)\|^2\leq   \sqrt{\frac{C}{N}\sum_i\| \widehat   {\mathbb A}_{1i}- \mathbb A_{1i}\|^4 }\sqrt{\frac{C}{N}\sum_i\sum_{k=1}^{\dim(\beta_i)}\| \widehat M_{2,k}^i(y)\|^4} \cr
	&\leq& O_P(T^{-1}) \cr
	b_2&=& \frac{1}{N}\sum_i\left\|   \mathbb A_{1i}  \begin{pmatrix}
		\tr(M^i_{2,1}(y)-\widehat M^i_{2,1}(y))\\
		\vdots\\
	\tr(M^i_{2,\dim(\beta_i)}(y)-\widehat M^i_{2,1}(y))
\end{pmatrix}\right\|^2\leq\max_k \frac{C}{N}\sum_i\left\|   M_{2,k}^i(y)-\widehat M_{2,k}^i(y)\right\|^2\cr
&\leq&\max_k \frac{C}{N}\sum_i\left\|   \frac{1}{T}\sum_{t=1}^T \ell_{it}  \ell_{it} '
+\frac{1}{T}\sum_{h=1}^{L}\sum_{t>h}[ \ell_{it}  \ell_{i(t-h)} '+ \ell_{i(t-h)}  \ell_{it} ' ]- \Var(\frac{1}{\sqrt{T}}\sum_t\ell_t) \right\|^2\cr
&&+\max_k \frac{C}{N}\sum_i\| J_{it}(y)  J_{it}(y) '-\widehat J_{it}(y) \widehat J_{it}(y) '\|^2\cr
&&+\max_k \frac{C}{N}\sum_i\|\frac{1}{T}\sum_{h=1}^{L}\sum_{t>h}[ J_{it} J_{i(t-h)} '-  \widehat J_{it} \widehat  J_{i(t-h)} ']\|^2 =O_P(L^2/T).
\end{eqnarray*}
where $J_{it}(y):=\mathbb A_{1i} \psi_{it}(y)\varpi^2_{it,k}(y) $ and $\widehat   J_{it}(y) $ is its estimator by replacing $\mathbb A_{1i}$, $\psi_{it}(y)$ and $\varpi_{it}^2(y)$ with their estimates.

\end{proof}

\section{A high-level weak convergence result}

\subsection{Outline of the proof}\label{sec:outline}
We consider a generic functional $\vartheta(y)$ and its generic estimator $\widehat\vartheta(y)$. When the support $\mathcal Y$ is continuous,  we proceed as follows.

\begin{enumerate}
    \item[Step I.]   We first assume a high-level functional  expansion as follows.
      \begin{equation}\label{eqb.1}
      \widehat\vartheta(y)-\vartheta(y)=\frac{1}{N}\sum_{i=1}^N\left[\frac{1}{\sqrt{T}}d_{\psi,i}(y)+d_{\vgamma,i}(y)\right] + o_P(\zeta_{NT}(y))
      \end{equation}
   where    $\zeta_{NT}(y)= (NT)^{-1/2}\Var(d_{\psi,i}(y))+N^{-1/2} \Var(d_{\vgamma,i}(y))$. We 
     make  high-level assumptions about this expansion in Assumption \ref{assb.1}. 

\item[Step II.] Section \ref{secd.2} establishes  Proposition \ref{probb.1}, which shows the weak convergence of the rescaled $\widehat\vartheta(\cdot)-\vartheta(\cdot)$ under Assumption \ref{assb.1}, assuming expansion (\ref{eqb.1}). 
Hence the main theorems are proved by applying  Proposition \ref{probb.1}. To do so, it suffices to verify Assumption \ref{assb.1} and  expansion (\ref{eqb.1}).  We verify them  in two settings, respectively in Step III and IV below. 

\item[Step III.] Consider  the functional taking the form 
\begin{equation}\label{eqd.2gener}
    \vartheta(y)=\mathbb E_tf(\vbeta_i(y), \vtheta(y), D_{it})
\end{equation}
for some known function $f$ and ``data" $D_{it}$. This includes:
\begin{eqnarray*}
    f(\vbeta_i(y), \vtheta(y), D_{it})&=&\vtheta(y)\quad \text{the coefficient}\cr 
    f(\vbeta_i(y), \vtheta(y), D_{it})&=& \Lambda (-\vx_{it}'\vbeta_i(y))\quad \text{the actual distribution  }\cr 
    f(\vbeta_i(y), \vtheta(y), D_{it})&=& \Lambda (-h_{it}(\vx_{it})'\vbeta_i^g(y))\quad \text{ the counterfactual distribution }.
\end{eqnarray*}
  where $ \vbeta_i^g(y)= \vtheta(y)[g(\vz_{i})-\vz_{i}] +\vbeta_i(y)$.  
We   estimate it by the debiased estimator
\begin{eqnarray}\label{eq.2geneestor}
    \widehat\vartheta ^g(y)&=&  \frac{1}{N}\sum_{i} f(\widehat\vbeta_i(y), \widehat\vtheta(y), D_{it})\cr 
    &&- \frac{1}{2NT}\sum_i\tr \left[\partial^2_\beta f(\widehat \beta_{i}(y), \widehat\vtheta(y), D_{it}) \widehat \Sigma_i(y) ^{-1}\right]
\end{eqnarray}
where
$
\widehat \Sigma_i(y)= -\nabla^2 Q_{y,i}(\widehat\vbeta_i(y)).$

\begin{itemize}
    \item  To verify the expansion (\ref{eqb.1}) and Assumption \ref{assb.1}, we  show  in  Sections \ref{sec:d.3} and \ref{sec:d.4} that they hold  under a ``lower-level" Assumption \ref{assc.1}.

  \item  We then verify  the ``lower-level" Assumption \ref{assc.1} in 
    Section \ref{sec:verifyV} under the settings of    Theorems \ref{th4.1} and   \ref{th4.20}. This then proves these two theorems. 
\end{itemize}

\item[Step IV.] In Section \ref{sec:proqte}, we consider the case $\vartheta(y) = \QTE$, the quantile effect. This functional does not take the form (\ref{eqd.2gener}) with a smooth function ``$f$". Hence we proceed to directly verifying  Assumption \ref{assb.1} and  expansion (\ref{eqb.1}). 
 
\end{enumerate}

All the above arguments require establishing weak convergence of a process in $y\in\mathcal Y$, the support of $y_{it}$. We verify it when both $\mathcal Y$ is continuous and discrete. When $\mathcal Y$ is continuous, we impose Assumption \ref{ass4.7} as the continuity assumption. 

When $\mathcal Y$ is discrete with finitely many possible outcomes,  we directly prove 
 $$
      \widehat\vartheta(y)-\vartheta(y)=\frac{1}{N}\sum_{i=1}^N\left[\frac{1}{\sqrt{T}}d_{\psi,i}(y)+d_{\vgamma,i}(y)\right] + o_P(\zeta_{NT}(y))
$$
for each $y\in\mathcal Y$ in Lemma \ref{l5fun}.  Then establishing the convergence in distribution under  finite dimensional distribution   would be sufficient. 
 
Define   
       \begin{eqnarray*}
       \bar V_{\psi}(y_k, y_l)&=&\E d_{\psi, i}(y_k)d_{\psi, i}(y_l), \quad  \bar V_{\vgamma}(y_k, y_l)=\E d_{\vgamma, i}(y_k)d_{\vgamma, i}(y_l)\cr
\sigma_{T}^2(y_k, y_l)&=& \frac{1}{ T}\bar V_{\psi}(y_k, y_l)+ \bar  V_{\vgamma}(y_k,y_l)\cr
\sigma_{T}^2(y)&=& \sigma_{T}^2(y, y),\quad  s_{NT}^2(y)= \frac{1}{N} \sigma_{T}^2(y)\cr
\bar V_{\psi}(y)&=& \bar V_{\psi}(y, y),\quad    \bar V_{\vgamma}(y) = \bar V_{\vgamma}(y, y)\cr
 H&=&  \lim_{T}( \frac{  \sigma_{T}^2(y_k, y_l)}{\sigma_{T}(y_k)\sigma_{T}(y_l)    } )_{M\times M} 
 \end{eqnarray*}

 Let $\partial_\beta f_i(y)=\partial_\beta f(\beta_i(y),\vtheta(y),y)$, $  \ddot f_{i,\beta}:=\partial^2_\beta f(\beta_{i}(y), \vtheta(y), D_{it}),$ $
  \ddot f_{i,\vtheta}:=\partial^2_{\vtheta} f(\beta_{i}(y), \vtheta(y), D_{it}),$, and $
  \ddot f_{i,\beta \vtheta}:=\partial^2_{\beta\vtheta} f(\beta_{i}(y), \vtheta(y), D_{it})
 $.  
In addition, let    $\bar G(y)= \E_t \partial_{\vtheta} f(\beta_{i}(y), \vtheta(y), D_{it})'$, where $\partial_{\vtheta}$ is taken with respect to the coordinates of $\vecc(\vtheta). $

 Define $\rho(y_1,y_2)=C|y_1-y_2|^{1/4}$ for some $C>0.$
 \begin{assumption}\label{assb.1}
       (i)  $\E d_{\psi, i}(y)=\E d_{\vgamma, i}(y)=0, $  $\E d_{\psi, i}(y_k) d_{\vgamma,i}(y_l)=0$  for all  $y, y_k, y_l.$
       
       (ii)  We have
  $0<c<\inf_y\bar V_\psi(y)<C$. In addition, $\bar V_{\vgamma}(y)\in[0, C]$, with zero as a feasible value for $\bar V_{\vgamma}(y)$.
       
       
       (iii)  $\E\sup_y |d_{\psi,i}(y)|^{2+a}+ \E\sup_y|\frac{d_{\vgamma,i}(y)^2}{\bar V_{\vgamma}(y)}|^{a}<C$ for some $a\geq2$.
       
       (iv)  For any $\delta>0$, 
 \begin{eqnarray*}
       \frac{1}{N}\sum_i   \E  \sup_{\rho(y_1,y_2)<\delta}   |d_{\psi,i}(y_1)-d_{\psi,i}(y_2)  |^2 
 +  
 \sup_{\rho(y_1,y_2)<\delta}   |\bar V_{\psi}(y_1) -   \bar V_{\psi}(y_2)|^2  &\leq&\delta^2\cr 
 \frac{1}{N}\sum_i \E\sup_{\rho(y_1, y_2)< \delta } \left|  \frac{  d_{\vgamma,i}(y_1) }{\sigma_T(y_1) }-  \frac{  d_{\vgamma,i}(y_2) }{\sigma_T(y_2) } \right|^2&\leq& \delta^2.
 \end{eqnarray*}
 \end{assumption}


In the assumption below, for any random variable $\vx_{it}$, let
$$
\mathbb   Z_t(\vx_{it}):=\frac{\vx_{it}-\E_t \vx_{it}}{\sqrt{\Var_t(\vx_{it})}}
$$
where $\E_t$ and $\Var_t$ are the expectation and variance operators with respect to the cross-sectional distribution of $\vx_{it}$ given $t$. 

\begin{assumption}\label{assc.1}
(i) $\max_i\E\sup_ y \|\nabla{f}_i\|^{8}+\max_i\E\sup_ y \|\nabla^2{f}_i\|^{4}<C.$\\	
(ii) $\E[\psi_{it}(y_k)| \beta_i(y_l), D_{it}]=0$ and $\Var_t(d_{\psi,i}(y))>c>0$. Also  $\Var_t(d_{\vgamma,i}(y))\in[0, C]$, with zero as a admissible value. \\
(iii)  $ \E\sup_y\left[ \mathbb Z_{t}(\vw_{i}'   S_{wz} \bar G(y)\vgamma_i(y) + f(\vbeta_i(y), \vtheta(y),  D_{it}) )\right]^{4}  <C$ and  $\sup_y\|\bar G(y)\|<C$.\\
(iv) Write $f(y)= f(\vbeta_i(y), \vtheta(y), D_{it})$ for simplicity.  	There is $C>0$, for all $y_1, y_2$, and $i$, 
$$ \mathbb E  |\partial_\beta f_i(y_1)-\partial_\beta {f}_i(y_2) |^4+\E|\ddot{f}_{i,\beta}(y_1)-\ddot{f}_{i,\beta}(y_2) |^4\leq C|y_1-y_2|^4.$$
\begin{eqnarray*}
    \frac{1}{N}\sum_i \E\sup_{\rho(y_1, y_2)< \delta } \left|  \frac{  d_{\vgamma,i}(y_1) }{\sigma_T(y_1) }-  \frac{  d_{\vgamma,i}(y_2) }{\sigma_T(y_2) } \right|^2&\leq& \delta^2\cr 
    \|\bar G(y_1)-
 \bar G(y_2)\|&<&C|y_1-y_2|
\end{eqnarray*}

In the above $\sigma_{T}^2(y)=\frac{1}{ T}\bar V_{\psi}(y)+ \bar  V_{\vgamma}(y)$, 
	 \begin{eqnarray*}
      \bar V_{\psi}(y  )&=&\E_t[\partial_\beta f_i (y)'   \mathbb A_{1i}(y) \frac{1}{T}\sum_{t} \psi_{it}(y)\psi_{it}(y)' \mathbb A_{1i}(y)\partial_\beta f_i(y)  ], \cr
       \bar V_{\vgamma}(y)&=&\Var_t [\vw_{i}' S_{wz} \bar  G(y)  \vgamma_i(y) ]+ \Var_t [ f(\beta_i(y),\vtheta(y),  D_{it}) ].   
	  \end{eqnarray*}
and $S_{wz}:= C_{1}^{-1}C_{2}(C_{2}'C_{1}^{-1}C_{2})^{-1}$ where $C_{1}=\E \vw_{i}\vw_{i}'$ and $C_{2}= \E \vw_{i}\vz_{i}'$.

	 \end{assumption}

\medskip

     \subsection{Step II. Generic weak convergence when $\mathcal Y$ is continuous}\label{secd.2}

\begin{prop}\label{probb.1} Suppose  expansion (\ref{eqb.1}) holds. Suppose   $\{d_{\psi, i}(y), d_{\vgamma,i}(y): y\in\mathcal T\}$ are i.i.d. across $i$. Assumption \ref{assb.1} holds. 
Then
$$
 \frac{\widehat\vartheta(\cdot)-\vartheta(\cdot)}{s_{NT}(\cdot)} \Rightarrow\mathbb G(\cdot)
 $$ 
 where $\mathbb G(\cdot)$ is a centered Gaussian process with covariance kernel 
 $$
 H(y_k, y_l)= \lim_{T}\frac{  \sigma_{T}^2(y_k, y_l)}{\sigma_{T}(y_k)\sigma_{T}(y_l)    }.
 $$
      \end{prop}
      
      \begin{proof}
 By expansion (\ref{eqb.1})    $ \widehat\vartheta(y)-\vartheta(y)= \sum_{i=1}^N \alpha_i(y) +  o_P(\zeta_{NT}(y))
$
where $$\alpha_i(y)=\frac{1}{N}  \frac{1}{\sqrt{T}} d_{\psi,i}(y)  + \frac{1}{N}d_{\vgamma,i}(y) .$$

 Below we  prove the weak convergence of $\sum_i\alpha_i(.)/s_{NT}(.)$.
 
 (i) show the  fidi of $ \sum_i\alpha_i(.)/s_{NT}(.)$. For any finite integer  $M>0$,  and any $y_1,...,y_M.$ Let
 $A_i=(\alpha_i(y_1)/s_{NT}(y_1),...,\alpha_i(y_M)/s_{NT}(y_M))'.$ We shall show
$$
\frac{g'\sum_iA_i}{\sqrt{g'  H g}}\to ^d\mathcal N(0,1),
$$  for any $g\neq 0$ as an $M$-dimensional fixed vector. Here
$$
 H_T=\Var(\sum_iA_i)= ( \frac{  \sigma_{T}^2(y_k, y_l)}{\sigma_{T}(y_k)\sigma_{T}(y_l)    } )_{M\times M} ,\quad H=  \lim_{T}H_T,
$$
  Then the fidi follows from the Cramer-Wold theorem.

  We  proceed by verifying the Lindeberg condition. First,  we  bound $\sum_i\E\left( (g'A_i)^4 \right)$. 
    \begin{eqnarray}\label{eqa.3add}
 \sum_i\E\left( (g'A_i)^4 \right)&\leq&  M\|g\|^4\sum_i\E\left( \sum_{m=1}^M\frac{\alpha_i(y_m) ^4}{s_{NT}^4(y_m)} \right)\cr
 &\leq& M\|g\|^4\frac{1}{N^4 } \sum_i \E \left( \sum_{m=1}^M\frac{\|  d_{\psi,i}(y_m)   \| ^4 }{s_{NT}^4(y_m)} \frac{1}{T^2}+ \frac{\|  d_{\vgamma,i}(y_m)\| ^4 }{s_{NT}^4(y_m)} \right) \cr
 &\leq& C\|g\|^4\frac{1}{N  } \sum_{m=1}^M\E  \left[\frac{ d_{\psi, i}(y_m)^4 }{    \bar V_{\psi}(y_m)^2 }  +   \frac{ d_{\vgamma, i}(y_m) ^4 }{ \bar V_{\vgamma}(y_m)^2} \right] =O(\frac{\|g\|^4}{N}).
  \end{eqnarray}

 In addition,    $\lambda_{\min}(   H_{T})  >\lambda_{\min}(   H)- o(1)>c $ for large $T$.
 Therefore, for all $\epsilon>0,$ we use the inequality that  $\E |Y|1\{|X|>a\}\leq \E |YX^2|/a^2$, 
    \begin{eqnarray*}
&&\frac{1}{g'     H_{T}g}\sum_i\E\left( (g'A_i)^21\{|g'A_i|>\epsilon \sqrt{g'     H_{T}g}\}\right)
   \cr
   &\leq& \frac{1}{ (g'  H_{T}g)^2\epsilon^2}\sum_i\E\left(  (g'A_i)^4 \right)\leq O(N^{-1}).
 \end{eqnarray*}

By Lindeberg's central limit theorem,
$$
 Y_{NT}:= g'\sum_iA_i/\sqrt{g'  H_{T}g}\to^d \mathcal N(0,1).
$$  
Therefore,
$$
\frac{ g'\sum_iA_i}{\sqrt{g'   Hg}} =  Y_{NT}+ Y_{NT}\left(\sqrt{\frac{ {g'  H_{T}g}}{ {g'  Hg}}}-1\right) = Y_{NT}+o_P(1)\to^d\mathcal N(0,1).
$$

(ii)   Define $\rho(y_1,y_2)=C|y_1-y_2|^{1/4}$ for some $C>0.$ Let $\ell^{\infty}(\mathcal Y)$ be the set of all uniformly bounded real functions on $\mathcal Y.$
We show $\sum_i\alpha_i(.)/s_{NT}(.)$ is asymptotically tight in $\ell^{\infty}(\mathcal Y)$,  by verifying the three conditions of Theorem 2.11.11 in \cite{VW}.  Let 
 \begin{eqnarray}\label{eqa.4add}
 b_i (y)  &=& 
 \frac{ \frac{1}{\sqrt{T}}  d_{\psi,i}(y) }{\sigma_T(y)}
,\quad c_i(y)=  \frac{  d_{\vgamma,i}(y) }{\sigma_T(y) } .\cr
 \bar b_i (y)  &=& 
 \frac{ \frac{1}{\sqrt{T}}  d_{\psi,i}(y) }{\left[ \frac{1}{T}\bar V_{\psi}(y) \right]^{1/2}}
,\quad  \bar c_i(y)=  \frac{ d_{\vgamma, i}(y)}{\bar V_{\vgamma}(y)^{1/2}} .
  \end{eqnarray}
 Let $F_i(y)=\frac{ a_i}{s_{NT}}=\frac{1}{\sqrt{N}}(b_i(y)+c_i(y)). $

Condition (1).  For every $\eta>0$, and an arbitrarily small $a>0$,  \begin{eqnarray*}
  &&\sum_i\E\sup_y|F_i(y)|1\{\sup_y|F_i(y)|>\eta\}\cr
  &\leq&\eta^{-1}\frac{1}{N}\sum_i\E\sup_y|  b_i(y)+c_i(y)|^21\{\sup_y|b_i(y)+c_i(y)|>\sqrt{N}\eta\}\cr
   &\leq&\frac{C}{\eta^{a+1}N^{a/2}} \frac{1}{N}\sum_i\E\sup_y|\bar   b_i(y) |^{2+a}
   +\frac{1}{\eta^{a+1}N^{a/2}} \frac{1}{N}\sum_i\E\sup_y| \bar   c_i(y)|^{2+a} =o(1).
  \end{eqnarray*}

Condition (2):  For every $y_1, y_2\in\mathcal Y$,   \begin{eqnarray*}
	&&\sum_i \E |F_i(y_1)-F_i(y_2)|^2\leq
	C\frac{1}{N}\sum_i\E|b_i(y_1)-b_i(y_2) |^2+C\frac{1}{N}\sum_i \E |c_i(y_1)-c_i(y_2) |^2\cr
	&\leq&C     \E   [d_{\psi,i}(y_1)-d_{\psi,i}(y_2)]^2
	+C  |\bar V_{\psi}(y_1) -   \bar V_{\psi}(y_2)|^2 
    +C\frac{1}{N}\sum_i \E |c_i(y_1)-c_i(y_2) |^2\cr
    &\leq&  C|y_1-y_2|^{1/2}\leq  \rho(y_1,y_2)^2.
\end{eqnarray*}
where the second last inequality follows from Assumption \ref{assb.1} (iv) combined with Lemma \ref{la.5sim}.

Condition (3): By Assumption \ref{assb.1}(iii), for every $\delta>0$,  
\begin{eqnarray}\label{eqa.5}
&&\sup_{\eta>0}\sum_i\eta^2 P\left(\sup_{\rho(y_1,y_2)<\delta}|F_i(y_1)-F_i(y_2)|>\eta\right)\cr
&\leq&\sum_i  \E \left(\sup_{\rho(y_1,y_2)<\delta}|F_i(y_1)-F_i(y_2)|^2\right)
\cr
&\leq& \frac{1}{N}\sum_i  \E \left(\sup_{\rho(y_1,y_2)<\delta}| b_i(y_1)-b_i(y_2)|^2\right)+  \frac{1}{N}\sum_i  \E \left(\sup_{\rho(y_1,y_2)<\delta}| c_i(y_1)-c_i(y_2)|^2\right)\cr
&\leq&C       \E  \sup_{\rho(y_1,y_2)<\delta}   |d_{\psi,i}(y_1)-d_{\psi,i}(y_2)  |^2 
+C \sup_{\rho(y_1,y_2)<\delta}   |\bar V_{\psi}(y_1) -   \bar V_{\psi}(y_2)|^2  \E
 \sup_y d_{\psi,i}(y)^2
\cr
&&+   \frac{1}{N}\sum_i  \E  \sup_{\rho(y_1,y_2)<\delta}| c_i(y_1)-c_i(y_2)|^2 
\leq\delta^2
\end{eqnarray}
Thus all conditions are satisfied;  $\frac{\sum_i \alpha_i}{s_{NT}}$ is asymptotically tight.

 Together,   the process $ \sum_i\alpha_i(.)/s_{NT}(.)$   weakly converges to a centered Gaussian process, with covariance kernel $$
 H(y_k, y_l)=\lim_T \frac{  \sigma_T^2(y_k, y_l)}{\sigma_T(y_k)\sigma_T(y_l)    } .
 $$
 
 (iii) Next, we show that $o_P(1)\sup_y \zeta_{NT}(y)s_{NT}^{-1}(y)=o(1)$. 
We have 
\begin{eqnarray*}
&&o(\frac{1}{\sqrt{NT}})\frac{1}{\inf_ys_{NT}(y)}=
 \frac{1}{\inf_y\bar V_\psi(y)}
 o(1)=o(1)\cr
 &&
 o(\frac{1}{ \sqrt{N}})\sup_y  \|\bar V_{\vgamma}(y)\|^{1/2}s_{NT}^{-1}(y)
\leq o(1)\sup_y(\frac{\bar V_{\vgamma}(y) }{\bar V_{\vgamma}(y)})^{1/2}=o(1).
 \end{eqnarray*}
 
 Hence uniformly in $y$, 
 $$
 \frac{\widehat\vartheta(y)-\vartheta(y)}{s_{NT}(y)} = \frac{\sum_i\alpha_i(y)}{s_{NT}(y)}+o_P( 1)\Rightarrow\mathbb G
 $$
This implies the weak convergence .

      \end{proof}

  \begin{lem}\label{la.5sim}
Let $X_i(y_1,y_2)$ be a random variable so that there are $C,c>0$, for all $\epsilon>0$ $\frac{1}{N}\sum_i\E\sup_{|y_1-y_2|\leq \epsilon}\|X_i(y_1,y_2)\|<C\epsilon^c$. Then for all $y_1\neq y_2$,
$$
\frac{1}{N}\sum_i\E \|X(y_1,y_2)\|<C|y_1-y_2|^c.
$$
  \end{lem}
  \begin{proof}
  \begin{eqnarray*}
  \frac{1}{N}\sum_i\E |X(y_1,y_2)|  \leq \sup_{\epsilon>0}\frac{1}{\epsilon^c}\frac{1}{N}\sum_i\E \sup_{|y_1-y_2|=\epsilon} \|X(y_1,y_2)\||y_1-y_2|^c\leq  C |y_1-y_2|^c.
  \end{eqnarray*}
  
  \end{proof}

      \subsection{Step III. Expansion (\ref{eqb.1})  when $\mathcal Y$ is continuous}\label{sec:d.3}

Consider 
\begin{equation*} 
    \vartheta(y)=\mathbb E_tf(\vbeta_i(y), \vtheta(y), D_{it}).
\end{equation*}
	Recall     $\bar G(y)= \E_t \partial_{\vtheta} f(\beta_{i}(y), \vtheta(y), D_{it})'$.

\begin{lem}\label{lc.2expfun} Suppose Assumption \ref{assc.1} holds.
Uniformly in $y$, 
\begin{eqnarray*}
 \widehat\vartheta (y)-\vartheta (y)&=& o_P(\zeta_{NT}(y))+
 \frac{1}{NT}\sum_{it}   (   \vw_{i}'   S_{wz} \bar G(y) -\partial_\beta f                 _i(y)' )   \mathbb A_{1i}(y)   \psi_{it}(y)
  \cr
 && +  \frac{1}{N}\sum_{i=1}^N\left( \vw_{i}' S_{wz} \bar G(y)\vgamma_i(y)
 	 +  \left[ f(\vbeta_i(y), \vtheta(y), D_{it}) - \mathbb E_tf(\vbeta_i(y),\vtheta(y), D_{it})\right]\right).
 \end{eqnarray*}
  \end{lem}
	 
	 \begin{proof} Write $  \partial_\beta f                 _i(y):=  \partial_\beta f                 (\vbeta_i(y),\vtheta(y), D_{it})$. Let $  G_N(y):=  \frac{1}{ N} \sum_i\partial_{\vtheta} f                 (\vbeta_i(y),\vtheta(y), D_{it})$, where $\partial_{\vtheta}$ is taken with respect to the coordinates of $\vecc(\vtheta). $ 
By the Taylor expansion up to the second order, (for the first term involving $\widehat\vtheta-\vtheta$, use the identity $\tr(A'B)=\vecc(A)'\vecc(B)$):  
 \begin{eqnarray}\label{eqc.3}
 && \frac{1}{N}\sum_{i} f(\widehat\vbeta_i(y), \widehat\vtheta(y), D_{it})-  f(\vbeta_i(y), \vtheta(y), D_{it})= D_0+...+D_3+ R_1
 \cr
D_0 &:=&  \frac{1}{N}\sum_i  \partial_\beta f                 _i(y)'(\widehat\vbeta_i(y)-\vbeta_i(y))
 \cr
 D_1&:=&\frac{1}{2N}\sum_i\tr \left[
 \ddot f_{i,\beta}(\widehat\vbeta_i(y)-\vbeta_i(y)) (\widehat\vbeta_i(y)-\vbeta_i(y))' \right]    
 \cr
D_2 &:=& \tr[G_N(y)(\widehat\vtheta(y)-\vtheta(y))],\cr
&=& \tr[ \frac{1}{NT}\sum_{it}  G_N(y)  \mathbb A_{1i}(y)   \psi_{it}(y)   \vw_{i}' S_{wz} ]
 +\tr[\frac{1}{N}\sum_{i=1}^N G_N(y) \vgamma_i(y)\vw_{i}' S_{wz}]+o_P(\zeta_{NT}(y))\cr
 &=&  \frac{1}{NT}\sum_{it} \vw_{i}' S_{wz} \bar G(y)  \mathbb A_{1i}(y)   \psi_{it}(y)    
 +\frac{1}{N}\sum_{i=1}^N \vw_{i}' S_{wz} \bar G(y) \vgamma_i(y)+o_P(\zeta_{NT}(y)) \cr
    D_3&:=&\frac{1}{N}\sum_i(\widehat\vbeta_i(y)-\vbeta_i(y))' \ddot f_{i,\beta\vtheta}(\widehat\vtheta-\vtheta) +  ( \widehat\vtheta-\vtheta )'\frac{1}{2N}\sum_i\ddot f_{i,\vtheta}     (\widehat\vtheta-\vtheta)=o_P(\zeta_{NT}(y))\cr
 \end{eqnarray}
 where for some $a_i$, 
$$
R_1=\frac{1}{6N}\sum_i\partial^3_\beta f(  a_i, D_{it}) (\widehat\vbeta_i(y)-\vbeta_i(y))\otimes (\widehat\vbeta_i(y)-\vbeta_i(y))\otimes (\widehat\vbeta_i(y)-\vbeta_i(y)).
$$
We have
$$\sup_y R_1\leq \sup_y\frac{C}{N}\sum_i\|\widehat\beta_i-\beta_i\|^3=O_P(\frac{1}{T^{3/2}})=o_P(\frac{1}{\sqrt{NT}}).
$$

To analyze $D_0+D_1$, by Lemma \ref{la.2anadebias}, 
	$
	\widehat\beta_i-\beta_i=- \mathbb A_{1i}  \frac{1}{T}\sum_t\psi_{it}(y)
	+  R_{i,4} +R_{i,5}  +\widetilde\Delta_i
	$
	where $\sup_y\frac{1}{N}\sum_i\| \widetilde\Delta_i\|^2=O_P(L^2T^{-3})$.
  Substituting this to the above expression, 
	\begin{eqnarray*}
 &&D_0+D_1\cr 
 &=&
-	\frac{1}{N}\sum_i \partial_\beta f_i(y)'   \mathbb A_{1i}  \frac{1}{T}\sum_t\psi_{it}(y)	
+ \frac{1}{2NT}\sum_i\tr \left[\ddot{f}_{i,\beta}  \E v_i(y) \right]
+\sum_{d=1}^3H_d
\cr
H_1&:=&		\frac{1}{2NT}\sum_i\tr \left[\ddot{f}_{i,\beta}   ( v_i(y) -\E v_i(y)) \right] \cr
H_2&:=&
 	\frac{1}{N}\sum_i \partial_\beta f_i(y)'   (R_{i,4} +R_{i,5} )	\cr
 	H_3&:=&
	\frac{1}{N}\sum_i \partial_\beta f_i(y)'  \widetilde\Delta_i	-	\frac{1}{2N}\sum_i\tr \left[\ddot{f}_{i,\beta}     \mathbb A_{1i}  \frac{1}{T}\sum_t\psi_{it}(y)  (R_{i,4} +R_{i,5})'  \right]\cr
	&&
	-	\frac{1}{2N}\sum_i\tr \left[\ddot{f}_{i,\beta}     \mathbb A_{1i}  \frac{1}{T}\sum_t\psi_{it}(y)  \widetilde\Delta_i'  \right]\cr
	&&
	+	\frac{1}{2N}\sum_i\tr \left[\ddot{f}_{i,\beta}    (R_{i,4} +R_{i,5})  (\widehat\vbeta_i(y)-\vbeta_i(y))'  \right]
	+	\frac{1}{2N}\sum_i\tr \left[\ddot{f}_{i,\beta}     \widetilde\Delta_i  (\widehat\vbeta_i(y)-\vbeta_i(y))'  \right]\cr
	v_i(y)&:=&  \mathbb A_{1i}  \frac{1}{\sqrt{T}}\sum_t\psi_{it}(y)   \frac{1}{\sqrt{T}}\sum_s\psi_{is}(y)' \mathbb A_{1i} .
		\end{eqnarray*}
We proceed with the following steps. Step 1, show $\sum_{d=1}^3H_d $ is negligible. Step 2, estimate the bias $\E v_i(y) $ by $\widehat \Sigma_i(y)^{-1}$ and compute the debiased estimator, 
and show that the bias estimation is negligible.

\textbf{Step 1(a).}  Write $F_i(y)= 	\frac{1}{2\sqrt{N}}\tr \left[\ddot{f}_{i,\beta}   ( v_i(y) -\E v_i(y)) \right]$. Then $H_1=\frac{1}{T\sqrt{N}}\sum_i F_i(y)$. We now show $\sum_i F_i(y)=O_P(1)$ uniformly in $y$ by showing it is asymptotically tight. For notational simplicity, we focus on an arbitrary element of $\ddot{f}_{i,\beta} B_i(y)$ and continue using $\ddot{f}_{i,\beta}B_i(y)$ to denote this element with abuse of notation. Since the dimension of $\vbeta_i(y)$ is fixed, this does not affect the asymptotic behavior.  For any $\eta>0$, and  $a>0$,
\begin{eqnarray*}
&&	\sum_i\E \sup_y|F_i(y)|1\{\sup_y|F_i(y)|>\eta\}
\leq	\frac{1}{\eta}\sum_i\E \sup_y|F_i(y)|^21\{\sup_y|F_i(y)|>\eta\}
\cr
&=&\frac{1}{4N	\eta}\sum_i\E \sup_y \left[\ddot{f}_{i,\beta}   ( v_i(y)-\E v_i(y)) \right]^21\{\sup_y|	 \left[\ddot{f}_{i,\beta}   ( v_i(y)-\E v_i(y)) \right]|>2\sqrt{N}\eta\}\cr
&\leq& \frac{1}{4 	N^{a/2}\eta^{1+a}}\frac{1}{N}\sum_i\E \sup_y \left[\ddot{f}_{i,\beta}   ( v_i(y)-\E v_i(y)) \right]^{2+a}
\cr
&\leq&  \frac{C}{ 	N^{a/2}\eta^{1+a}}\frac{1}{N}\sum_i[\E \sup_y \left[  v_i(y)-\E v_i(y) \right]^{4}]^{(2+a)/4}=o(1)
	\end{eqnarray*}
provided that $\E \sup_y \left[  v_i(y)-\E v_i(y) \right]^{4}\leq C\E \sup_y v_i(y)^4$ and $\max_i\E\sup_ y \|\ddot{f}\|^{4/3}<C.$

We recall that $\ddot{f}_{i,\beta}$ depends on $y$ through $\vbeta_i(y)$.  For every $y_1, y_2\in\mathcal Y$, by Assumption \ref{assemp}
and Lemma \ref{la.5sim},
$
\frac{1}{N}\sum_i\mathbb E (v_i(y_1)-v_i(y_2))^4
\leq C|y_1-y_2|.
$
Hence
\begin{eqnarray*}
	&&\sum_i\E|F_i(y_1)-F_i(y_2)|^2\leq\frac{C}{N}\sum_i(\mathbb E  |\ddot{f}_{i,\beta}(y_1)-\ddot{f}_{i,\beta}(y_2) |^4)^{1/2} (\E  v_i(y_1)  ^4)^{1/2}\cr
	&&
	+[\frac{C}{N}\sum_i\mathbb E (v_i(y_1)-v_i(y_2))^4]^{1/2}
	\leq C|y_1-y_2|^2+ C|y_1-y_2|^{1/2}\leq C|y_1-y_2|^{1/2}.
	\end{eqnarray*}
For every $\delta>0$, and  $\rho(y_1, y_2)=\bar C|y_1-y_2|^{1/4}$, for sufficiently large $\bar C$, 
\begin{eqnarray*}
	&&\sup_{\eta>0}\sum_i\eta^2 P\left(\sup_{\rho(y_1,y_2)<\delta}|F_i(y_1)-F_i(y_2)|>\eta\right)\leq \sum_i \E\left(\sup_{\rho(y_1,y_2)<\delta}|F_i(y_1)-F_i(y_2)|^2\right)
	\cr
	&\leq& \frac{C}{N}\sum_i(\mathbb E  \sup_{\rho(y_1,y_2)<\delta}|\ddot{f}_{i,\beta}(y_1)-\ddot{f}_{i,\beta}(y_2) |^4)^{1/2} (\E\sup_{y}  v_i(y)  ^4)^{1/2}\cr
	&&
	+[\frac{C}{N}\sum_i\mathbb E \sup_{\rho(y_1,y_2)<\delta}(v_i(y_1)-v_i(y_2))^4]^{1/2}
	\leq \delta^2.
\end{eqnarray*}
 Hence all  conditions of  Theorem 2.11.11 in \cite{VW} are verified. Thus $\sum_iF_i(y)=O_P(1)$ uniformly in $y$.  This implies $\sup_yH_1=O_P(\frac{1}{T\sqrt{N}}).$

\textbf{Step 1(b).}  Show    $\sup_yH_2=O_P(\frac{1}{T\sqrt{N}}).$  This follows from  Lemma \ref{la.2}.

 \textbf{Step 1(c).}   Analyze $H_3$. 
By Cauchy-Schwarz inequality, uniformly in $y$,
 \begin{eqnarray*}
 	H_3^2&\leq& O_P(1)\frac{1}{N}\sum_i\|\widetilde\Delta_i\|^2+O_P(1)\left(\frac{1}{N}\sum_i\left[   \frac{1}{T}\sum_t\psi_{it}(y)  \right]^4\right)^{1/2}\frac{1}{N}\sum_i\left[  \|R_{i,4} +R_{i,5}\|^2 +\|\widetilde\Delta_i\|^2\right]\cr
 	&&+O_P(1)\left(\frac{1}{N}\sum_i\left[ \widehat\beta_i-\beta_i \right]^4\right)^{1/2}\frac{1}{N}\sum_i\left[  \|R_{i,4} +R_{i,5}\|^2 +\|\widetilde\Delta_i\|^2\right]=O_P(\frac{1}{T^3}+\frac{L^2}{T^4}).
\end{eqnarray*}	
Together, provided that $N=o(T^2)$ and $NL^2=o(T^3)$,
 \begin{eqnarray*}
&&\sup_y|H_1+H_2+H_3|=O_P(\frac{1}{T\sqrt{N}}+\frac{1}{T^{3/2}}+\frac{L}{T^2})=o_P(\frac{1}{\sqrt{NT}}).
\end{eqnarray*}

\textbf{Step 2.} Bias correction. Because $\psi_{is}(y)$ is a martingale difference, and the loss function is the log-likelihood, 
$$\E v_i(y)=  \mathbb A_{1i}   \frac{1}{T}\sum_t\E \psi_{it}(y))\psi_{it}(y))' \mathbb A_{1i} =-\mathbb A_{1i}.$$

The effect of bias correction is: uniformly in $y$, 
 \begin{eqnarray*}
&& \frac{1}{2NT}\sum_i\tr \left[\partial^2_\beta f(  \widehat \beta_{i}(y) , D_{it}) \widehat \Sigma_i(y) ^{-1}\right]- \frac{1}{2NT}\sum_i\tr \left[\partial^2_\beta f(\beta_{i}(y), D_{it}) \E v_i(y) \right]\cr
&\leq& \frac{1}{2NT}\sum_i\tr \left[\partial^2_\beta f(\widehat\beta_{i}(y), D_{it})-\partial^2_\beta f(\beta_{i}(y), D_{it})    \right]   \widehat \Sigma_i(y)^{-1}\cr
&&+\frac{1}{2NT}\sum_i\tr \partial^2_\beta f(\beta_{i}(y), D_{it})\left[ \widehat \Sigma_i(y)^{-1}-\E v_i(y)\right]\cr
&\leq&\frac{C}{T}(\frac{1}{N}\sum_i\|\widehat\vbeta_i(y)-\vbeta_i(y)\|^2  )^{1/2}
(\frac{1}{N}\sum_i \|\widehat \Sigma_i(y)\|^2 )^{1/2}
+\frac{C}{T}(\frac{1}{N}\sum_i \|\widehat \Sigma_i(y)^{-1}-\E v_i(y)\|^2 )^{1/2} \cr
&=&O_P(\frac{1}{T^{3/2}})=o_P(\frac{1}{\sqrt{NT}}),
\end{eqnarray*}
where we used 
 \begin{eqnarray*}
&&\frac{1}{N}\sum_i \|\widehat \Sigma_i(y)^{-1}-\E v_i(y)\|^2
\leq O_P(1)\frac{1}{N}\sum_i \|[\nabla^2 Q_i(\widehat\vbeta_i(y))]^{-1} -[\nabla^2 \E Q_i(\vbeta_i(y))]^{-1}\|^2 =O_P(\frac{1}{T}).
\end{eqnarray*}

 So
 
 \begin{eqnarray*}
 && \frac{1}{N}\sum_{i} f(\widehat\vbeta_i(y), \widehat\vtheta(y), D_{it})-  \frac{1}{2NT}\sum_i\tr \left[
 \partial^2_\beta f(\widehat \beta_{i}(y), \widehat\vtheta(y), D_{it})  \widehat \Sigma_i(y)^{-1} \right]   -  f(\vbeta_i(y), \vtheta(y), D_{it})\cr
 &=& \frac{1}{NT}\sum_{it}  [  \vw_{i}' S_{wz} G (y) -\partial_\beta f                 _i(y)' ]  \mathbb A_{1i}(y)   \psi_{it}(y) +\frac{1}{N}\sum_{i=1}^N \vw_{i}' S_{wz} G (y) \vgamma_i(y)+o_P(\zeta_{NT}(y)).
 \end{eqnarray*}

\end{proof}

  \begin{lem}\label{la.2} Suppose Assumption \ref{assc.1} holds. Uniformly in $y\in\mathcal Y$,

(i)  $\frac{1}{N}\sum_i\partial_\beta f_i(y)'R_{i,4}  =O_P(\frac{1}{T\sqrt{N}}),$  

(ii)  $\frac{1}{N}\sum_i \partial_\beta f_i(y)' R_{i,5} =O_P(\frac{1}{T\sqrt{N}}).$ 
  \end{lem}
  \begin{proof}
  
 (i)  
 Recall that 
  $
\frac{1}{N}\sum_iR_{i,4}\partial_\beta f_i(y)'=  \frac{1}{N}\sum_i [\E (M_i(y))- M_i(y)]\partial_\beta f_i(y)' 
  $ where  by   $(AB)\otimes (AB)=(A\otimes A)(B\otimes B)$, 
  \begin{eqnarray*}
  M_i(y)&=&\frac{1}{2}\mathbb A_{1i} \mathbb A_{2i}( \mathbb A_{1i}\nabla Q_{i}(\beta_i)
)\otimes ( \mathbb A_{1i}\nabla Q_{i}(\beta_i))    \cr
&=&\frac{1}{2}\mathbb A_{1i} \mathbb A_{2i}( \mathbb A_{1i}\otimes \mathbb A_{1i})(\nabla Q_{i}(\beta_i
)\otimes \nabla Q_{i}(\beta_i))   
\end{eqnarray*}
  While this is a random matrix, its dimension is fixed. Hence we consider the one-dimensional case without loss of generality. In this case, $  M_i(y)$ is a scalar variable, which depends on $y$ through $\beta_i.$ Let 
  $b_i(y)=\frac{1}{2}\mathbb A_{1i} \mathbb A_{2i}( \mathbb A_{1i}\otimes \mathbb A_{1i})$ and 
  $$
  a_i(y)=T\nabla Q_{i}(\beta_i
)\otimes \nabla Q_{i}(\beta_i)-T\E[ \nabla Q_{i}(\beta_i
)\otimes \nabla Q_{i}(\beta_i)].
  $$
  Also,  let  $F_i(y)=-\frac{1}{\sqrt{N}}a_i(y)b_i(y)\partial_\beta f_i(y)'$. Then 
  
  $\frac{T\sqrt{N}}{N}\sum_iR_{i,4}\partial_\beta f_i(y)'=\sum_iF_i(y)$ and $\E F_i(y)=0$.

  Let $\ell^{\infty}(\mathcal Y)$ be the set of all uniformly bounded read functions on $\mathcal Y.$  It suffices to show that   $\sum_iF_i(y)$   is asymptotically tight in  $\ell^{\infty}(\mathcal Y)$,  by verifying conditions of Theorem 2.11.11 in \cite{VW}.

   Define a semi-metric $\rho(y_1,y_2)=\bar C|y_1-y_2|^{1/4}$ for all $y\in\mathcal Y$ and some large $\bar C>0$. To verify Condition (1) of the cited theorem, note for every $\eta>0$, and fix some $0<a<2$, we use the inequality $x^{b}1\{x>\eta\}\leq x^{b+a}\eta^{-a}$ for $x>0$ to have:
  \begin{eqnarray*}
  &&\sum_i \E \sup_y|F_i(y)|1\{\sup_y|F_i(y)|>\eta\}\cr
  &\leq&\eta^{-1}\frac{1}{N}\sum_i \E \sup_y|  a_i(y)b_i(y)\partial_\beta f_i(y)|^21\{\sup_y| a_i(y)b_i(y) \partial_\beta f_i(y)|>\sqrt{N}\eta\}\cr
   &\leq&\frac{1}{\eta^{a+1}N^{a/2}} \frac{1}{N}\sum_i \E \sup_y|   a_i(y)b_i(y) \partial_\beta f_i(y)|^{2+a}
    \cr
     &\leq&\frac{C}{\eta^{a+1}N^{a/2}} \frac{1}{N}\sum_i[ \E \sup_y|   a_i(y)  |^{m} ]^b \|\partial_\beta f_i(y)\|^4 
  \end{eqnarray*}
  for some constants $b >0$ and $m=4(2+a)/(2-a)$ using Holder's inequality. 
  
 By Assumption \ref{assemp}, for  some $c>0$,
  $\E[\sup_y(\frac{1}{\sqrt{T}}\sum_t\psi_{it}^j(y))^{8+c}|W]<C$.  Note that without loss of generality, we can write 
$$a_i(y)=\frac{1}{\sqrt{T}}\sum_t\psi_{it}^1(y)\frac{1}{\sqrt{T}}\sum_t\psi_{it}^2(y)-\E[\frac{1}{\sqrt{T}}\sum_t\psi_{it}^1(y)\frac{1}{\sqrt{T}}\sum_t\psi_{it}^2(y)]$$
for some functions $\E\psi_{it}^1(y)=\E\psi_{it}^2(y)=0.$  This implies $\frac{1}{N}\sum_i\E\sup_y|   a_i(y)  |^{m}<C$. Also $\max_i\sup_y\|\partial_\beta f_i(y)\|^4<C.$
      This verifies Condition (1).

Condition (2): 
For every $y_1, y_2\in\mathcal Y$, 
 \begin{eqnarray*}
	&&\sum_i\E|F_i(y_1)-F_i(y_2)|^2\leq
	\frac{1}{N}\sum_i(\E|a_i(y_1)b_i(y_1)-a_i(y_2)b_i(y_2)|^4)^{1/2}(\E\|\partial_\beta f_i(y_1)\|^4)^{1/2}\cr
 &&+\frac{1}{N}\sum_i(\E|a_i(y_2)b_i(y_2)|^4)^{1/2}(\E\|\partial_\beta f_i(y_1)-\partial_\beta f_i(y_2)\|^4)^{1/2}\cr \cr 
	&\leq&C	\frac{1}{N}\sum_ib_i(y_1)^2\frac{1}{N}\sum_i(\E |a_i(y_1)-a_i(y_2)|^4)^{1/2}
	+C\frac{1}{N}\sum_i|b_i(y_1)-b_i(y_2)|^2\frac{1}{N}\sum_i(\E a_i(y_2)^4)^{1/2}\cr
 &&+\frac{1}{N}\sum_i(\E\|\partial_\beta f_i(y_1)-\partial_\beta f_i(y_2)\|^4)^{1/2}\cr 
	&\leq&C	\frac{1}{N}\sum_i(\E |a_i(y_1)-a_i(y_2)|^4)^{1/2}
	+C\frac{1}{N}\sum_i|b_i(y_1)-b_i(y_2)|^2+\frac{1}{N}\sum_i(\E\|\partial_\beta f_i(y_1)-\partial_\beta f_i(y_2)\|^4)^{1/2}.
\end{eqnarray*}
 
First,  $	\frac{1}{N}\sum_i(\E |a_i(y_1)-a_i(y_2)|^4)^{1/2}$
 is bounded by $ I_1+I_2$ where for $A^{\otimes 2}:=A\otimes A$, 
\begin{eqnarray*}
	I_1&=&\left[\frac{1}{N}\sum_i |  \E  (\frac{1}{\sqrt{T}}\sum_t\psi_{it}(y_1) )^{\otimes 2}-  \E (\frac{1}{\sqrt{T}}\sum_t\psi_{it}(y_2) )^{\otimes 2} |^4\right]^{1/2} \leq C|y_1-y_2|^2\cr
	I_2&=& \left[\frac{1}{N}\sum_i\E |  (\frac{1}{\sqrt{T}}\sum_t\psi_{it}(y_1) )^{\otimes 2}- (\frac{1}{\sqrt{T}}\sum_t\psi_{it}(y_2) )^{\otimes 2} |^4 \right]^{1/2}\leq C|y_1-y_2|^{1/2}.
\end{eqnarray*}
The bound for $I_1$ is due to    Assumption \ref{ass4.7} (iii) combined with Lemma \ref{la.5sim}. To bound $I_2$, we fix any two elements of $\psi_{it}(y)$: $\psi_{it}(y_1)^1$ and $\psi_{it}(y_1)^2$, and let $f_j(y)=\frac{1}{\sqrt{T}}\sum_t\psi_{it}(y)^j$ for $j=1,2. $ Then
\begin{eqnarray*}
&&\left[\frac{1}{N}\sum_i\E |  f_1(y_1)  f_2(y_1) - f_1(y_2)  f_2(y_2) |^4\right]^{1/2}
\cr
&\leq& C(\max_{j=1,2}\frac{1}{N}\sum_i\E |  f_1(y_1) -f_1(y_2)    |^8)^{1/4} (\max_{j=1,2}\sup_y\frac{1}{N}\sum_i\E |  f_j(y)    |^8)^{1/4}\cr
&\leq& C|y_1-y_2|^{1/2}
\end{eqnarray*}
due to Assumption \ref{assemp}. This shows  $	\frac{1}{N}\sum_i(\E |a_i(y_1)-a_i(y_2)|^4)^{1/2}<C|y_1-y_2|^{1/2}$.

Next,   $\frac{1}{N}\sum_i|b_i(y_1)-b_i(y_2)|^2<C|y_1-y_2|^2$ since $\mathbb A_{1i}(y)$ and $\mathbb A_{2i}(y)$ are Lipschitz continuous with universal constants. 

Finally, $\frac{1}{N}\sum_i(\E\|\partial_\beta f_i(y_1)-\partial_\beta f_i(y_2)\|^4)^{1/2}\leq C|y_1-y_2|^{1/2}$ by Assumption \ref{assc.1}(ii).

 This verifies Condition (2) that 
 $$
 \sum_i\E|F_i(y_1)-F_i(y_2)|^2\leq C|y_1-y_2|^{1/2}\leq C\rho(y_1,y_2)^2.
 $$

Condition (3): For every $\delta>0$,
\begin{eqnarray*}
&&\sup_{\eta>0}\sum_i\eta^2 P\left(\sup_{\rho(y_1,y_2)<\delta}|F_i(y_1)-F_i(y_2)|>\eta\right)\cr
&\leq&\sum_i \E\left(\sup_{\rho(y_1,y_2)<\delta}|F_i(y_1)-F_i(y_2)|^2\right)
\cr
&\leq& \frac{1}{N}\sum_i \E\left(\sup_{\rho(y_1,y_2)<\delta}|a_i(y_1)b_i(y_1)-a_i(y_2)b_i(y_2)|^2\|\partial_\beta f_i(y_1)\|^2\right)\cr
&&+  \frac{1}{N}\sum_i \E\left(\sup_{\rho(y_1,y_2)<\delta}|a_i(y_2)b_i(y_2)|^2\|\partial_\beta f_i(y_1)-\partial_\beta f_i(y_2)\|^2\right)\cr  
&\leq&C	\frac{1}{N}\sum_i(\E\sup_{|y_1-y_2|<(\delta/\bar C)^4} |a_i(y_1)-a_i(y_2)|^4)^{1/2}
	+C\frac{1}{N}\sum_i\sup_{|y_1-y_2|<(\delta/\bar C)^4}|b_i(y_1)-b_i(y_2)|^2\cr
&& +C\frac{1}{N}\sum_i\sup_{|y_1-y_2|<(\delta/\bar C)^4}|\partial_\beta f_i(y_1)-\partial_\beta f_i(y_2)|^2\leq\delta^2
\end{eqnarray*}
by choosing a sufficiently large $\bar C$ in the definition of $\rho.$  
In the above, to bound 
$\frac{1}{N}\sum_i(\E\sup_{|y_1-y_2|<(\delta/\bar C)^4} |a_i(y_1)-a_i(y_2)|^4)^{1/2}$, note that a similar argument as verifying Condition (2) yields, by Assumption \ref{assemp},
\begin{eqnarray*}
&&\frac{1}{N}\sum_i(\E\sup_{|y_1-y_2|<(\delta/\bar C)^4} |f_1(y_1)f_2(y_1)-f_1(y_2)f_2(y_2)|^4)^{1/2}
\cr
&\leq &C\frac{1}{N}\sum_i(\E\sup_{|y_1-y_2|<(\delta/\bar C)^4} |f_1(y_1)-f_1(y_2)|^8)^{1/4}\cr
&\leq& C (\delta/\bar C)^2<\delta^2.
\end{eqnarray*}
 
 Hence all sufficient conditions of  Theorem 2.11.11 in \cite{VW} are verified. Thus $\sum_iF_i(y)=O_P(1)$ uniformly in $y$.

   (ii)  \textbf{Term $\frac{1}{N}\sum_iR_{i,5}\partial_\beta f_i(y)'$}. 
    Recall that 
    \begin{eqnarray*}
\frac{1}{N}\sum_iR_{i,5}\partial_\beta f_i(y)'&=&  \frac{1}{T\sqrt{N}}   \sum_i   F_i(y)\cr
 F_i(y)&=&\frac{1}{\sqrt{N}}  \mathbb  A_{1i}  M_i(y)\partial_\beta f_i(y)'\cr
 M_i(y)&=&T(A_{1i}^{-1}-\mathbb  A_{1i}  ^{-1} )\mathbb  A_{1i} \nabla Q_{i}(\beta_i)
  -T\E((A_{1i}^{-1}-\mathbb  A_{1i}  ^{-1} )\mathbb  A_{1i} \nabla Q_{i}(\beta_i))
  \end{eqnarray*}
We note $\E F_i(y)=0$.  It remains to show  $\sum_iF_i(y)$ to be asymptotically tight by verifying the conditions of Theorem 2.11.11 in \cite{VW}.

    Condition (1): for every $\eta>0,$ fix $0<a<2$, by the same argument as for $\frac{1}{N}\sum_iR_{i,4}\vw_{i}'$, 
    \begin{eqnarray*}
  &&\sum_i\E\sup_y|F_i(y)|1\{\sup_y|F_i(y)|>\eta\}\leq\frac{C}{\eta^{a+1}N^{a/2}} \frac{1}{N}\sum_i\E\sup_y|    M_i(y) |^{2+a}\|\partial_\beta f_i(y)\|^{2+a} \cr
    &\leq&\frac{C}{\eta^{a+1}N^{a/2}} \frac{1}{N}\sum_i[\E\sup_y| \sqrt{ T}  (\mathbb  A_{1i}^{-1} - A^{-1}_{1i})|^{4+2a}]^{1/2}| [\E\sup_y |\sqrt{ T} \nabla Q_{i}(\beta_i) |^{8+4a} ]^{1/4}+o(1)\cr
        &=&o(1).
  \end{eqnarray*}
 
    Condition (2). Define $a_i(y)=  \sqrt{T}   \mathbb  A_{1i} [A_{1i}^{-1}-\mathbb  A_{1i}^{-1} ] \mathbb A_{1i}$ and $b_i(y)=\sqrt{T} \nabla Q_{i}(\beta_i)$. Then $F_i(y)= \frac{1}{\sqrt{N}}   [ a_i(y)b_i(y)- \E a_i(y)b_i(y)   ]\partial_\beta f_i(y)'$,  
     \begin{eqnarray*}
	&&\sum_i\E|F_i(y_1)-F_i(y_2)|^2\leq
	C\frac{1}{N}\sum_i\E|a_i(y_1)b_i(y_1)-a_i(y_2)b_i(y_2)|^2 \|\partial_\beta f_i(y_1)\|^2\cr
	&&+  C\frac{1}{N}\sum_i\| \E  a_i(y_1)b_i(y_1)- \E   a_i(y_2)b_i(y_2)\|^2 \E\|\partial_\beta f_i(y)\|^2 \cr
 &&+C\frac{1}{N}\sum_i\E|a_i(y_2)b_i(y_2)|^2 \|\partial_\beta f_i(y_1)-\partial_\beta f_i(y_2)\|^2\cr
	&\leq& C\frac{1}{N}\sum_i[\E\|a_i(y_1) 
	-a_i(y_2)\|^4]^{1/2}  +C \frac{1}{N}\sum_i[\E \|b_i(y_1)
	-b_i(y_2)\|^4]^{1/2 } \cr 
 &&+C \frac{1}{N}\sum_i[\E \| \partial_\beta f_i(y_1)- \partial_\beta f_i(y_2)\|^4]^{1/2 } 
\end{eqnarray*}
where we used assumption $\Var(a_i(y)|\vw_{i})<\infty. $
The second term is bounded by $C|y_1-y_2|^{1/2}=C\rho(y_1,y_2)^2$.
 We now work on the  first term. Let $c_i(y)=\sqrt{T}[A_{1i}^{-1}(y) -\mathbb  A_{1i}^{-1}(y)]$.
      \begin{eqnarray*}
 a_i(y_1) 
	-a_i(y_2)
	&=& \mathbb  A_{1i}(y_1) c_i(y_1)\mathbb A_{1i} (y_1) - \mathbb  A_{1i}(y_2) c_i(y_2)\mathbb A_{1i} (y_2)  \cr
&=&[ \mathbb  A_{1i}(y_1)-  \mathbb  A_{1i}(y_2)]  c_i(y_1) \mathbb A_{1i} (y_1)
+\mathbb  A_{1i}(y_2)[ c_i(y_1)  - c_i(y_2)]   \mathbb A_{1i} (y_1)\cr
&&
+\mathbb  A_{1i}(y_2)  c_i(y_2)  [ \mathbb A_{1i} (y_1) -  \mathbb A_{1i} (y_2)  ].
	\end{eqnarray*}
	Hence  
   \begin{eqnarray*}
	 [  \frac{1}{N}\sum_i \E\|a_i(y_1) 
	-a_i(y_2)\|^4 ]^{1/2}
	&\leq&  C     \max_i \|   \mathbb  A_{1i}(y_1)-  \mathbb  A_{1i}(y_2)\|^ 2 [ \frac{1}{N}\sum_i \E\| c_i(y_1) \|^8 ]^{1/4}\cr
	 &&+ [  \frac{1}{N}\sum_i \E\|  c_i(y_1)  - c_i(y_2)\|^4 ]^{1/2}\cr
	&\leq& C|y_1-y_2|^2 + [  \frac{1}{N}\sum_i \E\|  c_i(y_1)  - c_i(y_2)\|^4 ]^{1/2}\cr
	&\leq& C|y_1-y_2|^{1/2},
	\end{eqnarray*}	
where the bound for terms involving $  \mathbb  A_{1i}(y_1)-  \mathbb  A_{1i}(y_2)$  simply follows from the fact that $ \mathbb  A_{1i}$  is  continuously differentiable with respect to $y$, with gradients uniformly bounded in $(y,i)$ (almost surely).

Condition (3): For every $\delta>0$,  for sufficiently large $\bar C$, 
\begin{eqnarray*}
	&&\sup_{\eta>0}\sum_i\eta^2 P\left(\sup_{\rho(y_1,y_2)<\delta}|F_i(y_1)-F_i(y_2)|>\eta\right)\cr
	&\leq& \frac{1}{N}\sum_i \E\left(\sup_{\rho(y_1,y_2)<\delta}\left|a_i(y_1)b_i(y_1)-a_i(y_2)b_i(y_2)
	- \E [a_i(y_1)b_i(y_1)-a_i(y_2)b_i(y_2)]
	\right|^2\|\partial_\beta f_i(y_1)\|^2\right)\cr
	&\leq&C	\frac{1}{N}\sum_i(\E\sup_{|y_1-y_2|<(\delta/\bar C)^4} |a_i(y_1)-a_i(y_2)|^4)^{1/2}
	+C	\frac{1}{N}\sum_i(\E\sup_{|y_1-y_2|<(\delta/\bar C)^4} |b_i(y_1)-b_i(y_2)|^4)^{1/2}\cr
	&&+C\frac{1}{N}\sum_i \E\left(\sup_{\rho(y_1,y_2)<\delta}  \E \| a_i(y_1)-a_i(y_2)\|^2\|\partial_\beta f_i(y_1)\|^2\right)\sup_y \E \|b_i(y)\|^2\cr
	&&+C\frac{1}{N}\sum_i \E\left(\sup_{\rho(y_1,y_2)<\delta}  \E \| b_i(y_1)-b_i(y_2)\|^2\|\partial_\beta f_i(y_1)\|^2\right)\sup_y \E \|a_i(y)\|^2\cr
 &&+C	\frac{1}{N}\sum_i(\E\sup_{|y_1-y_2|<(\delta/\bar C)^4} |\partial_\beta f_i(y_1)-\partial_\beta f_i(y_2)|^4)^{1/2} \cr 
	&\leq & (C/\bar C)\delta^2+C [    \frac{1}{N}\sum_i \E\sup_{|y_1-y_2|<(\delta/\bar C)^4}\|  c_i  (y_1)- c_i(y_2)\|^4]^{1/2}	\cr
	&&+C	\frac{1}{N}\sum_i(\E\sup_{|y_1-y_2|<(\delta/\bar C)^4} |b_i(y_1)-b_i(y_2)|^4)^{1/2} \cr
	&&+ C\sup_{\vw_{i}} \sup_{|y_1-y_2|<(\delta/\bar C)^4} [ \E \|  c_i  (y_1)- c_i(y_2)\|^2+ \|  b_i  (y_1)- b_i(y_2)\|^2]\leq\delta^2.
\end{eqnarray*}

  \end{proof}
  
\subsection{Step IV. Verify Assumption \ref{assb.1} using Assumption \ref{assc.1}   when $\mathcal Y$ is continuous}\label{sec:d.4}

 \begin{lem}\label{lemverd1}
Assumption \ref{assc.1} implies  Assumption \ref{assb.1}  when $ \vartheta(y)=\mathbb E_tf(\vbeta_i(y), \vtheta(y), D_{it})$.
 \end{lem}

  \begin{proof}
    
 By Lemma \ref{lc.2expfun}, 
\begin{eqnarray*} 
\widehat \vartheta(y)- \vartheta(y)&=&  \frac{1}{N}\sum_{i=1}^N\left[\frac{1}{\sqrt{T}}d_{\psi,i}(y)+d_{\vgamma,i}(y)\right] +o_P(\zeta_{NT}(y)),
 \end{eqnarray*}

	 where $  \partial_\beta f                 _i(y):=  \partial_\beta f                 (\vbeta_i(y),\vtheta(y), D_{it})$, $\zeta_{NT}=\frac{1}{\sqrt{NT}}+ \bar V_{\vgamma}(y),$
	 \begin{eqnarray*}
	 d_{\psi,i}(y)&=& \frac{1}{\sqrt{T}} \sum_{t=1}^T     (   \vw_{i}'   S_{wz} \bar G(y) -\partial_\beta f                 _i(y)' )   \mathbb A_{1i}(y)   \psi_{it}(y) \cr
	 d_{\vgamma,i}(y)&=&  \vw_{i}'   S_{wz} \bar G(y)\vgamma_i(y) + f(\vbeta_i(y),\vtheta(y), D_{it})-\E_t f(\vbeta_i(y),\vtheta(y), D_{it}).
	 \end{eqnarray*}

 \textit{Verifying Assumption \ref{assb.1} (i)(ii).} They follow from   Assumption \ref{assc.1} (i)(ii).

 \textit{Verifying Assumption \ref{assb.1}  (iii).} 
 \begin{eqnarray*}
 \E\sup_y |d_{\psi,i}(y)|^{2+a}&\leq& C  \E \sup_y   \|\frac{1}{\sqrt{T}} \sum_{t=1}^T  \psi_{it}(y) \vw_{i}'  \|^{2+a}  + C  \E \sup_y   \|\frac{1}{\sqrt{T}} \sum_{t=1}^T  \psi_{it}(y)\partial_\beta f_i(y)' \|^{2+a}  \cr
 &<& C\cr
 \E\sup_y|\frac{d_{\vgamma,i}(y)^2}{\bar V_{\vgamma}(y)}|^{a}&\leq&C\E\sup_y\left[ \mathbb Z_{t}(\vw_{i}'   S_{wz} \bar G(y)\vgamma_i(y) + f(\vbeta_i(y), \vtheta(y),  D_{it}) )\right]^{2a}<C.
 \end{eqnarray*}
 
 \textit{Verifying Assumption \ref{assb.1}  (iv).} This condition is verified using the triangular inequality  and   Assumption \ref{assc.1}.
 
  \end{proof}

      \subsection{When $\mathcal Y$ is discrete with finite support}\label{sec:d.5}

\begin{lem}\label{l5fun} Suppose Assumption \ref{assc.1} (i)-(iii) hold.
For any  $y\in\mathcal Y$, 
\begin{eqnarray*}
 \widehat\vartheta (y)-\vartheta (y)&=& o_P(\zeta_{NT}(y))+
 \frac{1}{NT}\sum_{it}   (   \vw_{i}'   S_{wz} \bar G(y) -\partial_\beta f                 _i(y)' )   \mathbb A_{1i}(y)   \psi_{it}(y)
  \cr
 && +  \frac{1}{N}\sum_{i=1}^N\left( \vw_{i}' S_{wz} \bar G(y)\vgamma_i(y)
 	 +  \left[ f(\vbeta_i(y), \vtheta(y), D_{it}) - \mathbb E_tf(\vbeta_i(y),\vtheta(y), D_{it})\right]\right).
 \end{eqnarray*}
  \end{lem}

	 \begin{proof} 
The proof is similar to that of Lemma \ref{lc.2expfun}, except that we only need to establish the pointwise convergence for $y\in\mathcal Y$. So we omit repetitions.  
     \end{proof}

\section{ Proof of Theorems \ref{th4.1} and   \ref{th4.20}}\label{sec:verifyV}

When $\mathcal Y$ is continuous,  the proof proceeds as follows. First, we verify 
  Assumption \ref{assc.1} respectively under the settings of Theorems \ref{th4.1} and   \ref{th4.20}.
Then by Lemmas \ref{lc.2expfun} and  \ref{lemverd1}, 
we have expansion (\ref{eqb.1}) and Assumption \ref{assb.1}  hold. Thus we can apply Proposition \ref{probb.1}.
This will prove the theorems. 

When $\mathcal Y$ is discrete with finite support,  we verify 
  Assumption \ref{assc.1} (i)-(iii).
Then  Lemma  \ref{l5fun} holds. Then we can establish the convergence in distribution using central limit theorem.

\subsection{Proof of Theorems \ref{th4.1}.}

         \begin{proof}
 In this case $f(\vbeta,\vtheta, D) = \vtheta$.
     Recall that $\vw_{i}$ is the exogenous variable and  
               $$
               \widehat\vtheta(y)= \frac{1}{N}\sum_{i=1}^N\widehat\vbeta_i(y) \vw_{i}' S_{wz,N},\quad S_{wz,N}:=(\frac{1}{N}W'W)^{-1} W'Z(Z'P_WZ)^{-1}.
               $$
               In this case,
                \begin{eqnarray*}
	 d_{\psi,i}(y)&=& \frac{1}{\sqrt{T}} \sum_{t=1}^T      \vw_{i}'   S_{wz} \bar G(y)   \mathbb A_{1i}(y)   \psi_{it}(y) \cr
	 d_{\vgamma,i}(y)&=&  \vw_{i}'   S_{wz} \bar G(y)\vgamma_i(y) .
	 \end{eqnarray*}

              We note that $\partial_{\beta} f=0$ and $\partial_{\vtheta}f=\bar G(y)= I.$

 \textit{Verifying Assumption \ref{assc.1}(i).}  $\max_i\E\sup_ y \|\nabla{f}_i\|^{8}+\max_i\E\sup_ y \|\nabla^2{f}_i\|^{4}<C.$

 \textit{Verifying Assumption \ref{assc.1}(ii).} It follows from Assumption \ref{dgpcs}(i). Also    $\Var_t(d_{\psi,i}(y))>c>0$ follows from Lemma \ref{la.4new}.  

  \textit{Verifying Assumption \ref{assc.1}(iii).} Let   $W(y)=\vw_{i}'   S_{wz} \bar G(y)\vgamma_i(y) + f(\vbeta_i(y), \vtheta(y),  D_{it}) $. Then by Assumption \ref{ass4.3}(i)
    \begin{eqnarray*}
      \E\sup_y\left[ \mathbb Z_{t}( W(y) )\right]^{4}\leq C\E\left[ \sup_{y\in\mathcal Y}\left(\frac{\|   \vgamma_i(y)\vw_{i}'  \|  }{ \lambda^{1/2} _{\min}(V_{\vgamma}(y))}\right)^{4}   \right]<C.
    \end{eqnarray*}  

    \textit{Verifying Assumption \ref{assc.1}(iv).}   $\partial_\beta f_i(y)=0$ and $\ddot{f}_{i,\beta}(y_1)=0$. Also, we have $    \bar V_{\vgamma}(y)=\Var_t [\vw_{i}' S_{wz} \bar  G(y)  \vgamma_i(y) ]   $ and   $V_{\vgamma}(y)= \E\left\{ (S_{wz}'\vw_{i}\vw_{i}'S_{wz})
 \otimes \E(\vgamma_i(y)\vgamma_i(y)'     \mid \vw_{i})\right\}$.  Hence by Assumption \ref{ass4.7}(ii),
   \begin{eqnarray*}
    \frac{1}{N}\sum_i \E\sup_{\rho(y_1, y_2)< \delta } \left|  \frac{  d_{\vgamma,i}(y_1) }{\sigma_T(y_1) }-  \frac{  d_{\vgamma,i}(y_2) }{\sigma_T(y_2) } \right|^2&\leq&  \frac{1}{N}\sum_i \E\sup_{\rho(y_1, y_2)< \delta } \left|  \frac{ \vgamma_i(y_1)\vw_i }{\sigma_T(y_1) }-  \frac{ \vgamma_i(y_2)\vw_i }{\sigma_T(y_2) } \right|^2<\delta^2.
\end{eqnarray*}
    

Hence  Assumption \ref{assc.1} holds when $\mathcal Y$ is continuous.

When $\mathcal Y$  is discrete with finite support, by Lemma \ref{l5fun}, it suffices to verify Assumption \ref{assc.1} (i)-(iii). They can be verified using the same argument as above.
 
 \end{proof}

  \begin{lem}\label{la.4new} 
$\inf_{y}\lambda_{\min}(  V_\psi(y))>c>0$.   
  \end{lem}
  
  \begin{proof}

  We first define  some notation. For matrices we write  $A\geq 0$ if $A$ is   semipositive definite, and write  $A\geq B$ if $A-B\geq 0$.   
  Let  $ G_i (y) := \mathbb A_{1i}(y) \Var( \frac{1}{\sqrt{T}}\sum_{t\leq T}     \psi_{it}(y)  )\mathbb A_{1i}(y)  .$ 
Let     $S_i=S_{wz}'\vw_{i}\vw_{i}'S_{wz} $ and $S_{\psi,i}(y)= \Var( \frac{1}{\sqrt{T}}\sum_{t\leq T}     \psi_{it}(y) |\vw_{i})$.  Then almost surely
 $$
\inf_{Y_M} \min_i\lambda_{\min}(G_i (y))\geq \inf_{Y_M} \min_i\lambda_{\min}^2( {\mathbb A}_{i,y1})\min_i\lambda_{\min}(S_{\psi,i}(y) )>c.
 $$
  So
 $
  S_i\otimes  [G_i (y)-cI]\geq 0,
 $ which implies  $  S_i\otimes  G_i (y)) \geq   S_i\otimes   (cI)$.
Let  $v_y$ be the eigenvector of $ V_\psi(y)$ corresponding to its smallest eigenvalue, 
 \begin{eqnarray*}
 \inf_{Y_M} \lambda_{\min}( V_\psi(y))&=& \inf_{Y_M}  v_y'\E[S_i \otimes G_i (y)]v_y\geq  \inf_{Y_M} v_y'[ \E  S_i \otimes (c I)] v_y\cr
 &=& \inf_{Y_M} v_y'[(\E S_i)\otimes (cI )]v_y\geq \lambda_{\min}[(\E S_i)\otimes (cI )]\cr
& = & c\lambda_{\min}(\E S_i)
=c\lambda_{\min}(S_{wz}' \E  \vw_{i}\vw_{i}'S_{wz}) >c.
 \end{eqnarray*}

  \end{proof}

\subsection{Proof of Theorems \ref{th4.20}}\label{sec5.2prof}

\begin{proof}
    
\textbf{Verify Assumption \ref{assc.1} for $F_{t}$.}  In the case $F_t(y)=\vartheta (y) $, then we can write 
$$F_t(y)= \mathbb E_tf(\vbeta_i(y),  \vtheta(y), D_{it})$$ with $f(b, \vtheta,  D_{it})= \Lambda(-  \vx_{it}'b)$ which does not depend on $\vtheta$.
We have  $\partial_\beta f_i= -\dot  \Lambda(- \vx_{it}'\vbeta_i(y)) \vx_{it}$, and $\ddot f_i=\ddot  \Lambda(- \vx_{it}'\vbeta_i(y)) \vx_{it} \vx_{it}'$.

\textit{Verifying Assumption \ref{assc.1}(i)}. 
We have 
$$\max_i\E\sup_ y \|\partial_\beta {f}_i\|^{8}+\max_i\E\sup_ y \|\ddot{f}_{i,\beta}\|^{4}
\leq C\max_i\E \|   \vx_{it}\|^8<C
$$

  \textit{Verifying Assumption \ref{assc.1}(ii)}. 
 This holds given $\E[\psi_{it}(y_k)| \beta_i(y_l),    \vx_{it}]=0$ and 
\begin{eqnarray}\label{eqe.1c} \Var_t(d_{\psi,i}(y))
& =&\E_t\dot  \Lambda(-   \vx_{it}'\vbeta_i(y))^2   \vx_{it}'   \mathbb A_{1i}\Var_t(\frac{1}{\sqrt{T}}\sum_s \psi_{is}(y)|   \vx_{it},\vbeta_i(y))
 \mathbb A_{1i}    \vx_{it}\cr
 &\geq& \lambda_{\min}(\Var_t(\frac{1}{\sqrt{T}}\sum_s \psi_{is}(y)|   \vx_{it},\vbeta_i(y))) 
 \lambda_{\min}(\mathbb A_{1i}(y)^2)
 \E_t\dot  \Lambda(-   \vx_{it}'\vbeta_i(y))^2\|   \vx_{it}\|^2\cr 
 &>&c.
 \end{eqnarray}

  \textit{Verifying Assumption \ref{assc.1}(iii)}. 
  This holds since $\E\sup_y\left[\mathbb   Z_{t}(\Lambda(-   \vx_{it}'\vbeta_i(y)) )\right]^{4}<C.$

 \textit{Verifying Assumption \ref{assc.1}(iv)}. By Assumption \ref{ass4.7}(iv), for $k\geq 4$, 
\begin{eqnarray*}
&&
 \mathbb E  \|\ddot{f}_{i,\beta}(y_1)-\ddot{f}_{i,\beta}(y_2) \|^4
 \leq \mathbb E  | \ddot  \Lambda(-   \vx_{it}'\beta_i(y_1))-\ddot  \Lambda(-   \vx_{it}'\beta_i(y_2))   |^4 \|   \vx_{it}\|^{8}
 \leq C|y_1-y_2|^4\cr
&&\E|\partial_\beta {f}_i(y_1)-\partial_\beta {f}_i(y_2) |^4
\leq \mathbb E  | \dot  \Lambda(-   \vx_{it}'\beta_i(y_1))-\dot  \Lambda(-   \vx_{it}'\beta_i(y_2))   |^4 \|   \vx_{it}\|^{4}
 \leq C|y_1-y_2|^4 .
 \end{eqnarray*}
 The rest of the inequalities in this condition follow from Assumption \ref{ass4.7}.

\textbf{Verify Assumption \ref{assc.1} for $G_{t}$.} 

In this case $$G_t(y)= \mathbb E_tf(\vbeta_i(y),  \vtheta(y), D_{it})$$
where 
$f(\vbeta_i(y), \vtheta(y), D_{it})= \Lambda(-  h_{it}(\vx_{it})'\vbeta_i^g(y))$, and 
 $ \vbeta_i^g(y)= \vtheta(y)[g(\vz_{i})-\vz_{i}] +\vbeta_i(y)$. 
  We have   $\partial_\beta f_{i} = -\dot  \Lambda(- h_{it}( \vx_{it})'\vbeta^g_i(y)) h_{it}( \vx_{it})$, and 
  $\ddot f_{i,\beta}=\ddot  \Lambda(- h_{it}( \vx_{it})'\vbeta^g_i(y)) h_{it}( \vx_{it})h_{it}( \vx_{it})'$.
  
  \textit{Verifying Assumption \ref{assc.1}(i)}.
\begin{eqnarray*}
&&\max_i\E_t\sup_ y \|\partial_\beta f_{i} \|^{8}
+\max_i\E_t\sup_ y \|\partial_{\vtheta} f_{i} \|^{8}
\leq C\E_t \|h_{it}( \vx_{it})\|^8+C\E \|h_{it}( \vx_{it})\|^8\|g(\vz_{i})-\vz_{i}\|^8<C\cr
&&\max_i\E_t\sup_ y \|\ddot{f}_{i,\beta}\|^{4}+\max_i\E_t\sup_ y \|\ddot{f}_{i,\vtheta}\|^{4}\leq C\E_t \|h_{it}( \vx_{it})\|^8+C\E \|h_{it}( \vx_{it})\|^8\|g(\vz_{i})-\vz_{i}\|^8<C\cr
&&\max_i\E_t\sup_ y \|\ddot{f}_{i,\beta\vtheta}\|^{4}
\leq C\E \|h_{it}( \vx_{it})\|^8\|g(\vz_{i})-\vz_{i}\|^4<C.
\end{eqnarray*}

\textit{Verifying Assumption \ref{assc.1}(ii)}.
This holds for $\E[\psi_{it}(y_k)| \beta_i(y_l), h_{it}( \vx_{it}), \vz_{i}, \vw_{i}]=0$ and 
\begin{eqnarray*} \Var_t(d_{\psi,i}(y))
 &\geq& \lambda_{\min}(\Var_t(\frac{1}{\sqrt{T}}\sum_s \psi_{is}(y)| h_{it}( \vx_{it}),\vz_{i}, \vbeta_i(y))) 
 \lambda_{\min}(\mathbb A_{1i}(y)^2)\cr
 &&\times 
 \E_t \|    \vw_{i}'   S_{wz} \bar G(y) + \dot\Lambda(-h_{it}( \vx_{it})'\vbeta_i^g(y))h_{it}( \vx_{it})'  \|>c.
 \end{eqnarray*}

  \textit{Verifying Assumption \ref{assc.1}(iii)}. 
  This holds since $$\E\sup_y\left[ \mathbb Z_{t}(\vw_{i}'   S_{wz} \bar G(y)\vgamma_i(y) + \Lambda(-h_{it}( \vx_{it})'\vbeta_i^g(y)) )\right]^{4}  <C.$$

\textit{Verifying Assumption \ref{assc.1}(iv)}.  By Assumption \ref{ass4.7}(iv), for $k\geq 4$, 
\begin{eqnarray*}
&&\mathbb E_t  | \ddot  \Lambda(- h_{it}( \vx_{it})'\vbeta^g_i(y_1))-\ddot  \Lambda(- h_{it}( \vx_{it})'\vbeta^g_i(y_2))   |^k \| h_{it}( \vx_{it})\|^{2k}
 \leq C|y_1-y_2|^k\cr
 && \mathbb E_t  | \dot  \Lambda(- h_{it}( \vx_{it})'\vbeta^g_i(y_1))-\dot  \Lambda(- h_{it}( \vx_{it})'\vbeta^g_i(y_2))   |^k \| h_{it}( \vx_{it})\|^{k}
 \leq C|y_1-y_2|^k.
 \end{eqnarray*}

Also note that 
$$\bar G(y)= -\E_t \dot\Lambda(-h_{it}( \vx_{it})'\vbeta^g_i(y))\vecc(h_{it}( \vx_{it})(g(\vz_{i})-\vz_{i})').$$ 
Hence $\sup_y\|\bar G(y)\|\leq C \E_t\|h_{it}( \vx_{it})\|\|g(\vz_{i})-\vz_{i}\|<C$, and 
$$\|\bar G(y_1)-
 \bar G(y_2)\|\leq  C \mathbb E_t  | \dot  \Lambda(- h_{it}( \vx_{it})'\vbeta^g_i(y_1))-\dot  \Lambda(- h_{it}( \vx_{it})'\vbeta^g_i(y_2))   | \| h_{it}( \vx_{it})[g(\vz_{i})-\vz_{i}]'\| 
 \leq C|y_1-y_2|.
 $$

Hence Assumption \ref{assc.1} has been verified. By  Lemma \ref{lc.2expfun}, 
\begin{eqnarray}\label{eqe.1} 
\widehat F(y)- F(y)&=&  \frac{1}{N}\sum_{i=1}^N\left[\frac{1}{\sqrt{T}}d_{\psi,i}(y)+d_{\vgamma,i}(y)\right] +o_P(\zeta_{NT}(y)),
 \end{eqnarray}
where $(d_{\psi,i}, d_{\vgamma,i})\in \{(d^0_{\psi,i}, d^0_{\vgamma,i}),(d^{II}_{\psi,i}, d^{II}_{\vgamma,i})\}$, corresponding to $ F\in\{F_t,  G_{t}\}$ as defined in Section \ref{sec:Define}. The desired theorem then follows from Proposition \ref{probb.1}.

  When $\mathcal Y$ is discrete with finite support, verifying  
Assumption \ref{assc.1} (i)-(iii)  for $F_t, G_t, F_{\infty} $  $G_{\infty}$ follows similarly so we omit its proof for brevity. 
\end{proof}

 \section{Proof of Theorem \ref{th4.30} }\label{sec:proqte}

  Consider a generic $F\in\{F_t,  G_{t}\}$. Let $\widehat{F}\in\{\widehat{F}_t, \widehat{G}_{t}\}$ be its estimator. 
The proof proceeds as follows. We respectively verify expansion (\ref{eqb.1}) and Assumption \ref{assb.1}. Then   we  apply Proposition \ref{probb.1} to prove the theorem. 
  
 To verify  expansion (\ref{eqb.1}), the goal is to obtain an expansion for $\phi(\widehat{F} ,\tau)- \phi({ F} ,\tau)$ uniformly in $\tau.$ The novelty of our analysis is that $\widehat{F} -{ F}$ does \text{not} weakly converge due to the   unknown rate of convergence we discussed earlier. Hence the usual  functional delta method is not directly applicable.
 Instead, we obtain an expansion for the standardized $\phi(\widehat{F} ,\tau)- \phi({F} ,\tau).$

 \subsection{Verify  expansion (\ref{eqb.1})}

\begin{proof}
    	 Lemma \ref{lc.30} below shows the uniform expansions of $\phi(\widehat{F} ,\tau)- \phi({ F} ,\tau)$ for $F\in\{F_t,  G_{t}\}$ and  $\widehat{F}\in\{\widehat{F}_t, \widehat{G}_{t}\}$. This implies
 	 \begin{eqnarray*}
 	  \widehat\QTE(\tau)- \QTE(\tau)&=&\frac{1}{N}\sum_{i=1}^N\left[\frac{1}{\sqrt{T}}p_{\psi,i}(\tau)+p_{\vgamma,i}(\tau)\right] +o_P(\frac{1}{\sqrt{NT}}+\sqrt{\frac{1}{N}\Var_t(p_{\vgamma,i}(\tau))}),
 	 \end{eqnarray*}
 where 	 for $q_0(\tau)=\phi(  F_{t}, \tau)$, $q_{II}(\tau)=\phi(  G_{t}, \tau),$
 	 \begin{eqnarray}\label{eqd.1adfa}
 	  p_{\psi,i}(\tau)&=&  \frac{-d_{\psi,i}^{II}(q_{II}(\tau))}{\dot G_{t}(q_I(\tau))}+\frac{d_{\psi,i}^0(q_0(\tau))}{\dot F_{t}(q_0(\tau))},\quad 	 p_{\vgamma,i}(\tau) =   \frac{-d_{\vgamma,i}^{II}(q_{II}(\tau))}{\dot G_{t}(q_{II}(\tau))}+\frac{d_{\vgamma,i}^0(q_0(\tau))}{\dot F_{t}(q_0(\tau))}.\cr
 	  \end{eqnarray}
In the $o_P(.)$ term, we used the Assumption \ref{ase.48} that $\Var_t(d^0_{\vgamma,i})+\Var_t(d^{II}_{\vgamma,i})=O(\Var_t(p_{\vgamma,i}^{II}))$.  Then   expansion (\ref{eqb.1}) has been verified. 
\end{proof}

  \begin{lem}\label{lc.30}
Let $\dot{F}$ be the density of $ F\in\{F_t,  G_{t}\}$. Let $q(\tau)=\phi(F, \tau)$, $z(y)= (NT)^{-1/2}+ N^{-1/2}\Var_t(d_{\vgamma,i}(y))^{1/2}.$  Uniformly in $\tau$, we have 
$$ 	\phi(\hF,\tau)- \phi( F,\tau)=\frac{-1}{	\dot  F(  q(\tau))}\frac{1}{N}\sum_i \left[\frac{1}{\sqrt{T}}d_{\psi,i}(q(\tau))+d_{\vgamma,i}(q(\tau))\right]+ o_P(z(q(\tau))), $$
where $(d_{\psi,i}, d_{\vgamma,i})\in \{(d^0_{\psi,i}, d^0_{\vgamma,i}),(d^{II}_{\psi,i}, d^{II}_{\vgamma,i})\}$, corresponding to $ F\in\{F_t,  G_{t}\}$ as defined in Section \ref{sec:Define}.

 \end{lem}
\begin{proof}
	
Consider a generic $ F\in\{F_t,  G_{t}\}$. Let $\widehat{F}\in\{\widehat{F}_t, \widehat{G}_{t}\}$ be its estimator. 
		Note that $ F(\phi( F, \tau)) = \widehat{F}(\phi(\widehat{F}, \tau))=\tau $, we have
	\begin{equation}\label{eqc.4}
 F(\phi(\hF,\tau)) -  F(\phi( F,\tau)) = -[\hF(\phi(\hF,\tau)) - F(\phi(\hF,\tau))  ].
	\end{equation}
	Applying the mean value theorem to the left hand side, there is $\widetilde q_\tau$ so that 
	$$
	\phi(\hF,\tau)-\phi( F,\tau)=\frac{-1}{\dot  F(\widetilde q_\tau)}[\hF(\phi(\hF,\tau)) - F(\phi(\hF,\tau))  ].
	$$
	We have proved that $\sup_y|\widehat {  F}-  F|=o_P(1)$ in  Lemma \ref{lc.2expfun}. By the continuous mapping theorem $\sup_\tau |\phi(\widehat{F} , \tau)-\phi(F , \tau)|=o_P(1)$. Hence $1/\dot  F(\widetilde q_\tau)<C$ uniformly in $\tau$. This implies $
	|\phi(\hF,\tau)-\phi( F,\tau)|\leq C|\Delta_{ F}(\phi(\hF,\tau))|
	$
	where $$
	\Delta_{ F}(y):= \hF(y) - F(y).
	$$
	Applying the second-order mean value theorem to the left hand side of (\ref{eqc.4}), there is $c_\tau$ so that, for $q(\tau):=\phi( F,\tau)$,
	$$
	\dot  F(q(\tau))(\phi(\hF,\tau)- q(\tau)) + \frac{1}{2}\frac{d^2 F(c_\tau)}{dy}(\phi(\hF,\tau)-q(\tau))^2=  -[\hF(\phi(\hF,\tau)) - F(\phi(\hF,\tau))  ].
	$$
	Rearranging and applying $
	|\phi(\hF,\tau)-\phi( F,\tau)|\leq C|\Delta_{ F}(\phi(\hF,\tau))|
	$, we have 
	\begin{eqnarray}\label{eqc.5}
		\phi(\hF,\tau)-q(\tau)&=&\frac{-1}{	\dot  F(q(\tau))} \Delta_{ F}(\phi(\hF,\tau)) + M_1(\tau)\cr
		&=&\frac{-1}{	\dot  F(q(\tau))} \Delta_{ F}(q(\tau)) + M_1(\tau)+ M_2(\tau)\cr
		M_1(\tau)&\leq& C\Delta_{ F}(\phi(\hF,\tau))^2
		\leq C|\Delta_{ F}(\phi(\hF,\tau))- \Delta_{ F}(q(\tau))|^2+C\Delta_{ F}(q(\tau))^2
		\cr
			M_2(\tau)&=& -\frac{ \Delta_{ F}(\phi(\hF,\tau))- \Delta_{ F}(q(\tau))}{	\dot  F(q(\tau))}
			\leq C|\Delta_{ F}(\phi(\hF,\tau))- \Delta_{ F}(q(\tau))|.\cr
	\end{eqnarray}

By (\ref{eqe.1})  for all $F\in\{F_t, G_{t}\}$,  
$$
\Delta_{ F}(y)=
		 \frac{1}{N}\sum_i \left[\frac{1}{\sqrt{T}}d_{\psi,i}(y)+d_{\vgamma,i}(y)\right]+
			o_P(\zeta_{NT}(y)).
 $$

 By Lemma \ref{lc.3}, $\frac{1}{\sqrt{N}}\sum_{i}d_{\psi,i}(y) $  
 and $V_F(y)^{-1/2}\frac{1}{\sqrt{N}}\sum_{i} d_{\vgamma,i}(y)$  are  stochastically equicontinuous  in  $\ell^{\infty}(\mathcal Y)$, where $V_F(y):=\frac{1}{T}\Var_t(d_{\psi,i}(y))+\Var_t(d_{\vgamma,i}(y))$.    Then for $q(\tau)=\phi( F,\tau)$, $\widehat q(\tau)=\phi(\hF,\tau)$, 
 	\begin{eqnarray*}
&&	|\Delta_{ F}(\phi(\hF,\tau))- \Delta_{ F}(q(\tau))|\leq \frac{1}{\sqrt{NT}}\left|\frac{1}{\sqrt{N}}\sum_{i}d_{\psi,i}(\widehat q(\tau))- d_{\psi,i}( q(\tau))\right|\cr
	&&+\frac{1}{\sqrt{N}}\left|\frac{1}{\sqrt{N}}\sum_i d_{\vgamma,i}(\widehat q(\tau))-d_{\vgamma,i}(q(\tau)) \right|
	+	o_P(\zeta_{NT}(q(\tau)))\cr
	&=&
 \frac{ 	V_F(\widehat q(\tau))^{1/2}}{\sqrt{N}} \left|   	V_F(\widehat q(\tau))^{-1/2}    \frac{1}{\sqrt{N}}\sum_{i}d_{\vgamma,i}(\widehat q(\tau))-   	V_F(q(\tau))^{-1/2}    \frac{1}{\sqrt{N}}\sum_{i}d_{\vgamma,i}(q(\tau))\right|\cr
 &&+\frac{1}{\sqrt{N}} \left| 	V_F(\widehat q(\tau))^{1/2}   	V_F(q(\tau))^{-1/2}    - 1\right| \left|\frac{1}{\sqrt{N}}\sum_{i}d_{\vgamma,i}(q(\tau))\right| \cr
	&=&o_P(\zeta_{NT}(q(\tau)))\cr
&&		M_1(\tau)+	M_2(\tau)=o_P(\zeta_{NT}(q(\tau))).
	\end{eqnarray*}

	The desired expansion then follows from  (\ref{eqc.5}).

\end{proof}

 \begin{lem}\label{lc.3}
   Let $V_F(y):=\frac{1}{T}\Var_t(d_{\psi,i}(y))+\Var_t(d_{\vgamma,i}(y))$.  Then 
 $\frac{1}{\sqrt{N}}\sum_{i}d_{\psi, i}(y)$,    and $V_F(y)^{-1/2}  \frac{1}{\sqrt{N}}\sum_{i}d_{\vgamma, i}(y)$   are  asymptotically stochastically equicontinuous (ASE).
 \end{lem}
 
 \begin{proof}
 	We show respectively that  both $\frac{1}{\sqrt{N}}\sum_{i}d_{\psi, i}(y)$,    and $\Var_t(d_{\vgamma,i}(y))^{-1/2}  \frac{1}{\sqrt{N}}\sum_{i}d_{\vgamma, i}(y)$ are asymptotically tight under the metric $\rho(y_1,y_2)= \bar C|y_1-y_2|^{1/4}$ for some large $\bar C.$
 	
 	(i)  For any $\eta,\delta>0$,  by Assumption \ref{assb.1}, 
\begin{eqnarray*}
&&	\sum_i\E \sup_y|N^{-1/2}d_{\psi, i}(y)|1\{\sup_y|N^{-1/2}d_{\psi, i}(y)|>\eta\}
\cr
&\leq&	\frac{1}{N\eta}\sum_i\E \sup_y|d_{\psi, i}(y)|^21\{\sup_y|d_{\psi, i}(y)|>\sqrt{N}\eta\}
\cr
&\leq & C \sqrt{P(\sup_y|d_{\psi, i}(y)|^2> N\eta^2)}
\leq \frac{C}{N}\sqrt{\E\sup_y|d_{\psi, i}(y)|^2}=o(1).
\cr
	&&\frac{1}{N}\sum_i\E|d_{\psi, i}(y_1)-d_{\psi, i}(y_2)|^2\leq C|y_1-y_2|^{1/2}.
\cr
	&&\sup_{\eta>0}\sum_i\eta^2 P\left(\sup_{\rho(y_1,y_2)<\delta}|d_{\psi, i}(y_1)-d_{\psi, i}(y_2)|>\sqrt{N}\eta\right)\leq  \E\left(\sup_{\rho(y_1,y_2)<\delta}|d_{\psi, i}(y_1)-d_{\psi, i}(y_2)|^2\right)
	\cr
	&\leq& \delta^2.
\end{eqnarray*}
 Hence all  conditions of  Theorem 2.11.11 in \cite{VW} are verified. This implies the ASE of  $\frac{1}{\sqrt{N}}\sum_{i}d_{\psi, i}(y)$.

(ii) Write $v(y)= V_F(y)^{1/2} $. Suppose $V_F(y_1)\geq V_F(y_2)$.     Still by Assumption \ref{assb.1}, 
    \begin{eqnarray*}
&&	\sum_i\E \sup_y|N^{-1/2}V_F(y)^{-1/2}d_{\vgamma, i}(y)|1\{\sup_y|N^{-1/2}V_F(y)^{-1/2}d_{\vgamma, i}(y)|>\eta\}
\cr
&\leq&	\frac{1}{N\eta}\sum_i\E \sup_y[\frac{|d_{\vgamma, i}(y)|}{V_F(y)^{1/2}}]^21\{\sup_yV_F(y)^{-1/2}|d_{\vgamma,  i}(y)|>\sqrt{N}\eta\}
\cr
&\leq & \frac{C}{N}\sqrt{\E\sup_y[\frac{|d_{\vgamma, i}(y)|}{V_F(y)^{1/2}}]^2}=o(1).\cr
	&&\frac{1}{N}\sum_i\E|d_{\psi, i}(y_1)-d_{\psi, i}(y_2)|^2
	\leq C|y_1-y_2|^{1/2}. \cr
&&	\sup_{\eta>0}\sum_i\eta^2 P\left(\sup_{\rho(y_1,y_2)<\delta}|V_F(y_1)^{-1/2}d_{\vgamma, i}(y_1)-V_F(y_2)^{-1/2}d_{\vgamma, i}(y_2)|>\sqrt{N}\eta\right)\cr
&\leq&  \E\left(\sup_{\rho(y_1,y_2)<\delta}|V_F(y_1)^{-1/2}d_{\vgamma, i}(y_1)-V_F(y_2)^{-1/2}d_{\vgamma, i}(y_2)|^2\right)
\leq  \delta^2.
 \end{eqnarray*}
 	 Hence all  conditions of  Theorem 2.11.11 in \cite{VW} are verified. This implies the ASE of  
 $V_F(y)^{-1/2}  \frac{1}{\sqrt{N}}\sum_{i}d_{\vgamma, i}(y)$.

 \end{proof}

  \bibliographystyle{ims}
\bibliography{liaoBib_newest}

\end{document}